\documentclass[twocolumn,pre,aps,showpacs,showkeys,amsmath]{revtex4}
\usepackage{graphicx}
\usepackage{adjustbox}
\usepackage{multirow}

\begin{document}

\title{Influence of Various Temporal Recoding on Pavlovian Eyeblink Conditioning in The Cerebellum}
\author{Sang-Yoon Kim}
\email{sykim@icn.re.kr}
\author{Woochang Lim}
\email{wclim@icn.re.kr}
\affiliation{Institute for Computational Neuroscience and Department of Science Education, Daegu National University of Education, Daegu 42411, Korea}

\begin{abstract}
We consider the Pavlovian eyeblink conditioning (EBC) via repeated presentation of paired conditioned stimulus (tone) and unconditioned stimulus (airpuff).
The influence of various temporal recoding of granule cells on the EBC is investigated in a cerebellar network where the connection probability $p_c$ from Golgi to granule cells is changed. In an optimal case of $p_c^*~(=0.029)$, individual granule cells show various well- and ill-matched firing patterns relative to the unconditioned stimulus. Then, these variously-recoded signals are fed into the Purkinje cells (PCs) through parallel-fibers (PFs), and
the instructor climbing-fiber (CF) signals from the inferior olive depress them effectively.
In the case of well-matched PF-PC synapses, their synaptic weights are strongly depressed through strong long-term depression (LTD). On the other hand, practically no LTD occurs for the ill-matched PF-PC synapses. This type of ``effective'' depression at the PF-PC synapses coordinates firings of PCs effectively, which then make effective inhibitory coordination on cerebellar nucleus neuron [which elicits conditioned response (CR; eyeblink)]. When the learning trial passes a threshold, acquisition of CR begins. In this case, the timing degree ${\cal T}_d$ of CR becomes good due to presence of the ill-matched firing group which plays a role of protection barrier for the timing. With further increase in the trial, strength $\cal S$ of CR (corresponding to the amplitude of eyelid closure) increases due to strong LTD in the well-matched firing group, while its timing degree ${\cal T}_d$ decreases. In this way, the well- and the ill-matched firing groups play their own roles for the strength and the timing of CR, respectively. Thus, with increasing the learning trial, the (overall) learning efficiency degree ${\cal L}_e$ (taking into consideration both timing and strength of CR) for the CR is increased, and eventually it becomes saturated. By changing $p_c$ from $p_c^*$, we also investigate the influence of various temporal recoding on the EBC. It is thus found that, the more various in temporal recoding, the more effective in  learning for the Pavlovian EBC.
\end{abstract}


\maketitle

\section{Introduction}
\label{sec:INT}
The cerebellum plays a crucial role in precise temporal and spatial motor control for coordination of voluntary movements (e.g., locomotion, balance, and posture),  leading to smooth and balanced body movement \cite{Ito1,Ito2,Ito3}. In addition, it also participate in higher cognitive functions (e.g., attention, language, and speech) \cite{Ito2,Ito3}. The purpose of cerebellar motor learning is to carry out precise timing (associated with temporal information of movement) and gain (related to spatial information of movement) control for movements \cite{Yama1}. Experimental studies on timing and gain control for eye movements have been done in the two kinds of paradigms; (1) timing control for the Pavlovian eyeblink conditioning (EBC) \cite{EB1,EB2,EB3,EB4,EB5,EB6,EB7} and (2) gain control for the vestibulo-ocular reflex and the optokinetic response \cite{Ito1,VOR1}.

Here, we are interested in the Pavlovian EBC [see Fig.~1(a)] which is a representative example for the classical conditioning \cite{Wagner}. An animal (e.g., rabbit, rat, or mouse) receives repeated presentations of paired  conditioned stimulus (CS; tone) and (eyeblink-eliciting) unconditioned stimulus (US; airpuff). When the training trial passes a threshold, the animal acquires the ability to elicit eyelid conditioned response (CR; acquisition of learned eyeblink) via learning representation of the time passage between the onsets of CS and US
(i.e., the animal becomes conditioned to closes its eye in response to the tone CS with a time delay equal to the inter-stimulus interval (ISI) between the CS and the US onsets). The CRs exhibit two distinct features: (1) gradual acquisition of CR (i.e., CRs are acquired gradually over many training trials of repeated CS-US pairings) \cite{GA1,GA2,GA3,GA4,GA5} and (2) adaptive timing of CR (i.e., CRs are well timed such that the time of peak eyelid closure matches well the ISI between the onsets of CS and US) \cite{T1,T2,T3,T4,T5}. Experimental works on EBC have been done in several animal species such as humans \cite{Men}, monkeys \cite{GA2}, dogs \cite{GA1}, ferrets \cite{GA5}, rabbits \cite{GA3}, rats \cite{GA4}, and mice \cite{Mice}. Particularly, since a series of groundbreaking experiments in rabbits \cite{Rabbit1,Rabbit2}, EBC in restrained rabbits has served as a good model for motor learning.

Marr \cite{Marr} and, later, Albus \cite{Albus} formulated their seminal theory for cerebellar motor learning on the basis of its structure.
Particularly, they paid attention to the recurrent network between the granule (GR) and the Golgi (GO) cells as a device of representation of spatial information (i.e., spatial coding). The input spatial patterns, conveyed via the mossy fibers (MFs), become more sparse and dissimilar to each other
(i.e., pattern separation occurs) through recoding procedure in the granular layer composed of GR and GO cells. These recoded inputs are conveyed into the Purkinje cells (PCs) through the parallel fibers (PFs) (corresponding to the axons of GR cells). In addition to the PF recoded signals, the PCs also receive the error-teaching signals through the climbing-fiber (CF) from the inferior olive (IO). We assume that the PF-PC synapses are the only synapses where motor learning takes place. Thus, synaptic plasticity (i.e., potentiation or depression of synaptic strengths) may occur at PF-PC synapses.
It is assumed by Marr \cite{Marr} that a Hebbian type of long-term potentiation (LTP) occurs at the PF-PC synapses when both the PF and the CF signals are conjunctively excited \cite{Hebb,Br}. In opposition to Marr's learning via LTP, it is assumed by Albus \cite{Albus} that an anti-Hebbian type of long-term depression (LTD) takes place in the case of conjunctive excitations of both the PF and the CF signals. In the case of Albus' learning via LTD, PCs learn when to stop their inhibition (i.e. when to disinhibit) rather than when to fire, because their firing activities become reduced. Several later experimental works have provided the support for the Albus' learning via LTD \cite{Ito4,Ito5,Sakurai}. Thus, LTD became established as a kind of synaptic plasticity for motor learning in the cerebellum \cite{Ito6,Ito7,Ito8,Ito9}.

A number of computational works for the EBC have been done. Several artificial models have been proposed
for the time-passage representation (i.e., time coding) in the cerebellum \cite{AN1,AN2,AN3,AN4,AN5,AN6}. However, these artificial models lacked biological plausibility. A realistic cerebellar model, based on many biological properties, has been built by focusing on the recurrent loop between the GR and the GO cells in the granular layer as a time-coding device \cite{BN1}. Then, the realistic model generated a temporal code based on the population of active GR cells, and
also, it was extended to reproduce the experimental results of the Pavlovian EBC \cite{BN2,BN3}. However, the computational mechanism to generate such a temporal code was unclear mainly due to complexity of the realistic model. To understand the computational mechanism for the time coding, a rate-coding model was developed in a simple recurrent inhibitory network, and its dynamics was analyzed in both the numerical and analytical way \cite{BN4}. This rate-coding model generated a non-recurrent sequence of active neurons, corresponding to representation of a time-passage. Due to randomness in the recurrent connections, individual neurons exhibited random repetition of transitions between the active (bursting) and the inactive (silent) states which were persistent long-lasting ones.
However, this rate-coding model is free of actual time scales. A spiking neural network model (with actual time scales) was built to examine representation of time passage in the cerebellum \cite{BN5}, and a large-scale computational simulation was also performed to reproduce some features of the EBC in the
experiments.

However, the influence of various temporal recoding of GR cells on the Pavlovian EBC remains to be necessary for more clarification in several dynamical aspects, although much understanding on the representation of time passage in the cerebellum was achieved in previous computational works.
As in the case of spatial coding for the optokinetic response \cite{OKR}, variously-recoded PF signals are dynamically classified
for clear understanding of their matching with the instructor CF signals.
Then, the dynamical classification of various firing patterns of GR cells leads us  to clearly understand synaptic plasticity at PF-PC synapses and the following learning procedure. Consequently, we expect understanding on the rate of acquisition and the timing and strength of CR to be so much enhanced.

To this purpose, we employ a cerebellar ring network for the Pavlovian EBC, the basic framework of which was developed in the case of optokinetic response
\cite{OKR}. In such a cerebellar ring network, we vary the connection probability $p_c$ from GO to GR cells and make a dynamical classification of
various firing patterns of GR cells. GR cells in the whole population are divided into GR clusters. These GR clusters show various well- and ill-matched firing patterns with respect to the US (i.e., airpuff unconditioned stimulus) signal. Each firing pattern may be characterized in terms of the matching index
between the firing pattern and the US. For quantifying the degree of various temporal recoding of GR cells, we introduce the variety degree $\cal{V}$, given by the relative standard deviation in the distribution of matching indices of the firing patterns in the GR clusters, similar to the case of spatial coding for the optokinetic response \cite{OKR}. We pay main attention to an optimal case of $p_c^*=0.029$ where the firing patterns of GR clusters are the most various. In the optimal case, ${\cal{V}^*} \simeq 1.842$ which is a quantitative measure for various temporal recoding of GR cells. Dynamical origin of these various firing  patterns of GR cells is also investigated. It is thus found that, various total synaptic inputs (including both the excitatory inputs via MFs and the inhibitory inputs from the pre-synaptic GO cells) into the GR clusters lead to generation of various firing patterns (i.e. outputs) in the GR clusters.

Based on dynamical classification of various firing patterns in the GR clusters, we employ a refined rule for synaptic plasticity (developed from the experimental result in \cite{Safo}), and investigate intensively the influence of various temporal recoding of GR cells on synaptic plasticity at PF-PC synapses and subsequent learning process. PCs (corresponding to the cerebellar output) receive both the variously-recoded PF signals from the GR cells and the error-teaching
CF signals from the IO neuron; the CF signals are well-matched with the US signal. In this case, CF and PF signals may be considered as ``instructors'' and ``students,'' respectively. Then, well-matched PF student signals are strongly depressed via strong LTD by the CF instructor signals. In contrast, practically no LTD occurs for the ill-matched PF student signals because most of them have no associations with the US signals which are strongly localized in the middle of each trial. In this way, the student PF signals are effectively depressed by the instructor CF signals.

During learning trials with repeated presentation of CS-US pairs, the ``effective'' depression at PF-PC synapses coordinates activations of PCs effectively, which then make effective inhibitory coordination on the cerebellar nucleus (CN) neuron [which elicits CR (i.e., learned eyeblink)]. In the optimal case of $p_c^*=0.029,$ when passing the 141th threshold trial, acquisition of CR starts. In this optimal case, the timing degree ${\cal T}_d$ of the CR becomes good because of presence of ill-matched firing group which plays a role of protection barrier for the timing. With further increase in the trial, strength $\cal S$ of CR [denoting the amplitude of eyelid closure (measured by the electromyography (EMG))] increases due to strong LTD in the well-matched firing group, while its timing degree ${\cal T}_d$ decreases. In this way, the well- and the ill-matched firing groups play their own roles for the strength and the timing of CR, respectively. Thus, the (overall) learning efficiency degree ${\cal L}_e$ (considering both timing and strength of CR) for the CR increases with the training trial, and eventually it gets saturated.

Saturation in the learning procedure may be clearly shown in the inferior olive (IO) system. During the learning trials, the IO neuron receives both the excitatory US signal and the inhibitory input from the CN neuron. In this case, the learning progress degree ${\cal{L}}_p$ is given by the ratio of the trial-averaged inhibitory input from the CN neuron to the magnitude of the trial-averaged excitatory input of the US signal. With increasing trial from the threshold, the trial-averaged inhibitory input (from the CN neuron) is increased, and approaches the constant trial-averaged excitatory input (through the US signal). Thus, at about the 250th trial, the learning progress degree becomes saturated at ${\cal{L}}_p =1$. In this saturated case, the trial-averaged excitatory and inhibitory inputs to the IO neuron get balanced, and we obtain the saturated learning efficiency degree ${\cal{L}}_e^*~(\simeq 11.19)$ of the CR in the CN.

Finally, we vary $p_c$ from $p_c^*$, and investigate the influence of various temporal recoding of GR cells on the Pavlovian EBC.
The distributions of both the variety degree $\cal V$ of firing patterns of GR cells and the saturated learning efficiency degree ${\cal L}_e^*$ of CR versus $p_c$ are found to form bell-shaped curves with peaks at $p_c^*$. Moreover, both $\cal V$ and ${\cal L}_e^*$ have a strong correlation.
As a result, the more various in temporal recoding of GR cells, the more effective in learning for the Pavlovian EBC.

This paper is organized as follows. In Sec.~\ref{sec:CRN}, a cerebellar ring network for the Pavlovian EBC is introduced. The governing equations for the population dynamics in the ring network are also given, together with a refined rule for the synaptic plasticity at the PF-PC synapses.
Then, in the main Sec.~\ref{sec:MS}, we investigate the influence of various temporal recoding of GR cells on learning for the Pavlovian EBC
by varying $p_c$. Finally, summary and discussion are given in Sec.~\ref{sec:SUM}.

\begin{figure}
\includegraphics[width=0.85\columnwidth]{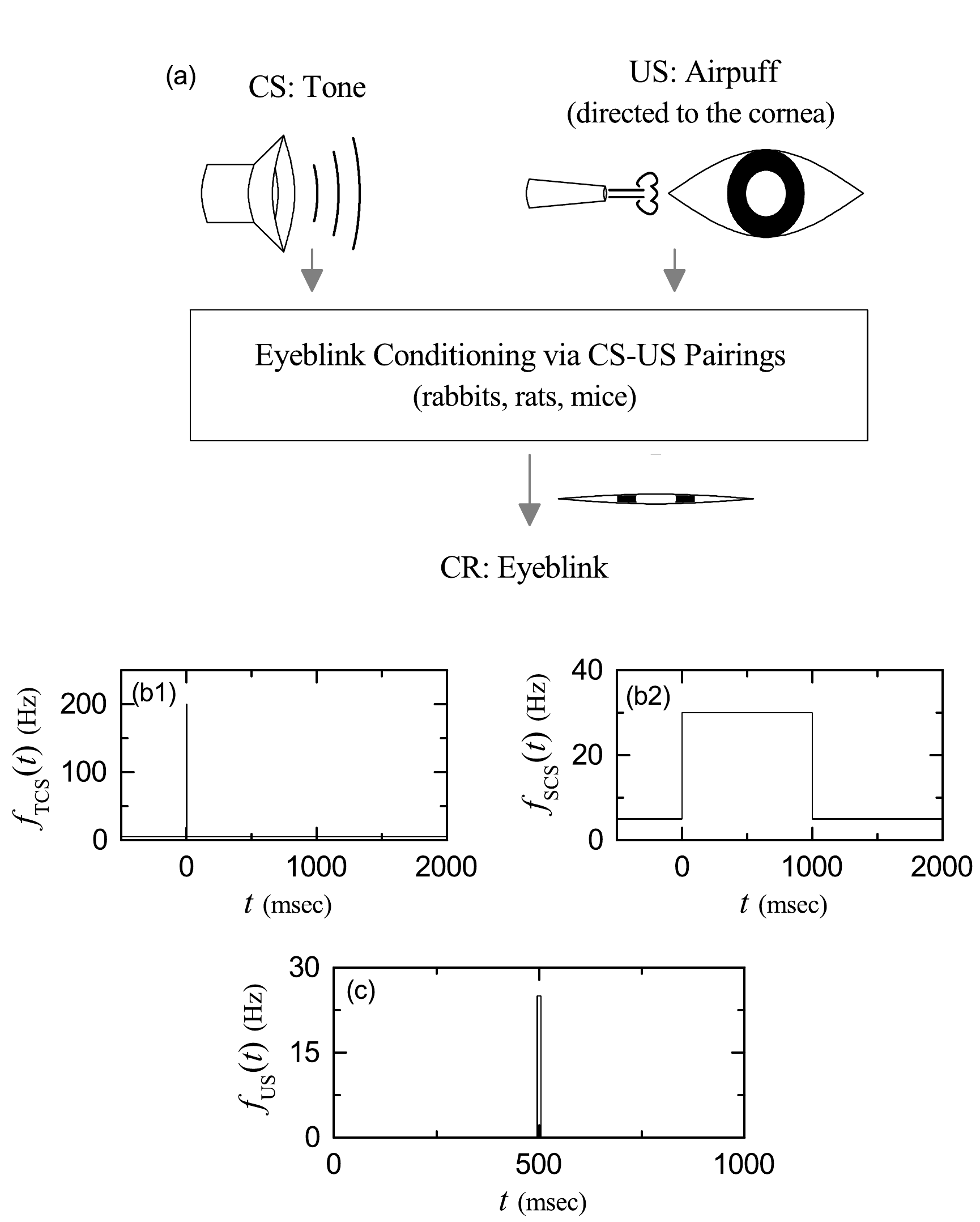}
\caption{Pavlovian eyeblink conditioning (EBC). (a) Eyelid conditioned response (CR) (i.e., learned eyeblink) via repeated presentation of paired CS (conditioned stimulus) and US (unconditioned stimulus). Firing rates of (b1) transient conditioned stimulus (TCS) for resetting and (b2) sustained
conditioned stimulus (SCS) (EBC signal). (c) Firing rate of transient unconditioned stimulus (US) for timing (eliciting unconditioned response).
}
\label{fig:EBC}
\end{figure}

\section{Cerebellar Ring Network for The Pavlovian Eyeblink Conditioning}
\label{sec:CRN}
In this section, we describe our cerebellar ring network for the Pavlovian EBC.
The basic framework of such a ring network was developed in the case of optokinetic response \cite{OKR}.
The ring network for the EBC is essentially the same as that for the optokinetic response, except for the number of
glomeruli in each GR cluster which is 4 and 2 for the EBC and the optokinetic response, respectively.
Also, the external input signals (i.e., the context MF signal and the desired IO signal) for the EBC are different from
those for the optokinetic response. For the sake of completeness, we also include a detailed explanation on the cerebellar ring network within this paper.

\begin{figure*}
\includegraphics[width=1.45\columnwidth]{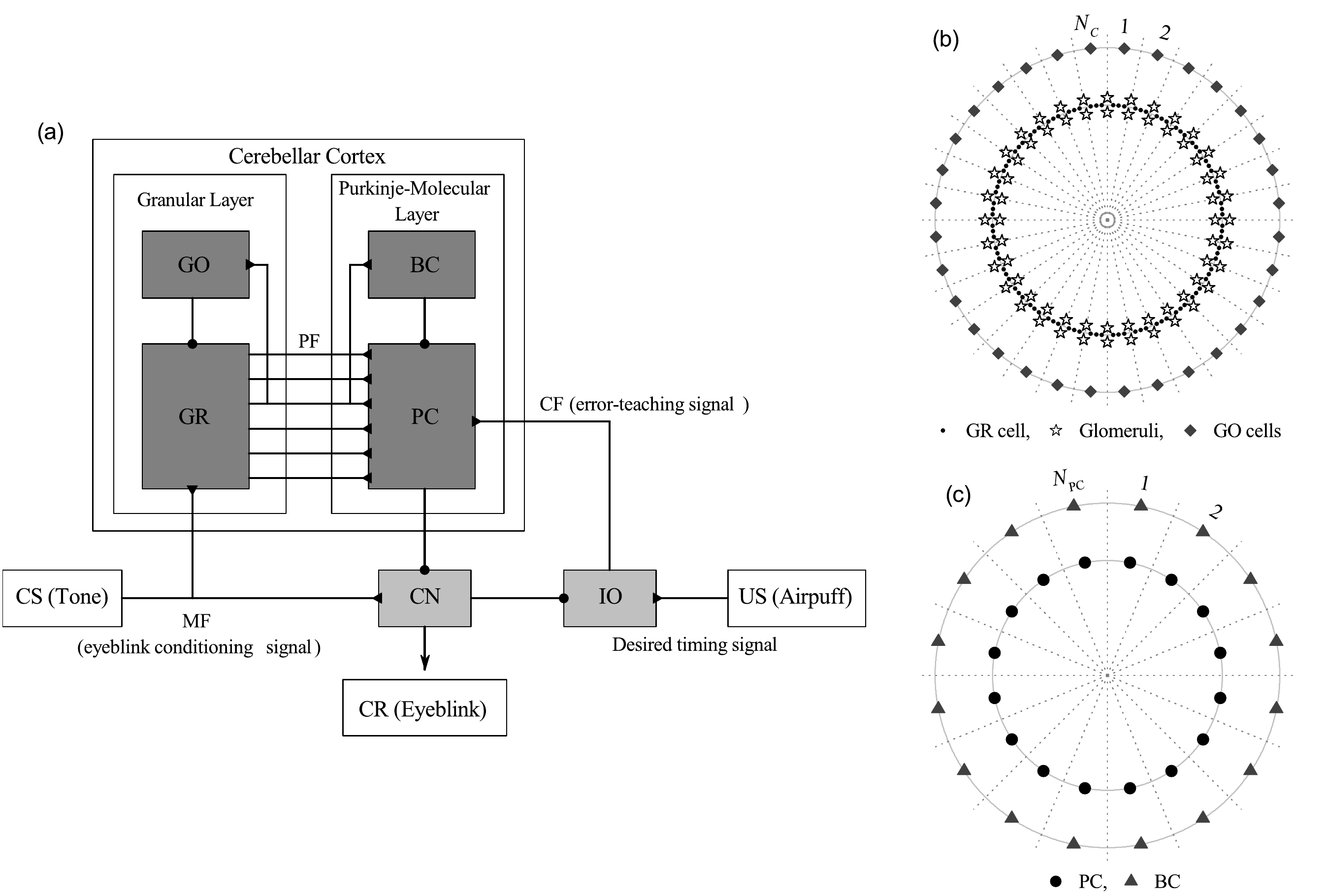}
\caption{Cerebellar ring network for the EBC. (a) Box diagram for the cerebellar network. Lines with triangles and circles represent excitatory and inhibitory synapses, respectively. GR (granule cell), GO (Golgi cell), and PF (parallel fiber) in the granular layer, PC (Purkinje cell) and BC (basket cell) in the Purkinje-molecular layer, and other parts for CN (cerebellar nuclei), IO(inferior olive), MF (mossy fiber), and CF (climbing fiber). (b) Schematic diagram for granular-layer ring network with concentric inner GR and outer GO rings. Numbers represent granular layer zones (bounded by dotted lines) for $N_C=32$. In each $I$th zone ($I=1,\cdots,N_C$), there exists the $I$th GR cluster on the inner GR ring. Each GR cluster consists of GR cells (solid circles), and it is bounded by 4 glomeruli (stars). On the outer GO ring in the $I$th zone, there exists the $I$th GO cell (diamonds). (c) Schematic diagram for Purkinje-molecular-layer ring network with concentric inner PC and outer BC rings. Numbers represent the Purkinje-molecular-layer zones (bounded by dotted lines) for $N_{\rm PC} =16$. In each $J$th zone, there exist the $J$th PC (solid circle) on the inner PC ring and the $J$th BC (solid triangle) on the outer BC ring.
}
\label{fig:RN}
\end{figure*}

\subsection{Conditioned Stimulus and Unconditioned Stimulus}
\label{subsec:Signals}
Figure \ref{fig:EBC}(a) shows the Pavlovian EBC.
During the training trials, repeated presentations of paired  tone CS and delayed airpuff US are made to an animal (e.g., rabbit, rat, or mouse).
As the training trial passes a threshold, the animal acquires the ability to elicit eyelid CR (i.e., acquisition of learned eyeblink)
through learning representation of the time passage between the CS and the US onsets. Accordingly, the animal gets conditioned to closes its eye
in response to the tone CS with a time delay equal to the ISI between the onsets of CS and US.

In this subsection, we give explanations on the two external input signals.
We first consider the CS for the EBC signal.
When the CS is a tone, the pontine nucleus in the pons receives the auditory information, and then it sends the ``context'' signal for the EBC via MFs to both the GR cells and the CN neuron. There are a transient CS for resetting and a sustained CS (representing the tone) \cite{BN5}.
Each step ($0 < t < 2000$ msec) for EBC learning consists of the trial stage ($0 < t < 1000$ msec) and the break stage ($1000 < t < 2000$ msec); $t$ denotes the time. In the trial stage, the transient CS is modeled in terms of strong and brief Poisson spike trains of 200 Hz for $0 < t < 5$ msec and the subsequent
background Poisson spike trains of 5 Hz for $5 < t < 1000$ msec. On the other hand, the sustained CS is modeled in terms of Poisson spike trains of 30 Hz
for $0 < t < 1000$ msec. In the following break stage, the CS is modeled in terms of the background Poisson spike trains of 5 Hz for $1000 < t < 2000$ msec.
The firing rates $f_{\rm TCS}(t)$ and $f_{\rm SCS}(t)$ of the transient CS and the sustained CS are shown in Figs.~\ref{fig:EBC}(b1) and \ref{fig:EBC}(b2), respectively. These figures also show the preparatory step for $-500 < t < 0$ msec where the CS is modeled in terms of the background Poisson spike trains of 5 Hz;
this preparatory step precedes just the 1st step for the EBC learning.

Next, we consider the US for the desired timing signal.
When an airpuff US is delivered to the cornea of an eye, sensory information is carried to the sensory trigeminal nucleus (which  extends through the whole of midbrain, pons, and medulla and into the high cervical spinal cord). Then, the trigeminal nucleus also sends the desired timing signal to the IO.
In the trial stage ($0<t<1000$ msec), the US (eliciting unconditioned response) is modeled in terms of the strong and brief Poisson spike trains of 25 Hz for a short interval in the middle of the trial stage, $t^* - \Delta t < t < t^* + \Delta t$ ($t^*=500$ msec and $\Delta t=$ 5 msec) \cite{BN5}.
The firing rate $f_{\rm US}(t)$ of the US is shown in Fig.~\ref{fig:EBC}(c).

\subsection{Framework for Cerebellar Ring Network for The Pavlovian EBC}
\label{subsec:ACRN}
A cerebellar ring network was introduced in the case of optokinetic response \cite{OKR}.
As in the famous small-world ring network \cite{SWN1,SWN2}, a one-dimensional ring network with a simple architecture was developed, which is in contrast to the
two-dimensional square-lattice network \cite{Yama1,BN5}. This kind of ring network has advantage for computational and analytical efficiency, and its
visual representation may also be easily made.

Here, we employ such a cerebellar ring network for the Pavlovian EBC.
Figure \ref{fig:RN}(a) shows the box diagram for the cerebellar network. The granular layer, corresponding to the input layer of the cerebellar cortex, is composed of the excitatory GR cells and the inhibitory GO cells. On the other hand, the Purkinje-molecular layer, corresponding to the output layer of the cerebellar cortex, consist of the inhibitory PCs and the inhibitory BCs (basket cells). The MF context signals for the EBC are fed from the pontine nucleus in the pons to the GR cells;
each GR cell receives two transient and two sustained CS signals via four MFs (i.e., two pairs of transient and sustained CS signals are fed into each GR cell).
Various temporal recoding is made in the granular layer via inhibitory coordination of GO cells on GR cells. Then, these various-recoded outputs are fed via PFs to the PCs and the BCs in the Purkinje-molecular layer.

The PCs receive another excitatory error-teaching CF signals from the IO, along with the inhibitory inputs from the BCs.
Then, depending on the type of PF signals (i.e., well- or ill-matched PF signals), various PF (student) signals are effectively depressed by the error-teaching (instructor) CF signals. Such ``effective'' depression at the PF-PC synapses coordinates firings of PCs effectively, which then exert effective inhibitory coordination on the CN neuron. The CN neuron also receives two excitatory signals; one transient and one sustained CS signals via MFs.
In the earlier trials, the CN neuron can not fire, due to strong inhibition from the PCs. As the learning trial passes a threshold, the CN neuron starts firing, and then it exerts excitatory projections onto the eyeblink pre-motoneurons in the midbrain which then supply motor commands to eyeblink motoneurons. Thus, acquisition of CR begins (i.e., acquisition of learned eyeblink starts). This CN neuron also provides inhibitory inputs to the IO neuron which also receives the excitatory signals for the desired timing from the trigeminal nucleus. Then, the IO neuron supplies excitatory error-teaching CF signals to the PCs.

Figure \ref{fig:RN}(b) shows a schematic diagram for the granular-layer ring network with concentric inner GR and outer GO rings.
Numbers represent granular-layer zones (bounded by dotted lines); the numbers 1, 2, $\cdots$, and $N_C$ represent the 1st, the 2nd, $\cdots$, and the $N_C$th granular-layer zones, respectively. Thus, the total number of granular-layer zones is $N_C$; Fig.~2(b) shows an example for $N_C=32$.
In each $I$th zone ($I=1,\cdots, N_C$), there exists the $I$th GR cluster on the inner GR ring.
Each GR cluster consists of $N_{\rm GR}$ excitatory GR cells (solid circles). Then, location of each GR cell may be denoted by the two indices $(I,i)$ which represent the $i$th GR cell in the $I$th GR cluster, where $i=1,\cdots,N_{\rm GR}$. Here, we consider the case of $N_C=2^{10}$ and $N_{\rm GR}=50$, and thus
the total number of GR cells is 51,200. In this case, the $I$th zone covers the angular range of $(I-1)~ \theta_{\rm GR}^* < \theta < I~ \theta_{\rm GR}^*$
($\theta_{\rm GR}^* = 0.35^{\circ}$). On the outer GO ring in each $I$th zone, there exists the $I$th inhibitory GO cell (diamond), and thus the total number of GO cells is $N_C$.

We note that each GR cluster is bounded by 4 glomeruli (corresponding to the axon terminals of the MFs) (stars) at both boundaries of the GR cluster;
at each boundary, a pair of glomeruli (upper and lower ones) exist. This is in contrast to the case of optokinetic response where 2 glomeruli bound each GR cluster;
at each left or right boundary, only one glomerulus is located.
GR cells within each GR cluster share the same inhibitory and excitatory synaptic inputs through their dendrites which contact the four glomeruli at both ends of the GR cluster. Each glomerulus receives inhibitory inputs from nearby 81 (clockwise side: 41 and counter-clockwise side: 40) GO cells with a random connection probability $p_c~(=0.029)$. Hence, on average, about 2 GO cell axons innervate each glomerulus. Thus, each GR cell receives about 9 inhibitory  inputs through 4 dendrites which synaptically contact the glomeruli at both boundaries. In this way, each GR cell in the GR cluster shares the same inhibitory synaptic inputs from nearby GO cells through the intermediate glomeruli at both ends.

In addition, each GR cell shares the same four excitatory inputs via the four glomeruli at both boundaries, because a glomerulus
receives an excitatory MF input. We note that transient CS signals are supplied via the two upper glomeruli, while sustained CS signals are fed through
the two lower glomeruli. Here, we take into consideration stochastic variability of synaptic transmission from a glomerulus to GR cells,
and supply independent Poisson spike trains with the same firing rate to each GR cell for the excitatory MF signals.
In this GR-GO feedback system, each GO cell receives excitatory synaptic inputs through PFs from GR cells in the nearby 49 (central side: 1, clockwise side: 24 and counter-clockwise side: 24) GR clusters with a random connection probability $0.1$. Hence, 245 PFs (i.e. GR cell axons) innervate a GO cell.

Figure \ref{fig:RN}(c) shows a schematic diagram for the Purkinje-molecular-layer ring network with concentric inner PC and outer BC rings.
Numbers represent the Purkinje-molecular-layer zones (bounded by dotted lines). In each $J$th zone ($J=1,\cdots, N_{\rm PC}$), there exist the $J$th PC (solid circles) on the inner PC ring and the $J$th BC (solid triangles) on the outer BC ring. Here, we consider the case of $N_{\rm PC}=16,$ and thus the total numbers of PC and BC are 16, respectively. In this case, each $J$th ($J=1,\cdots,N_{\rm PC}$) zone covers the angular range of $(J-1)~ \theta_{\rm PC}^* < \theta < J~ \theta_{\rm PC}^*,$ where $\theta_{\rm PC}^* \simeq 22.5^{\circ}$ (corresponding to about 64 zones in the granular-layer ring network).
We note that variously-recoded PFs innervate PCs and BCs. Each PC (BC) in the $J$th Purkinje-molecular-layer zone receives excitatory synaptic inputs via PFs from
all the GR cells in the 288 GR clusters (clockwise side: 144 and counter-clockwise side: 144 when starting from the angle $\theta = (J-1)~ \theta_{\rm PC}^*$ in the granular-layer ring network). Thus, each PC (BC) is synaptically connected via PFs to the 14,400 GR cells (which corresponds to about 28 $\%$ of the total GR cells).
In addition to the PF signals, each PC also receives inhibitory inputs from nearby 3 BCs (central side: 1, clockwise side: 1 and counter-clockwise side: 1) and
excitatory error-teaching CF signal from the IO.

Here, for simplicity, we consider just one CN neuron and one IO neuron. Both excitatory inputs (corresponding to one transient and one sustained CS signals) via 2 MFs and inhibitory inputs from all the 16 PCs are fed into the CN neuron. Then, the CN neuron provides excitatory input
to the eyeblink pre-motoneurons in the midbrain and also supplies inhibitory input to the IO neuron.
One additional excitatory desired timing signal from the trigeminal nucleus is also fed into the IO neuron. Then, through integration of both excitatory and inhibitory inputs, the IO neuron provides excitatory error-teaching CF signals to the PCs.

\subsection{Elements of The Cerebellar Ring Network}
\label{subsec:LIF}
As in the case of optokinetic response \cite{OKR}, we choose leaky integrate-and-fire (LIF) neuron models as elements of the cerebellar ring network \cite{LIF}. Here, the LIF neuron models incorporate additional afterhyperpolarization (AHP) currents that determine refractory periods. This LIF neuron model is one of the simplest spiking neuron models. Because of its simplicity, it may be easily analyzed and simulated. Hence, it has been very popularly employed as a neuron model.

Dynamics of states of individual neurons in the $X$ population are governed by the following equations:
\begin{equation}
C_{X} \frac{dv_{i}^{(X)}}{dt} = -I_{L,i}^{(X)} - I_{AHP,i}^{(X)} + I_{ext}^{(X)} - I_{syn,i}^{(X)}, \;\;\; i=1, \cdots, N_{X},
\label{eq:GE}
\end{equation}
where $N_X$ is the total number of neurons in the $X$ population, $X=$ GR and GO in the granular layer, $X=$ PC and BC in the Purkinje-molecular layer, and in the other parts $X=$ CN and IO.
The state of the $i$th neuron in the $X$ population at a time $t$ (msec) is characterized by its membrane potential $v_i^{(X)}$ (mV), and
$C_{X}$ (pF) denotes the membrane capacitance of the cells in the $X$ population. The time-evolution of $v_i^{(X)}(t)$ is governed by 4 types of currents (pA) into the $i$th neuron in the $X$ population; the leakage current $I_{L,i}^{(X)}$, the AHP current $I_{AHP,i}^{(X)}$, the external constant current $I_{ext}^{(X)}$ (independent of $i$), and the synaptic current $I_{syn,i}^{(X)}$.

We consider a single LIF neuron model [without the AHP current and the synaptic current in Eq.~(\ref{eq:GE})] which describes a simple
parallel resistor-capacitor circuit. Here, the leakage term is due to the resistor and the integration of the external current is due to the capacitor
which is in parallel to the resistor. Thus, in Eq.~(\ref{eq:GE}), the 1st type of leakage current $I_{L,i}^{(X)}$ for the $i$th neuron in the $X$ population
is given by:
\begin{equation}
I_{L,i}^{(X)} = g_{L}^{(X)} (v_{i}^{(X)} - V_{L}^{(X)}),
\label{eq:Leakage}
\end{equation}
where $g_L^{(X)}$ and $V_L^{(X)}$ are conductance (nS) and reversal potential for the leakage current, respectively.

The $i$th neuron fires a spike when its membrane potential $v_i^{(X)}$ reaches a threshold $v_{th}^{(X)}$ at a time $t_{f,i}^{(X)}$.
Then,the 2nd type of AHP current $I_{AHP,i}^{(X)}$ follows after firing (i.e., $t \geq t_{f,i}^{(X)}$):
\begin{equation}
I_{AHP,i}^{(X)} = g_{AHP}^{(X)}(t) ~(v_{i}^{(X)} - V_{AHP}^{(X)})~~~{\rm ~for~} \; t \ge t_{f,i}^{(X)},
\label{eq:AHP1}
\end{equation}
where $V_{AHP}^{(X)}$ is the reversal potential for the AHP current.
The conductance $g_{AHP}^{(X)}(t)$ is given by an exponential-decay function:
\begin{equation}
g_{AHP}^{(X)}(t) = \bar{g}_{AHP}^{(X)}~  e^{-(t-t_{f,i}^{(X)})/\tau_{AHP}^{(X)}} ,
\label{eq:AHP2}
\end{equation}
where $\bar{g}_{AHP}^{(X)}$ and $\tau_{AHP}^{(X)}$ are the maximum conductance and the decay time constant for the AHP current.
As $\tau_{AHP}^{(X)}$ increases, the refractory period becomes longer.

The 3rd type of external constant current $I_{ext}^{(X)}$ for spontaneous firing is provided to only PCs
because of their high spontaneous firing rate \cite{PC1,PC2}. In Appendix, Table \ref{tab:SingleParm} shows the parameter values for the capacitance $C_X$, the leakage current $I_L^{(X)}$, the AHP current $I_{AHP}^{(X)}$, and the external constant current $I_{ext}^{(X)}$. These values are adopted from physiological data \cite{BN5}.

\subsection{Three Kinds of Synaptic Currents}
\label{subsec:SC}
Here, we are concerned about the 4th type of synaptic current $I_{syn,i}^{(X)}$ into the $i$th neuron in the $X$ population in Eq.~(\ref{eq:GE}).
It is composed of the following 3 kinds of synaptic currents:
\begin{equation}
I_{syn,i}^{(X)} = I_{{\rm AMPA},i}^{(X,Y)} + I_{{\rm NMDA},i}^{(X,Y)} + I_{{\rm GABA},i}^{(X,Z)}.
\label{eq:ISyn1}
\end{equation}
Here, $I_{{\rm AMPA},i}^{(X,Y)}$ and $I_{{\rm NMDA},i}^{(X,Y)}$ are the excitatory AMPA ($\alpha$-amino-3-hydroxy-5-methyl-4-isoxazolepropionic acid) receptor-mediated and NMDA ($N$-methyl-$D$-aspartate) receptor-mediated currents from the pre-synaptic source $Y$ population to the post-synaptic $i$th neuron in the target $X$ population. In contrast, $I_{{\rm GABA},i}^{(X,Z)}$ is the inhibitory $\rm GABA_A$ ($\gamma$-aminobutyric acid type A) receptor-mediated current
from the pre-synaptic source $Z$ population to the post-synaptic $i$th neuron in the target $X$ population.

As in the case of the AHP current, the $R$ (= AMPA, NMDA, or GABA) receptor-mediated synaptic current $I_{R,i}^{(T,S)}$ from the pre-synaptic source $S$ population to the $i$th post-synaptic neuron in the target $T$ population is given by:
\begin{equation}
I_{R,i}^{(T,S)} = g_{R,i}^{(T,S)}(t)~(v_{i}^{(T)} - V_{R}^{(S)}),
\label{eq:ISyn2}
\end{equation}
where $g_{(R,i)}^{(T,S)}(t)$ and $V_R^{(S)}$ are synaptic conductance and synaptic reversal potential
(determined by the type of the pre-synaptic source $S$ population), respectively.
We obtain the synaptic conductance $g_{R,i}^{(T,S)}(t)$ from:
\begin{equation}
g_{R,i}^{(T,S)}(t) = \bar{g}_{R}^{(T)} \sum_{j=1}^{N_S} J_{ij}^{(T,S)}~ w_{ij}^{(T,S)} ~ s_{j}^{(T,S)}(t),
\label{eq:ISyn3}
\end{equation}
where $\bar{g}_{R}^{(T)}$  and $J_{ij}^{(T,S)}$ are the maximum conductance and the synaptic weight of the synapse
from the $j$th pre-synaptic neuron in the source $S$ population to the $i$th post-synaptic neuron in the target $T$ population, respectively.
The inter-population synaptic connection from the source $S$ population (with $N_s$ neurons) to the target $T$ population is given in terms of the connection weight matrix $W^{(T,S)}$ ($=\{ w_{ij}^{(T,S)} \}$) where $w_{ij}^{(T,S)}=1$ if the $j$th neuron in the source $S$ population is pre-synaptic to the $i$th neuron
in the target $T$ population; otherwise $w_{ij}^{(T,S)}=0$.

The post-synaptic ion channels are opened because of the binding of neurotransmitters (emitted from the source $S$ population) to receptors in the target
$T$ population. The fraction of open ion channels at time $t$ is represented by $s^{(T,S)}$. The time course of $s_j^{(T,S)}(t)$ of the $j$th neuron
in the source $S$ population is given by a sum of exponential-decay functions $E_{R}^{(T,S)} (t - t_{f}^{(j)})$:
\begin{equation}
s_{j}^{(T,S)}(t) = \sum_{f=1}^{F_{j}^{(S)}} E_{R}^{(T,S)} (t - t_{f}^{(j)}),
\label{eq:ISyn4}
\end{equation}
where $t_f^{(j)}$ and $F_j^{(S)}$ are the $f$th spike time and the total number of spikes of the $j$th neuron in the source $S$ population, respectively.
The exponential-decay function $E_{R}^{(T,S)} (t)$ (which corresponds to contribution of a pre-synaptic spike occurring at $t=0$ in the absence of synaptic delay)
is given by:
\begin{subequations}
\begin{eqnarray}
E_{R}^{(T,S)}(t) &=& e^{-t/\tau_{R}^{(T)}} \Theta(t)~~~{\rm or} \label{eq:ISyn5a} \\
 &=& (A_{1} e^{-t/\tau_{R,1}^{(T)}} + A_{2} e^{-t/\tau_{R,2}^{(T)}}) \Theta(t), \label{eq:ISyn5b}
\end{eqnarray}
\end{subequations}
where $\Theta(t)$ is the Heaviside step function: $\Theta(t)=1$ for $t \geq 0$ and 0 for $t <0$.
Depending on the source and the target populations, $E_{R}^{(T,S)} (t)$ may be a type-1 single exponential-decay function of
Eq.~(\ref{eq:ISyn5a}) or a type-2 dual exponential-decay function of Eq.~(\ref{eq:ISyn5b}). In the type-1 case, there exists one synaptic decay time constant $\tau_R^{(T)}$ (determined by the receptor on the post-synaptic target $T$ population), while in the type-2 case, two synaptic decay time constants, $\tau_{R,1}^{(T)}$ and $\tau_{R,2}^{(T)}$ appear.
In most cases, the type-1 single exponential-decay function of Eq.~(\ref{eq:ISyn5a}) appears, except for the two synaptic currents $I_{\rm GABA}^{\rm ( GR,GO)}$
and $I_{\rm NMDA}^{\rm (GO,GR)}$.

In Appendix, Table \ref{tab:SynParm} shows the parameter values for the maximum conductance $\bar{g}_{R}^{(T)}$, the synaptic weight $J_{ij}^{(T,S)}$, the synaptic reversal potential $V_{R}^{(S)}$, the synaptic decay time constant $\tau_{R}^{(T)}$, and the amplitudes $A_1$ and $A_2$ for the type-2 exponential-decay function in the granular layer, the Purkinje-molecular layer, and the other parts for the CN and IO, respectively. These values are adopted from physiological data \cite{BN5}.

\subsection{Refined Rule for Synaptic Plasticity}
\label{subsec:SP}
As in \cite{OKR}, we employ a rule for synaptic plasticity, based on the experimental result in \cite{Safo}.
This rule is a refined one for the LTD in comparison with the rule used in \cite{Yama1,BN5}, the details of which will be explained below.

The coupling strength of the synapse from the pre-synaptic neuron $j$ in the source $S$ population to the post-synaptic neuron $i$ in the target $T$ population is $J_{ij}^{(T,S)}$. Initial synaptic strengths for $J_{ij}^{(T,S)}$ are given in Table \ref{tab:SynParm}.
In this work, we assume that learning occurs only at the PF-PC synapses. Hence, only the synaptic strengths $J_{ij}^{\rm (PC,PF)}$ of PF-PC synapses
may be modifiable (i.e., they are depressed or potentiated), while synaptic strengths of all the other synapses are static.
[Here, the index $j$ for the PFs corresponds to the two indices $(M,m)$ for GR cells representing the $m$th ($1 \leq m \leq 50$) cell in the $M$th
($1 \leq M \leq 2^{10}$) GR cluster.] Synaptic plasticity at PF-PC synapses have been so much studied in many experimental \cite{Ito4,Ito5,Sakurai,Ito6,SPExp1,SPExp2,SPExp3,SPExp4,SPExp5,Safo,SPExp6,SPExp7,SPExp8,SPExp9} and computational \cite{Albus,BN5,Yama1,SPCom1,SPCom2,SPCom3,SPCom4,SPCom5,SPCom6,SPCom7} works.

As the time $t$ is increased, synaptic strength $J_{ij}^{\rm (PC,PF)}(t)$ for each PF-PC synapse is updated with the following multiplicative rule (depending on states) \cite{Safo,OKR}:
\begin{equation}
J_{ij}^{\rm (PC,PF)}(t) \rightarrow J_{ij}^{\rm (PC,PF)}(t) + \Delta J_{ij}^{\rm (PC,PF)}(t),
\label{eq:SP}
\end{equation}
where
\begin{widetext}
\begin{eqnarray}
\Delta J_{ij}^{\rm (PC,PF)}(t) &=& \Delta{\rm LTD}_{ij}^{(1)} + \Delta{\rm LTD}_{ij}^{(2)} + \Delta{\rm LTP}_{ij}, \label{eq:DeltaJ} \\
\Delta{\rm LTD}_{ij}^{(1)} &=& - \delta_{LTD} \cdot J_{ij}^{\rm (PC,PF)}(t) \cdot CF_{i}(t) \cdot \sum_{\Delta t=0}^{\Delta t_{r}^{*}} \Delta J_{LTD}(\Delta t),
\label{eq:LTD1} \\
\Delta{\rm LTD}_{ij}^{(2)} &=& - \delta_{LTD} \cdot J_{ij}^{\rm (PC,PF)}(t) \cdot [ 1- CF_{i}(t) ] \cdot PF_{ij}(t) \cdot D_{i}(t) \cdot \sum_{\Delta t=0}^{\Delta t_{l}^{*}} \Delta J_{LTD} (\Delta t),
\label{eq:LTD2} \\
\Delta{\rm LTP}_{ij} &=&  \delta_{LTP} \cdot [J_{0}^{\rm (PC,PF)} - J_{ij}^{\rm (PC,PF)}(t)] \cdot [1-CF_{i}(t)] \cdot PF_{ij}(t) \cdot [1-D_{i}(t)].
\label{eq:LTP}
\end{eqnarray}
\end{widetext}
Here, $J_{0}^{\rm (PC,PF)}$ is the initial value (=0.006) for the synaptic strength of PF-PC synapses.
Synaptic modification (LTD or LTP) occurs, depending on the relative time difference $\Delta t$ [= $t_{\rm CF}$ (CF activation time) - $t_{\rm PF}$ (PF activation time)] between the spiking times of the error-teaching instructor CF and the variously-recoded student PF.
In Eqs.~(\ref{eq:LTD1})-(\ref{eq:LTP}), $CF_i(t)$ denotes a spike train of the CF signal coming into the $i$th PC.
When $CF_i(t)$ activates at a time $t$, $CF_i(t)=1$; otherwise, $CF_i(t)=0$. This instructor CF activation gives rise to LTD at PF-PC synapses in conjunction with earlier ($\Delta t >0)$ student PF activations in the range of $t_{\rm CF} - \Delta t_r^* < t_{\rm PF} <t_{\rm CF}$ ($\Delta t_r^* \simeq 277.5$ msec), which corresponds to the major LTD in Eq.~(\ref{eq:LTD1}).

We next consider the case of $CF_i(t)=0$, which corresponds to Eqs.~(\ref{eq:LTD2}) and (\ref{eq:LTP}).
Here, $PF_{ij}(t)$ denotes a spike train of the PF signal from the $j$th pre-synaptic GR cell to the $i$th post-synaptic PC.
When $PF_{ij}(t)$ activates at time $t$, $PF_{ij}(t)=1$; otherwise, $PF_{ij}(t)=0$.
In the case of $PF_{ij}(t)=1$, PF firing may cause LTD or LTP, depending on the presence of earlier CF firings in an effective range.
If CF firings exist in the range of $t_{\rm PF} + \Delta t_l^* < t_{\rm CF} <t_{\rm PF}$ ($\Delta t_l^* \simeq -117.5$ msec), $D_i(t)=1$; otherwise $D_i(t)=0$.
When both $PF_{ij}(t)=1$ and $D_i(t)=1$, the PF activation causes another LTD at PF-PC synapses in conjunction with earlier ($\Delta t <0$) CF activations
[see Eq.~(\ref{eq:LTD2})]. The probability for occurrence of earlier CF firings within the effective range is very low because mean firing rates of the CF signals
(corresponding to output firings of individual IO neurons) are {\small $\sim$} 1.5 Hz \cite{IO1,IO2}. Hence, this 2nd type of LTD is a minor one.
In contrast, in the case of $D_i(t)=0$ (i.e., absence of earlier associated CF firings), LTP occurs because of the PF firing alone
[see Eq.~(\ref{eq:LTP})]. The update rate $\delta_{LTD}$ for LTD in Eqs.~(\ref{eq:LTD1}) and (\ref{eq:LTD2}) is 0.005, while the update rate $\delta_{LTP}$ for LTP
in Eqs.~(\ref{eq:LTP}) is 0.0005 (=$\delta_{LTD}/10$) \cite{Yama1}.

\begin{figure}
\includegraphics[width=0.7\columnwidth]{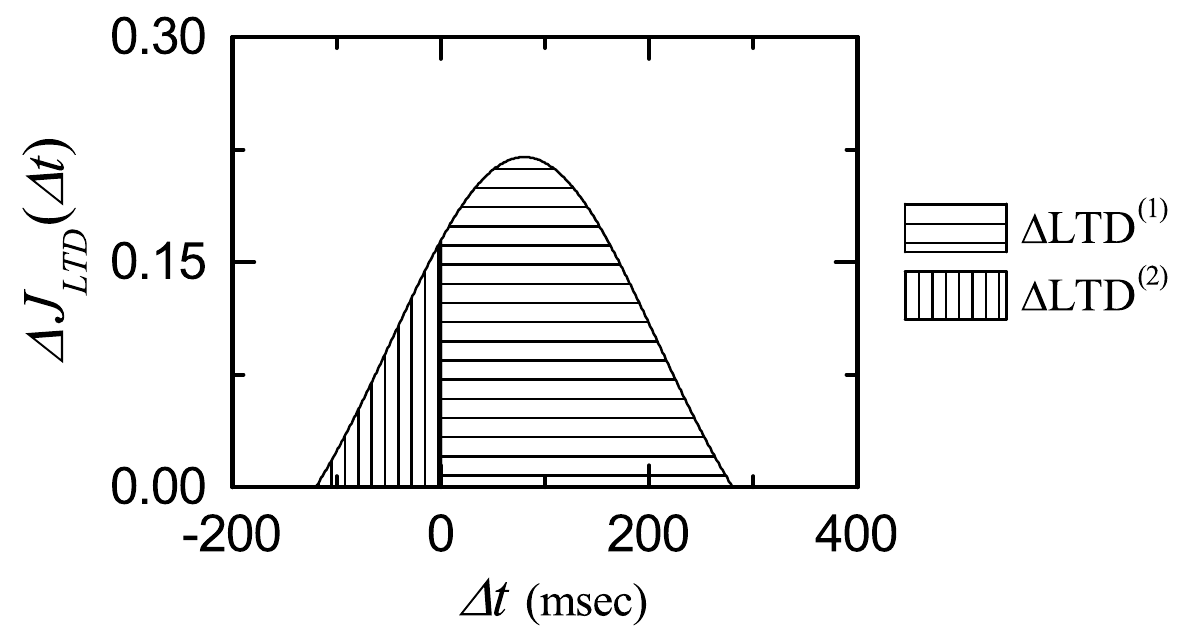}
\caption{Time window for the LTD at the PF-PC synapse. Plot of synaptic modification $\Delta J_{LTD}(\Delta t)$ for LTD versus $\Delta t.$
}\label{fig:TW}
\end{figure}

In the case of LTD in Eqs.~(\ref{eq:LTD1}) and (\ref{eq:LTD2}), the synaptic modification $\Delta J_{LTD} (\Delta t)$ changes depending on the relative time difference $\Delta t$ $(= t_{\rm CF} - t_{\rm PF}$).
We use the following time window for the synaptic modification $\Delta J_{LTD} (\Delta t)$ \cite{Safo,OKR}:
\begin{equation}
\Delta J_{LTD}(\Delta t) = A  + B \cdot e^{-(\Delta t -t_0)^2/\sigma^2},
\label{eq:TW}
\end{equation}
where $A=-0.12$, $B=0.4$, $t_0 = 80$, and $\sigma=180$.
Figure \ref{fig:TW} shows the time window for $\Delta J_{LTD} (\Delta t)$.
As shown well in Fig.~\ref{fig:TW}, LTD occurs in an effective range of $\Delta t_l^* < \Delta t < \Delta t_r^*$.
We note that a peak appears at $t_0=80$ msec, and hence peak LTD takes place when PF firing precedes CF firing by 80 msec.
A CF firing gives rise to LTD in association with earlier PF firings in the region hatched with horizontal lines ($0< \Delta t < \Delta t_r^*$), and
it also causes to another LTD in conjunction with later PF firings in the region hatched with vertical lines ($\Delta t_l^* < \Delta t <0$).
The effect of CF firing on earlier PF firings is much larger than that on later PF firings.
However, outside the effective range (i.e., $\Delta t > \Delta t_r^*$ or $< \Delta t_l^*$), PF firings alone results in occurrence of LTP,
because of absence of effectively associated CF firings.

Our refined rule for synaptic plasticity has the following advantages for the $\Delta {\rm LTD}$ in comparison with that in \cite{Yama1,BN5}.
Our rule is based on the experimental result in \cite{Safo}. In the presence of a CF firing, a major LTD  ($\Delta {\rm LTD}^{(1)}$) occurs in conjunction with earlier PF firings in the range of $t_{\rm CF} - \Delta t_r^* < t_{\rm PF} <t_{\rm CF}$ ($\Delta t_r^* \simeq 277.5$ msec), while a minor LTD ($\Delta {\rm LTD}^{(2)}$) takes place in conjunction with later PF firings in the range of $t_{\rm CF} < t_{\rm PF} <t_{\rm CF} - \Delta t_l^*$ ($\Delta t_l^* \simeq -117.5$ msec). The magnitude of LTD varies depending on $\Delta t$ (= $t_{\rm CF}$ - $t_{\rm PF}$); a peak LTD occurs when $\Delta t =80$ msec.
In contrast, the rule in \cite{Yama1,BN5} considers only the major LTD in association with earlier PF firings in the range of $t_{\rm CF} - 50 < t_{\rm PF} <t_{\rm CF}$, the magnitude of major LTD is equal, independently of $\Delta t$, and minor LTD in conjunction with later PF firings is not considered.
Outside the effective range of LTD, PF firings alone lead to LTP in both rules.

\subsection{Numerical Method}
\label{subsec:NM}
Numerical integration of the governing Eq.~(\ref{eq:GE}) for the time-evolution of states of individual neurons,
together with the update rule for synaptic plasticity of Eq.~(\ref{eq:SP}), is made by using the 2nd-order Runge-Kutta method
with the time step 1 msec. In each realization, we choose random initial points $v_i^{(X)}(0)$ for the $i$th neuron in the $X$ population
with uniform probability in the range of $v_i^{(X)}(0) \in (V_L^{(X)}-5.0, V_L^{(X)}+5.0)$; the values of $V_L^{(X)}$ are
given in Table \ref{tab:SingleParm}.

\section{Influence of Various Temporal Recoding in GR Clusters on Learning for The Pavlovian Eyeblink Conditioning}
\label{sec:MS}
In this section, we investigate the influence of various temporal recoding of GR cells on learning for the EBC
by changing the connection probability $p_c$ from the GO to the GR cells. We mainly focus on an optimal case
of $p_c^*=0.029$ where the firing patterns of GR clusters are the most various.
In this case, we first make dynamical classification of various firing patterns of the GR clusters. Next, we study the influence of various firing patterns
on the synaptic plasticity at the PF-PC synapses and the subsequent learning process in the PC-CN-IO system. Finally, we change
$p_c$ from the optimal value $p_c^*$, and investigate dependence of the variety degree $\cal{V}$ of firing patterns and the saturated learning efficiency degree ${\cal L}_e^*$ for the CR on $p_c$. Both $\cal{V}$ and ${\cal L}_e^*$ are found to form bell-shaped curves with peaks at $p_c^*$, and they have strong correlation
with the Pearson's coefficient $r\simeq 0.9982$ \cite{Pearson}. Consequently, the more various in temporal recoding in the GR clusters, the more effective in the
learning for the EBC.

\subsection{Collective Firing Activity in The Whole Population of GR Cells}
\label{subsec:WP}
Temporal recoding process is performed in the granular layer (corresponding to the input layer of the cerebellar cortex), composed of GR and GO cells (see Fig.~\ref{fig:RN}). GR cells (principal output cells in the granular layer) receive excitatory context signals for the EBC via the MFs [see Figs.~\ref{fig:EBC}(b1)
and \ref{fig:EBC}(b2)] and make various recoding of context signals through receiving effective inhibitory coordination of GO cells.
Thus, variously recoded signals are fed into the PCs (principal output cells in the cerebellar cortex) via PFs.

We first consider the firing activity in the whole population of GR cells for $p_c^*=0.029$.
Collective firing activity may be well visualized in the raster plot of spikes which is a collection of spike trains of individual neurons.
As a collective quantity showing whole-population firing behaviors, we use an instantaneous whole-population spike rate $R_{\rm GR}(t)$ which may be got from the raster plots of spikes \cite{RM,W_Review,Sparse1,Sparse2,Sparse3,Sparse4,Sparse5,Sparse6}.
To get a smooth instantaneous whole-population spike rate, we employ the kernel density estimation (kernel smoother) \cite{Kernel}.
Each spike in the raster plot is convoluted (or blurred) with a kernel function $K_{h}(t)$, and then a smooth estimate of instantaneous whole-population spike rate $R_{\rm GR}(t)$ is got by averaging the convoluted kernel function over all spikes of GR cells in the whole population:
\begin{equation}
R_{\rm GR}(t) = \frac{1}{N} \sum_{i=1}^{N} \sum_{s=1}^{n_i} K_h (t-t_{s}^{(i)}),
\label{eq:IWPSR}
\end{equation}
where $t_{s}^{(i)}$ is the $s$th spiking time of the $i$th GR cell, $n_i$ is the total number of spikes for the $i$th GR cell, and $N$ is the total number of
GR cells (i.e., $N = N_c \cdot N_{\rm GR} = 51,200$).
As a kernel function $K_{h}(t)$, we use a Gaussian kernel function of band width $h$:
\begin{equation}
K_h (t) = \frac{1}{\sqrt{2\pi}h} e^{-t^2 / 2h^2}, ~~~~ -\infty < t < \infty.
\label{eq:Gaussian}
\end{equation}
Throughout the paper, the band width $h$ of $K_h(t)$ is 10 msec.

\begin{figure}
\includegraphics[width=\columnwidth]{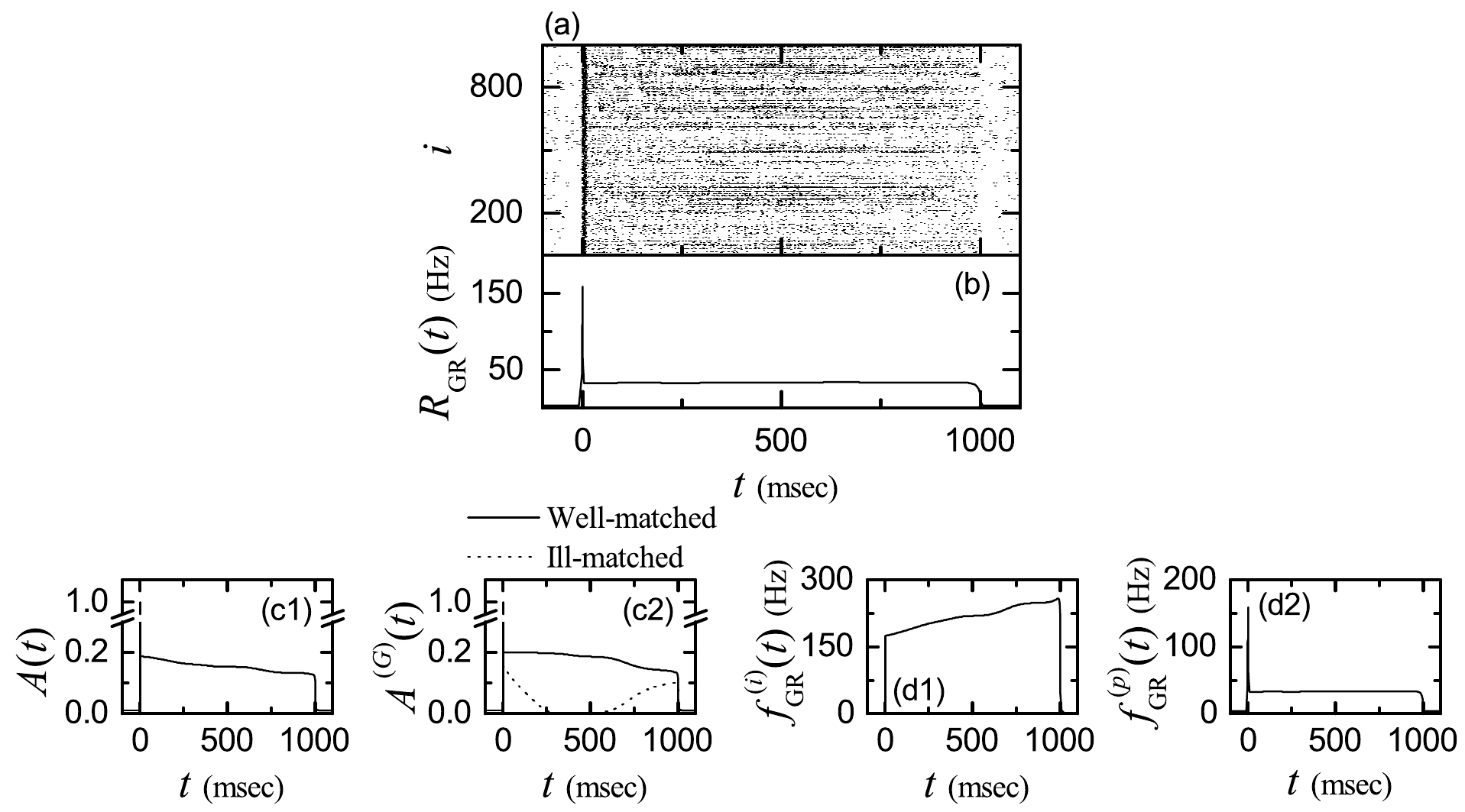}
\caption{Firing activity of GR cells in an optimal case of $p_c$ (connection probability from GO to GR cells) = 0.029.
(a) Raster plots of spikes of $10^3$ randomly chosen GR cells and (b) instantaneous whole-population spike rate $R_{\rm GR}(t)$ in the whole population of GR cells for the 1st step in the learning process for the EBC. Plots of the activation degrees (c1) $A(t)$ in the whole population of GR cells and (c2) $A^{(G)}(t)$ in the
$G$ firing group [$G~(=w):$ well-matched (solid curve) and $G~(=i):$ ill- matched (dotted curve)]. Plots of (d1) instantaneous individual firing rate $f_{\rm GR}^{(i)}(t)$ for the active GR cells and (d2) instantaneous population spike rate $f_{\rm GR}^{(p)}(t)$ in the whole population of GR cells
}
\label{fig:WP}
\end{figure}

Figure \ref{fig:WP}(a) shows a raster plot of spikes of $10^3$ randomly chosen GR cells.
At the beginning of trial stage ($0<t<7$ msec), all GR cells fire spikes due to the effect of strong transient CS signal of 200 Hz.
In the remaining part of the trial stage ($7 < t < 1000$ msec), GR cells make random repetition of transitions between active and inactive states because of
sustained CS signal of 30 Hz, and thus they seem to exhibit various spiking trains. Time passage from the CS onsets may be represented by the various firing
patterns of GR cells, which will be explained in details in Figs.~\ref{fig:DSP} and \ref{fig:Char}.
In the break stage ($1000 < t < 2000$ msec), GR cells fire very sparse
spikes. For simplicity, only the raster plot in the range of $1000 < t < 1100$ msec is shown; the raster plot in a part of the preparatory stage (essentially the same as the break stage) ($-100 < t < 0$ msec), just before the 1st trial stage, is also shown.
Figure \ref{fig:WP}(b) shows the instantaneous whole-population spike rate $R_{\rm GR}(t)$ in the whole population of GR cells.
$R_{\rm GR}(t)$ is basically in proportion to the transient and sustained CS inputs via MFs [see Figs.~\ref{fig:EBC}(b1)-\ref{fig:EBC}(b2)].
Thus, $R_{\rm GR}(t)$ is sharply peaked in the beginning of the trial stage due to the strong transient CS, and then it becomes nearly flat
in the remaining part of the trial stage where the sustained CS is present.
However, due to the inhibitory effect of GO cells, the overall firing rates are uniformly lowered as follows.
The time-averaged whole-population spike rates $\overline { R_{GR}(t) }$ in the time intervals of $0<t<5$ msec, $5<t<1000$ msec, and $1000<t<2000$ msec
are 155.4 Hz, 32.5 Hz, and 3.4 Hz, respectively.

We next consider the activation degree of GR cells. To examine it, we divide the whole learning step into bins.
In the beginning of the trial stage ($0 < t < 10$ msec), we divide the time interval into small bins (bin size: 1 msec) to properly take into consideration
the effect of strong transient CS; the effect of transient CS seems to persist until the 7th bin. Then, in the remaining trial stage ($10<t<1000$ msec), to use wide bins (bin size: 10 msec) seems to be sufficient for considering the effect of sustained CS.
Thus, we obtain the activation degree $A_i$ for the active GR cells in the $i$th bin:
\begin{equation}
A_i = \frac {N_{a,i}} {N}.
\label{eq:AD}
\end{equation}
Here, $N_{a,i}$ and $N~(=51,200)$ are the number of active GR cells in the $i$th bin and the total number of GR cells, respectively.
Figure \ref{fig:WP}(c1) shows a plot of the activation degree $A(t)$ in the whole population of GR cells.
In the initial 7 bins ($0<t<7$ msec), $A=1$ due to the effect of strong transient CS.
In the presence of sustained CS for $7 < t < 1000$ msec, the activation degree $A$ decreases monotonically from 0.189 to 0.131.
In this case, the time-averaged activation degree $\overline {A(t)}$ is 0.161.
Hence, representation of time passage from the onsets of CS can be made in a sparse recoding scheme
(i.e., MF inputs become more sparse via recoding in the granular layer).
In the break stage ($1000 < t < 2000$ msec), the time-averaged activation degree $\overline {A(t)}$ is 0.011 and small variations occur,
which may be regarded as nearly ``silent'' stage, in comparison with the trial stage.

The whole population of GR cells may be decomposed into two types of well-matched and ill-matched firing groups; details will be given
in Figs.~\ref{fig:DSP} and \ref{fig:Char}. Firing patterns in the well-matched (ill-matched) firing group are well (ill) matched with the airpuff US
signal. In this case, the activation degree $A_i^{(G)}$ of active GR cells in the $i$th bin in the $G$ firing group is given by:
\begin{equation}
A_i^{(G)} = \frac {N_{a,i}^{(G)}} {N^{(G)}}.
\label{eq:SAD}
\end{equation}
Here, $N_{a,i}^{(G)}$ and $N^{(G)}$ are the number of active GR cells in the $i$th bin and the total number of GR cells in the $G$ firing group, respectively
($G = w$ and $i$ for the well-matched and the ill-matched firing groups, respectively).
The number of clusters, belonging to the well- and the ill-matched firing groups are 841 and 183, respectively, and hence $N^{(w)}=42,050$ and $N^{(i)}=9,150$
because $N_{GR}=50$ (number of GR cells in each cluster).

Figure \ref{fig:WP}(c2) shows plots of activation degree $A^{(G)}(t)$ in the well-matched (solid line) and the ill-matched (dotted curve) firing groups.
In the beginning of the trial stage [i.e., in the initial 7 bins ($0<t<7$ msec)], $A^{(G)}=1$, independently of the firing groups, due to the effect of strong transient CS. On the other hand, in the remaining trial stage ($7<t<1000$ msec) where the sustained CS is present, $A^{(G)}(t)$ varies, strongly depending on the
type of firing groups. In the case of well-matched firing group, $A^{(w)}(t)$ decreases monotonically from 0.2 to 0.133, which is a little higher than $A(t)$ in the whole population. In contrast, in the case of ill-matched firing group, $A^{(i)}(t)$ forms a well-shaped curve with a central ``zero-bottom'' with
the time-averaged activation degree ${\overline {A^{(i)}(t)}} = 5.32 \cdot 10^{-4}$ for $330 < t < 580$ msec. Due to appearance of the central zero-bottom, contribution of the ill-matched firing group to $A(t)$ in the whole population may be negligible in the range of $330 < t < 580$ msec. In the break stage ($1000 < t < 2000$ msec), the time-averaged activation degree ${\overline {A^{(G)}(t)}} = 0.011$ ($G=w$ or $i$) with small variations, independently of the firing groups, which is the same as $\overline {A(t)}$ in the whole population.

In each $i$th bin, the contribution $C_i^{(G)}$ of each firing group to the activation degree $A_i$ in the whole population is given by the product of
the fraction $F^{(G)}$ and the activation degree $A_i^{(G)}$ of the firing group:
\begin{equation}
C_i^{(G)} = F^{(G)} \cdot A_i^{(G)} = {\frac {N_{a,i}^{(G)}} {N}},
\label{eq:Cont}
\end{equation}
where $F^{(G)} = N^{(G)} / N$; $F^{(w)}=0.821$ ($82.1 \%$)  and $F^{(i)}=0.179$ ($17.9 \%$) [see Fig.~\ref{fig:Char}(b)].
Hence, we can easily get the contribution $C_i^{(G)}$ of each firing group by just multiplying $A_i^{(G)}$ in Fig.~\ref{fig:WP}(c2) with the fraction $F^{(G)}$.
The sum of $C_i^{(G)}(t)$ over the well- and the ill-matched firing groups is just the activation degree $A_i(t)$ in the whole population
(i.e., $A_i = C_i^{(w)} + C_i^{(i)}$). In this case, contribution $C_i^{(i)}$ of the ill-matched firing group becomes small due to both low activation degree $A_i^{(i)}$ and small fraction $F^{(i)}$. Particularly, because of existence of the central zero-bottom, contribution $C_i^{(i)}$ is negligibly small
in the middle ($330 < t < 580$ msec) of the trial stage.

In the whole population, the activation degree $A(t)$ showing decreasing tendency is in contrast to the instantaneous whole-population spike rate $R_{\rm GR}(t)$ which is flat in the trial stage. To understand this discrepancy, we consider the bin-averaged instantaneous individual firing rate $f_{\rm GR}^{(i)}$ of active GR cells:
\begin{equation}
   f_{\rm GR}^{(i)} = \frac {N_{s,i}} {N_{a,i}~\Delta t}.
\label{eq:IFR}
\end{equation}
Here, $N_{s,i}$ is the number of spikes of GR cells in the $i$th bin, $N_{a,i}$ is the number of active GR cells in the $i$th bin, and
$\Delta t$ is the bin size. Figure \ref{fig:WP}(d1) shows a plot of $f_{\rm GR}^{(i)}(t)$ for the active GR cells.
In the initial 7 bins ($0<t<7$ msec) of the trial stage where $A(t)=1$, $f_{\rm GR}^{(i)}(t)$ decreases very slowly from 155.6 to 155.3 Hz with the time $t$
(i.e., the values of $f_{\rm GR}^{(i)}(t)$ are nearly the same). In the remaining part ($7 < t < 1000$ msec) of the trial stage, $f_{\rm GR}^{(i)}(t)$ increases monotonically from 173 to 258 Hz.
In this case, the bin-averaged instantaneous population spike rate $f_{\rm GR}^{(p)}$ is given by the product of the activation degree $A_i$ of Eq.~(\ref{eq:AD})
and the instantaneous individual firing rate $f_{\rm GR}^{(i)}$ of Eq.~(\ref{eq:IFR}):
\begin{equation}
f_{\rm GR}^{(p)} =  A_i~f_{\rm GR}^{(i)} = \frac {N_{s,i}} {N~\Delta t}.
\label{eq:PSR}
\end{equation}
Figure \ref{fig:WP}(d2) shows a plot of the instantaneous population spike rate $f_{\rm GR}^{(p)}(t)$.
It is flat except for the sharp peak in the beginning of the trial stage, as in the case of $R_{\rm GR}(t)$.
We note that both $f_{\rm GR}^{(p)}(t)$ and $R_{\rm GR}(t)$ correspond to bin-based estimate and kernel-based smooth estimate for the instantaneous
whole-population spike rate of the GR cells, respectively \cite{RM}.
Although the activation degree $A(t)$ of GR cells decreases with $t$, their population spike rate keeps the flatness (i.e., $f_{\rm GR}^{(p)}(t)$ becomes flat), because of the increase in the individual firing rate $f_{\rm GR}^{(i)}(t)$. As a result, the bin-averaged instantaneous population spike rate
$f_{\rm GR}^{(p)}(t)$ in Fig.~\ref{fig:WP}(d2) becomes essentially equal to the instantaneous whole-population spike rate $R_{\rm GR}(t)$ in Fig.~\ref{fig:WP}(b).

\subsection{Dynamical Classification and Dynamical Origin of Various Firing Patterns in The GR Clusters}
\label{subsec:DS}
There exist $N_C~(=2^{10})$ GR clusters in the whole population. $N_{\rm GR}~(=50)$ GR cells in each GR cluster share the same inhibitory and excitatory inputs via their dendrites which synaptically contact the four glomeruli (i.e., terminals of MFs) at both ends of the GR cluster [see Fig.~\ref{fig:RN}(b)]. Nearby inhibitory GO cell axons innervate the four glomeruli. Due to the shared inputs, GR cells in each GR cluster exhibit similar firing behaviors.

As in the case of $R_{\rm GR}(t)$ in Eq.~(\ref{eq:IWPSR}), the firing activity of the $I$th GR cluster is described in terms of its instantaneous cluster spike rate $R_{\rm GR}^{(I)}(t)$ ($I=1, \cdots, N_C$):
\begin{equation}
R_{\rm GR}^{(I)}(t) = \frac{1}{N_{\rm GR}} \sum_{i=1}^{N_{\rm GR}} \sum_{s=1}^{n_i^{(I)}} K_h (t-t_{s}^{(I,i)}),
\label{eq:ISPSR}
\end{equation}
where $t_{s}^{(I,i)}$ is the $s$th firing time of the $i$th GR cell in the $I$th GR cluster and $n_i^{(I)}$ is the total number of spikes for the $i$th GR cell
in the $I$th GR cluster.

We introduce the matching index ${\cal{M}}^{(I)}$ of each GR cluster between the firing behavior [$R_{\rm GR}^{(I)}(t)$] of each $I$th GR cluster and the airpuff US signal $f_{\rm US}(t)$ for the desired timing [see Fig.\ref{fig:EBC}(c)]. The matching index ${\cal{M}}^{(I)}$ is given by the cross-correlation at the zero-time lag [i.e., $Corr_{\rm GR}^{(I)}(0)$] between $R_{\rm GR}^{(I)}(t)$ and $f_{\rm US}(t)$:
\begin{equation}
Corr_{\rm GR}^{(I)} (\tau) = \frac{\overline{\Delta f_{\rm US}(t+\tau) \Delta R_{\rm GR}^{(I)}(t)}}{\sqrt{\overline{\Delta f_{\rm US}^{2}(t)}} \sqrt{\overline{{\Delta R_{\rm GR}^{(I)}}^2(t)}}},
\label{eq:CI}
\end{equation}
where $\Delta f_{\rm US}(t) = f_{\rm US}(t)-\overline{f_{\rm US}(t)}$, $\Delta R_{\rm GR}^{(I)}(t) = R_{\rm GR}^{(I)}(t)-\overline{R_{\rm GR}^{(I)}(t)}$, and the overline denotes the time average. We note that ${\cal{M}}^{(I)}$ represents well the phase difference between the firing patterns [$R_{\rm GR}^{(I)}(t)$] of GR clusters and the US signal [$f_{\rm US}(t)$].

\begin{figure}
\includegraphics[width=\columnwidth]{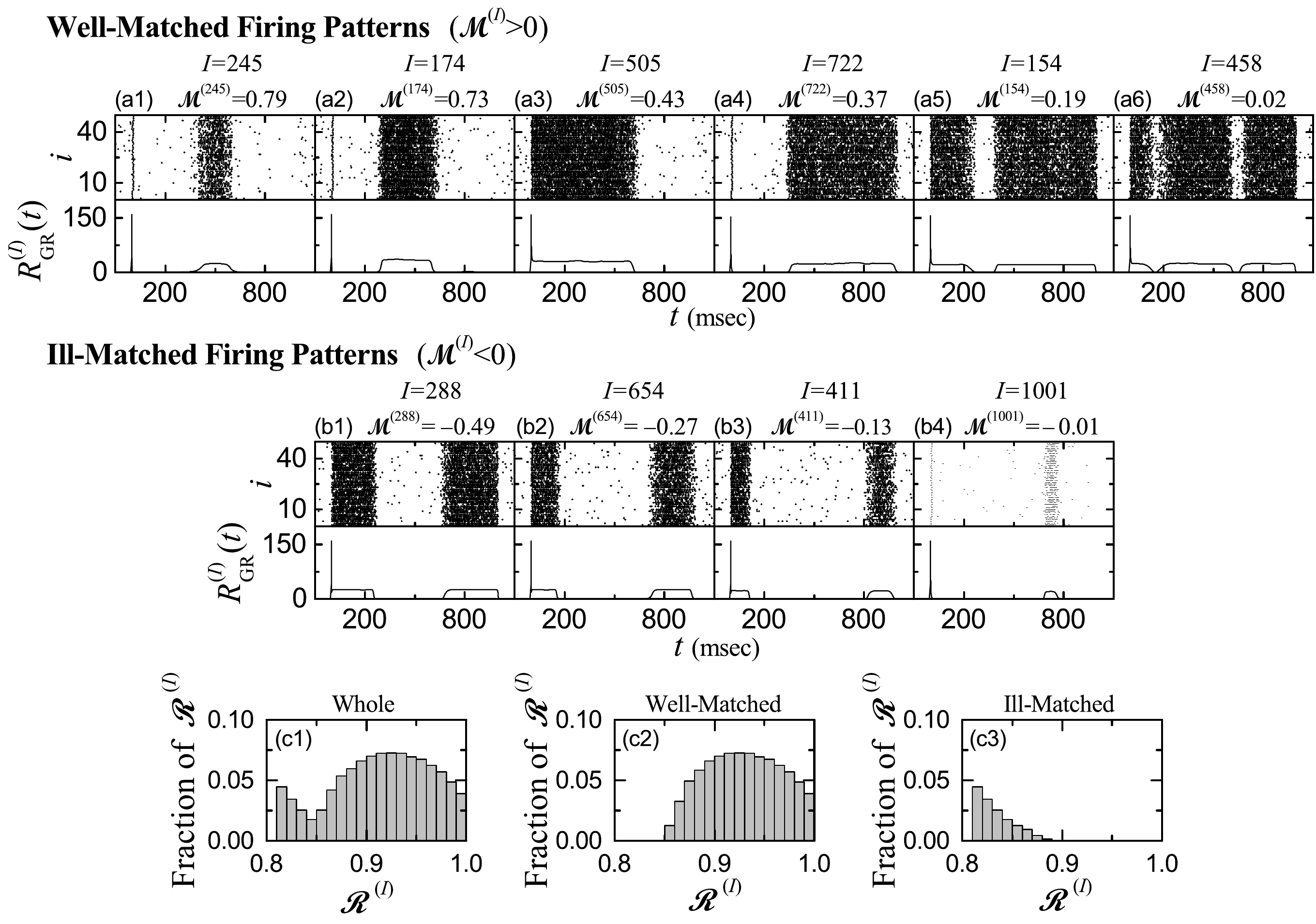}
\caption{Various firing patterns in the GR clusters in an optimal case of $p_c^* = 0.029$. Raster plots of spikes and instantaneous cluster spike rates
$R_{GR}^{(I)} (t)$ for various firing patterns. Six well-matched firing patterns in the $I$th GR clusters; $I=$ (a1) 245, (a2) 174, (a3) 505, (a4) 722, (a5) 154, and (a6) 458. Four ill-matched firing patterns in the $I$th GR cluster; $I=$ (b1) 288, (b2) 654, (b3) 411, and (b4) 1001. ${\cal M}^{(I)}$ represents the matching index of the firing pattern in the $I$th GR cluster. Distribution of the reproducibility degree ${\cal R}^{(I)}$ in the (c1) whole population and the (c2) well- and (c3) ill-matched firing groups. Bin size for the histograms in (c1)-(c3) is 0.01.
}
\label{fig:DSP}
\end{figure}

Figure \ref{fig:DSP} shows various firing patterns of GR clusters.
This type of variety results from inhibitory coordination of GO cells on the firing activity of GR cells in the GR-GO feedback loop in the granular layer.
Time passage from the CS onsets may be well represented by the various firing patterns of GR clusters because MF inputs become less similar (i.e., more orthogonal) to each other through recoding in the granular layer.

Six examples for the well-matched firing patterns in the $I$th ($I=$ 245, 174, 505, 722, 154, and 458) GR clusters are given in Figs.~\ref{fig:DSP}(a1)-\ref{fig:DSP}(a6), respectively. Raster plot of spikes of $N_{\rm GR}~(=50)$ GR cells and the corresponding instantaneous cluster spike rate $R_{\rm GR}^{(I)}(t)$ are shown, along with the value of the matching index ${\cal{M}}^{(I)}$ in each case of the $I$th GR cluster.
In all these cases, the instantaneous cluster spike rates $R_{\rm GR}^{(I)}(t)$ are well-matched with the US signal $f_{\rm US}(t)$, and hence these
well-matched GR clusters have positive matching indices (i.e., ${\cal{M}}^{(I)} > 0$).

In the 1st case of $I=245$ with the highest ${\cal{M}}^{(I)}$ $(=0.79)$, $R_{\rm GR}^{(245)}(t)$ is strongly localized around the middle of the trial stage (i.e. a central band of spikes is formed around $t=500$ msec), and hence it is the most well-matched with the US signal $f_{\rm US}(t)$. In the 2nd case of $I=174$ with ${\cal{M}}^{(I)}=0.73,$ $R_{\rm GR}^{(174)}(t)$ is also localized around $t=500$ msec, but its central firing band spreads a little more to the left side, in comparison with the case of $I=245$. Hence, its matching index relative to $f_{\rm US}(t)$ is a little decreased.

We note that LTD at the PF-PC synapses occurs within an effective range of $\Delta t_l^*~(\simeq -117.5) < \Delta t < \Delta t_r^*~(\simeq 277.5)$ (see Fig.~\ref{fig:TW}). Here, $\Delta t$ [= $t_{\rm CF}$ (CF activation time) - $t_{\rm PF}$ (PF activation time)] is the relative time difference between the
firing times of the error-teaching instructor CF and the variously-recoded student PF. The CF activation occurs approximately at $t_{\rm CF}=500$ msec due to
the strong brief US (strongly localized at $t=500$ msec). Then, LTD may occur when the PF activation time $t_{\rm PF}$ lies in an effective LTD range of
222.5 msec $< t_{\rm PF} < $ 617.5 msec. In the above two cases of GR clusters ($I=$ 245 and 174) with higher ${\cal{M}}^{(I)}$, their PF signals (corresponding to axons of the GR cells) are strongly depressed by the US instructor signal because most parts of their firing bands are well localized in the effective LTD range.

We next consider the 3rd and the 4th cases of the $I$th GR cluster ($I=$ 505 and 722) with intermediate ${\cal{M}}^{(I)}$.
In the cases of $I=505$ (722), the firing band in the raster plot extends to the left (right) until $t \simeq 0~(1000)$ msec. Thus, big left- and right-extended
firing bands appear for $I=505$ and 722, respectively. Some part of this big firing band lies inside the effective LTD range where LTD occurs in conjunction
with the CF firing. On the other hand, its remaining part lies outside the effective LTD range, and hence LTP occurs for the PF firings alone without association with the CF signal.

We also consider the case of lower ${\cal{M}}^{(I)}$ for $I=154$ and 458 (i.e., the 5th and the 6th cases).
In both cases, they have tendency to fill the raster plots with more spikes via appearance of two or more firing bands.
Thus, some central part of these bands lies inside the effective LTD range where LTD occurs.
In contrast, LTP occurs in the other left and right parts of the firing bands because they lie outside the effective LTD range;
in comparison with the case of intermediate ${\cal{M}}^{(I)}$, LTP region is extended.
In this way, as ${\cal{M}}^{(I)}$ is decreased toward the zero, the raster plot tends to be filled with more spikes (constituting firing bands), and hence
the region where LTP occurs is extended.

In addition to the well-matched firing patterns, ill-matched firing patterns with negative matching indices (i.e., ${\cal{M}}^{(I)} <0$)
also appear. Four examples for the ill-matched firing patterns in the $I$th ($I=$ 288, 654, 411, and 1001) GR clusters are given in
Figs.~\ref{fig:DSP}(b1)-\ref{fig:DSP}(b4), respectively.
We first consider the case of $I=288$ with the lowest ${\cal{M}}^{(I)}~(=-0.49)$ (i.e., its magnitude $| {\cal{M}}^{(I)} |$: largest).
This lowest case corresponds to the ``opposite'' case of the highest one for $I=245$ with ${\cal{M}}^{(I)}=0.79$ in the following sense.
A central gap with negligibly small number of spikes (i.e., practically no spikes) appears around $t=500$ msec, in contrast to the highest case
where a central firing band exists. Hence, in this lowest case, occurrence of LTD in the central gap may be practically negligible.
On the other hand, mainly LTP occurs in the left and right firing bands, most of which lie outside the effective LTD range.
The right firing band lies completely outside the effective LTD range, and hence no LTD occurs.
The width of the central gap is larger than the width of the effective LTD range. However, since the gap is shifted a little to the right,
a small part near the right boundary ($t \simeq 261$ msec) of the left firing band overlaps with a small region near the left boundary ($t \simeq 222.5$ msec) of the effective LTD region. In this small overlapped region of $239 < \Delta t < 277.5$ msec, the values of the synaptic modification $\Delta J_{\rm LTD}$
(i.e., the average synaptic modification $\langle \Delta J_{\rm LTD} \rangle \simeq 0.031$) are very small, and hence very weak LTD may occur.

As the magnitude $| {\cal{M}}^{(I)} |$ is decreased, the central gaps becomes widened, and the widths of the left and the right firing bands
also get decreased, as shown in the cases of $I=654$, 411, and 1001. In these cases, the two left and right firing bands lie completely outside the effective LTD range, and hence only LTP occurs for the PF signals alone without conjunction with the CF signal. In this way, as $| {\cal{M}}^{(I)} |$ approaches the zero from the negative side, spikes become more and more sparse, which is in contrast to the well-matched case where more and more spikes fill the raster plot as ${\cal{M}}^{(I)} $ goes to the zero from the positive side.

The above firing patterns are shown in the 1st learning step [consisting of the 1st trial stage ($0<t<1000$ msec) and the 1st break stage ($1000<t<2000$ msec)].
For simplicity, they are shown in the range of $1000 < t < 1100$ msec in the break stage, and a part of the preliminary stage ($-100 < t < 0$ msec), preceding the 1st learning step, is also shown.
We examine the reproducibility of the firing patterns across the learning steps.
To this end, we consider the cross-correlation between the instantaneous cluster spike rates $R_{\rm GR}^{(I,l)}(t)$ in the $I$th GR cluster for the $k$th
($l=k$) and the $(k+1)$th ($l=k+1$) learning steps;
\begin{equation}
Corr_{\rm GR}^{(I,k)} (\tau) = \frac{\overline{\Delta R_{\rm GR}^{(I,k)}(t+\tau) \Delta R_{\rm GR}^{(I,k+1)}(t)}}
{\sqrt{\overline{ {\Delta R_{\rm GR}^{(I,k)}}^{2}(t)}} \sqrt{\overline{{\Delta R_{\rm GR}^{(I,k+1)}}^2(t)}}},
\label{eq:R}
\end{equation}
where $\Delta R_{\rm GR}^{(I,l)}(t) = R_{\rm GR}^{(I,l)}(t)-\overline{R_{\rm GR}^{(I,l)}(t)}$ ($l=k$ and $k+1$) and the overline represents the time average.
Then, the reproducibility degree ${\cal R}^{(I)}$ of the $I$th GR cluster is given by the average value of the cross-correlations at the zero-time lag
between the instantaneous cluster spike rates $R_{\rm GR}^{(I,l)}(t)$ for the successive learning steps:
\begin{equation}
{\cal R}^{(I)} = {\frac {1} {N_{\rm step}-1}} \sum_{k=1}^{N_{\rm step}-1} Corr_{\rm GR}^{(I,k)} (0),
\label{eq:RD}
\end{equation}
where $N_{\rm step}$ is the total number of learning steps. Here, we consider the case of $N_{\rm step}=100$.

Figure \ref{fig:DSP}(c1) shows the distribution of the reproducibility degrees ${\cal R}^{(I)}$ for the whole GR clusters.
Double peaks appear; large broad peak and small sharp peaks at ${\cal R}^{(I)} = 0.925$ and 0.815, respectively.
The range of ${\cal R}^{(I)}$ is (0.812, 0.997). Hence, the firing patterns are highly reproducible across the learning steps.
Figures \ref{fig:DSP}(c2) and \ref{fig:DSP}(c3) also show the distributions of ${\cal R}^{(I)}$ of the GR clusters
in the well- and the ill-matched firing groups, respectively. In the case of well-matched firing group, the distribution of ${\cal R}^{(I)}$ has a broad peak and its range is (0.842, 0.997). On the other hand, in the ill-matched case, the distribution decreases from its maximum at ${\cal R}^{(I)} = 0.815$, and its range is (0.812, 0.879). We note that the average values of $\{ {\cal R}^{(I)} \}$ in the well- and the ill-matched firing group are 0.927 and 0.828, respectively. Hence, on average, the firing patterns of the GR clusters in the well-matched firing group may be more reproducible than those in the ill-matched firing group, because the average individual firing rate in the well-matched firing group is higher than that in the ill-matched firing group.

\begin{figure}
\includegraphics[width=\columnwidth]{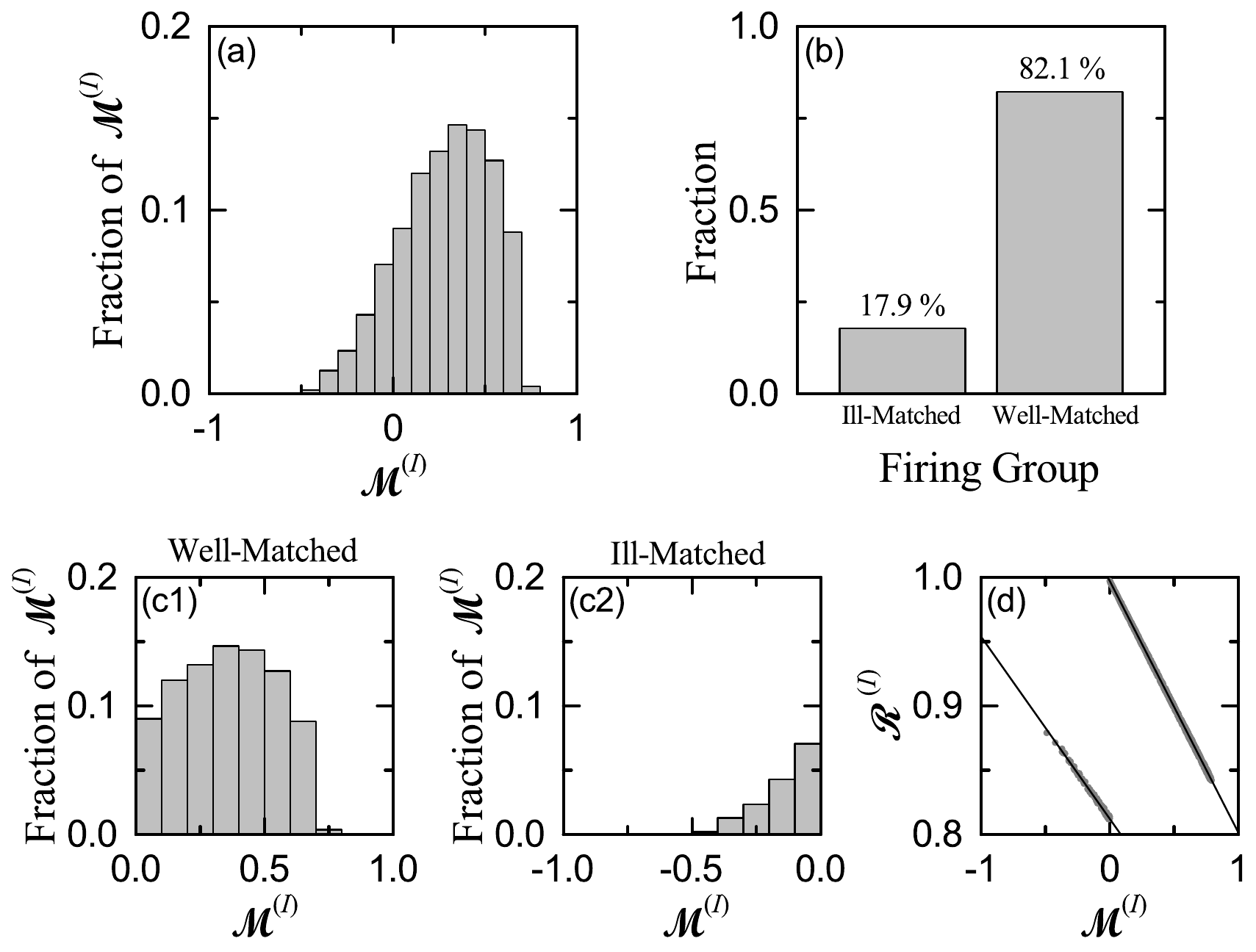}
\caption{Characterization of various firing patterns in the GR clusters in an optimal case of $p_c^* = 0.029$. (a) Distribution of matching indices
$\{ {\cal{M}}^{(I)} \}$ in the whole population. (b) Fraction of well-matched and ill-matched firing groups. Distribution of matching indices $\{ {\cal{M}}^{(I)} \}$  for the (c1) well- and (c2) ill-matched firing groups. Bin size for the histograms in (a) and in (c1)-(c3) is 0.1. (d) Plots of reproducibility degree ${\cal R}^{(I)}$ versus ${\cal{M}}^{(I)}$ in the well-matched (${\cal{M}}^{(I)} >0$) and the ill-matched ($ {\cal{M}}^{(I)} <0)$ firing groups.
}
\label{fig:Char}
\end{figure}

Results on characterization of the various well- and ill-matched firing patterns are given in Fig.~\ref{fig:Char}.
Figure \ref{fig:Char}(a) shows the plot of the fraction of matching indices ${\cal{M}}^{(I)}$ in the whole GR clusters.
${\cal{M}}^{(I)}$ increases slowly from the negative value to the peak at 0.35, and then it decreases rapidly.
For this distribution of $\{ {\cal{M}}^{(I)} \}$, the range is (-0.49, 0.79), the mean is 0.3331, and the standard deviation is 0.6135.
Then, we obtain the variety degree $\cal{V}$ for the firing patterns [$R_{\rm GR}^{(I)}(t)$] of all the GR clusters:
\begin{eqnarray}
   {\cal{V}} &=& {\rm Relative~ Standard ~ Deviation}\nonumber \\
& & ~{\rm for ~the~Distribution~of~} \{ {\cal{M}}^{(I)} \},
\label{eq:DD}
\end{eqnarray}
where the relative standard deviation is just the standard deviation divided by the mean.
In the optimal case of $p_c^*=0.029$, ${\cal{V}}^* \simeq  1.842$, which is just a quantitative measure for the various recoding made
through feedback cooperation between the GR and the GO cells in the granular layer.
It will be seen that ${\cal{V}}^*$ is just the maximum in Fig.~\ref{fig:Final}(b) for the plot of $\cal{V}$ versus $p_c$.
Hence, firing patterns of the GR clusters at $p_c^*$ is the most various.

We decompose the whole GR clusters into the well-matched ($\{ {\cal{M}}^{(I)} \} >0$) and the ill-matched ($\{ {\cal{M}}^{(I)} \} < 0$) firing groups.
Figure \ref{fig:Char}(b) shows the fraction of firing groups. The well-matched firing group is a major one with fraction 82.1$\%$,
while the ill-matched firing group is a minor one with fraction 17.9$\%$. In this case, the firing-group ratio ${\cal R}_{\rm sp}$, given by the ratio of the fraction of the well-matched firing group to that of the ill-matched firing group is 4.59. For this firing-group ratio, firing patterns of the GR clusters are the most various.

Figures \ref{fig:Char}(c1)-\ref{fig:Char}(c2) also show the plots of matching indices ${\cal{M}}^{(I)}$
of the GR clusters in the well- and the ill-matched firing groups, respectively.
In the case of well-matched firing group, the distribution of ${\cal{M}}^{(I)}$ with a peak at 0.35 has only positive values in the range of $(0.0, 0.79)$,
and its mean and standard deviations are 0.428 and 0.372, respectively. On the other hand, in the case of the ill-matched firing group,
the distribution of ${\cal{M}}^{(I)}$ with a maximum at -0.05 has only negative values in the range of $(-0.49, 0.0)$,
and its mean and standard deviations are -0.104 and 0.129, respectively. In this case, ${\cal{M}}^{(I)}$ increases slowly to the maximum.
As will be seen in the next subsection, these well- and the ill-matched firing groups play their own roles in the synaptic plasticity at PF-PC synapses
and the subsequent learning process for the EBC, respectively.

We also examine the correlation between the atching index ${\cal{M}}^{(I)}$ and the reproducibility degree ${\cal R}^{(I)}$.
Figure \ref{fig:Char}(d) show the plots of ${\cal{M}}^{(I)}$ versus ${\cal R}^{(I)}$ in the well- and the ill-matched firing groups.
In both cases, there appear strong negative correlations between ${\cal{M}}^{(I)}$ and ${\cal R}^{(I)}$; for the well-matched (ill-matched) firing group,
the Pearson's correlation coefficient is $r = -0.9999~(- 0.9978)$. When left-right reflections are made on Figs.~\ref{fig:Char}(c1)-\ref{fig:Char}(c2), shapes of the reflected ones are similar to the shapes of Figs.~\ref{fig:DSP}(c2)-\ref{fig:DSP}(c3), respectively, which implies the negative correlation between ${\cal{M}}^{(I)}$ and ${\cal R}^{(I)}$ in each firing group. As shown in Figs.~\ref{fig:DSP}(a1)-\ref{fig:DSP}(a6), as ${\cal{M}}^{(I)}$ decreases to the zero from the positive side, the raster plot tends to be filled with more spikes due to increased individual firing rates, which leads to increase in ${\cal R}^{(I)}$.
On the other hand, as ${\cal{M}}^{(I)}$ increases to the zero from the negative side, the raster plot of spikes tends to be more sparse because of decreased individual firing rates [see Figs.~\ref{fig:DSP}(b1)-\ref{fig:DSP}(b4)], which results in decrease in ${\cal R}^{(I)}$.
Consequently, there appears a gap at the limit ${\cal{M}}^{(I)}=0$.

\begin{figure}
\includegraphics[width=\columnwidth]{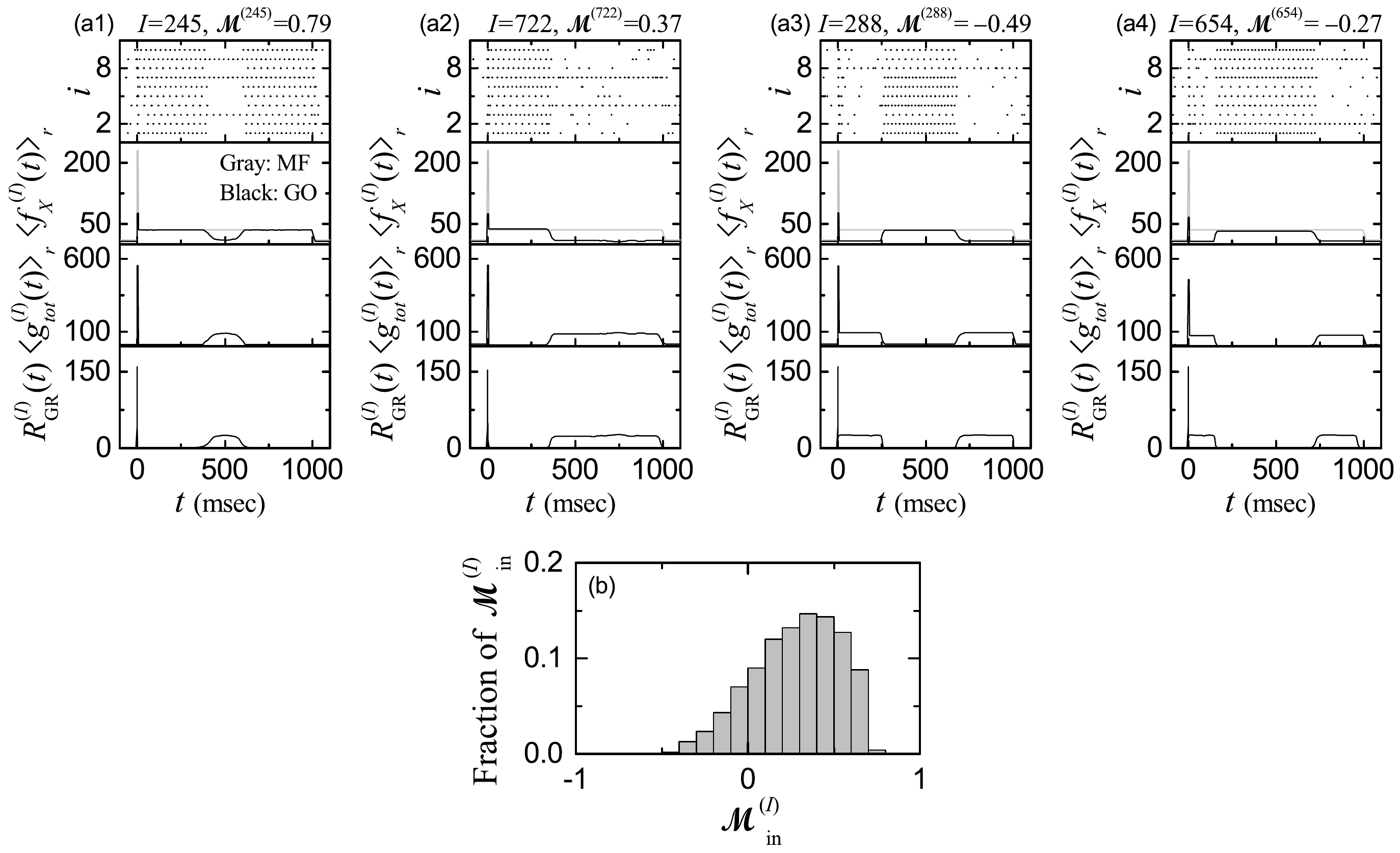}
\caption{Dynamical origin of various firing patterns in the GR clusters in an optimal case of $p_c^* = 0.029$.
Well-matched firing patterns for $I$= (a1) 245 and (a2) 722 and ill-matched firing patterns for $I$= (a3) 288 and (a4) 654. In (a1)-(a4),
top panel: raster plots of spikes in the sub-population of pre-synaptic GO cells innervating the $I$th GR cluster, 2nd panel: plots of $f_X^{(I)}(t):$ bin-averaged instantaneous spike rates of the MF signals ($X= {\rm MF}$) into the $I$th GR cluster (gray line) and bin-averaged instantaneous sub-population of pre-synaptic GO cells ($X={\rm GO}$) innervating the $I$th GR cluster (black line); $\langle \cdots \rangle_r$ represents the realization average (number of realizations is 100), 3rd panel: time course of $\langle g_{tot}^{(I)}(t) \rangle_r:$ conductance of total synaptic inputs (including both the excitatory and inhibitory inputs) into the $I$th GR cluster, and bottom panel: plots of $R_{\rm GR}^{(I)}(t):$ instantaneous cluster spike rate in the $I$th GR cluster.
(b) Distribution of matching indices $\{ {\cal M}_{\rm in}^{(I)} \}$ for the conductances of total synaptic inputs into the GR clusters.
}
\label{fig:Origin}
\end{figure}

Finally, we study the dynamical origin of various firing patterns in the $I$th GR clusters.
As examples, we consider two well-matched firing patterns for $I=245$ and 722 [see the firing patterns in Figs.~\ref{fig:DSP}(a1) and \ref{fig:DSP}(a4)] and two ill-matched firing patterns for $I=288$ and 654 [see the firing patterns Figs.~\ref{fig:DSP}(b1) and \ref{fig:DSP}(b2)]. In Fig.~\ref{fig:Origin}, (a1)-(a4) correspond to the cases of $I=$245, 722, 288, and 654, respectively.

Various recodings for the MF signals are made in the GR layer, consisting of excitatory GR and inhibitory GO cells (i.e., in the GR-GO cell feedback loop).
Thus, firing activities of GR cells are determined by two types of synaptic input currents (i.e., excitatory synaptic inputs via MF signals and inhibitory
synaptic inputs from randomly connected GO cells). Then, investigations on the dynamical origin of various firing patterns of the GR clusters are made via analysis of total synaptic inputs into the GR clusters. As in Eq.~(\ref{eq:ISyn2}), synaptic current is given by the product of synaptic conductance $g$ and potential difference. In this case, synaptic conductance determines the time-course of synaptic current. Hence, it is sufficient to consider the time-course of
synaptic conductance. The synaptic conductance $g$ is given by the product of synaptic strength per synapse, the number of synapses $M_{\rm syn}$, and the fraction $s$ of open (post-synaptic) ion channels [see Eq.~(\ref{eq:ISyn3})]. Here, the synaptic strength per synapse is given by the product of maximum synaptic conductance
$\bar{g}$ and synaptic weight $J$, and the time-course of $s(t)$ is given by a summation for exponential-decay functions over pre-synaptic spikes, as shown
in Eqs.~(\ref{eq:ISyn3}) and (\ref{eq:ISyn4}).

Here, we make an approximation of the fraction $s(t)$ of open ion channels (i.e., contributions of summed effects of pre-synaptic spikes) by the bin-averaged spike rate $f_X^{(I)}(t)$ of pre-synaptic neurons ($X=$ MF and GO); $f_{\rm MF}^{(I)}(t)$ is the bin-averaged spike rate of the MF signals into the $I$th GR cluster
and $f_{\rm GO}^{(I)}(t)$ is the bin-averaged spike rate of the pre-synaptic GO cells innervating the $I$th GR cluster.
In the case of MF signal, we get:
\begin{equation}
     f_{\rm MF}^{(I)}(t) = f_{\rm TCS}^{(I)}(t) + f_{\rm SCS}^{(I)}(t),
\label{fMF}
\end{equation}
where $f_{\rm TCS}^{(I)}(t)$ and $f_{\rm SCS}^{(I)}(t)$ are the bin-averaged spike rates of the transient and the sustained CS signals, respectively.

Then, the conductance $g_X^{(I)}(t)$ of synaptic input from $X$ (=MF or GO) into the $I$th GR cluster ($I=1, \cdots, N_C$) is given by:
\begin{equation}
 g_X^{(I)}(t) \simeq {\rm M}_f^{(R)} \cdot f_X^{(I)}(t).
\label{gX}
\end{equation}
Here, the multiplication factor ${\rm M}_f^{(R)}$ [= maximum synaptic conductance ${\bar g}_R$ $\times$ synaptic weight
$J^{({\rm GR},X)}$ $\times$ number of synapses $M_{\rm syn}^{({\rm GR},X)}$] varies depending on $X$ and the receptor $R$ on the post-synaptic GR cells.
In the case of excitatory synaptic currents into the $I$th GR cluster with AMPA receptors via TCS or SCS MF signal, ${\rm M}_f^{\rm (AMPA)} = 2.88;$
${\bar g}_{\rm AMPA}=0.18,$ $J^{\rm (GR,MF)}=8.0,$ and $M_{\rm syn}^{({\rm GR},X)} =2$ $(X$=TCS, SCS). On the other hand, in the case of the $I$th GR cluster with NMDA
receptors, ${\bar g}_{\rm NMDA}=0.025,$ and hence ${\rm M}_f^{(\rm NMDA)} = 0.4,$ which is much less than ${\rm M}_f^{(\rm AMPA)}$.
For the inhibitory synaptic current from pre-synaptic GO cells to the $I$th GR cluster with GABA receptors,
${\rm M}_f^{(\rm GABA)} = 2.63$; $\bar{g}_{\rm GABA}=0.028,$ $J^{\rm (GR,GO)}=10,$ and $M_{\rm syn}^{\rm (GR,GO)} = 9.4.$
Then, the conductance $g_{tot}^{(I)}$ of total synaptic inputs (including both the excitatory and the inhibitory inputs) into the $I$th GR cluster is given by:
\begin{eqnarray}
g_{tot}^{(I)}(t) &=& g_{\rm MF}^{(I)} - g_{\rm GO}^{(I)} = g_{\rm AMPA}^{(I)} + g_{\rm NMDA}^{(I)} - g_{\rm GO}^{(I)} \nonumber \\
             & = & 3.28~ f_{\rm MF}^{(I)}(t) - 2.63~ f_{\rm GO}^{(I)}(t).
\label{eq:gTOT}
\end{eqnarray}

In Figs.~\ref{fig:Origin}(a1)-\ref{fig:Origin}(a4), the top panels show the raster plots of spikes in the sub-populations of pre-synaptic
GO cells innervating the $I$th GR clusters. From these raster plots, bin-averaged (sub-population) spike rates $f_{\rm GO}^{(I)}(t)$ may be obtained.
The bin-averaged spike rate of pre-synaptic GO cells in the $i$th bin is given by $\frac {n_i^{(s)}} {N_{pre}~\Delta t}$, where $n_i^{(s)}$ is the
number of spikes in the $i$th bin, $\Delta t$ is the bin size, and $N_{pre}$ (=10) is the number of pre-synaptic GO cells.
As in Fig.~\ref{fig:WP}, in the beginning of the trial stage ($0 < t < 10$ msec), we employ the small bin-size of $\Delta t = 1$ msec to properly take into consideration the effect of strong transient CS, and in the remaining trial stage ($10<t<1000$ msec), we use the wide bin-size of $\Delta t = 10$ msec
for considering the effect of sustained CS. Through an average over 100 realizations, we obtain the realization-averaged (bin-averaged) spike rate of pre-synaptic GO cells $\langle f_{\rm GO}^{(I)}(t) \rangle_r$ because $N_{pre}~(=10)$ is small; $\langle \cdots \rangle_r$ represent a realization-average.
The 2nd panels show $\langle f_{\rm GO}^{(I)}(t) \rangle_r$ (black line) and $\langle f_{\rm MF}^{(I)}(t) \rangle_r$ (gray line).
We note $\langle f_{\rm GO}^{(I)}(t) \rangle$ varies depending on $I$, while $\langle f_{\rm MF}^{(I)}(t) \rangle$ is independent of $I$.
Then, we obtain the realization-averaged conductance $\langle g_{tot}^{(I)}(t) \rangle_r$ of total synaptic inputs in Eq.~(\ref{eq:gTOT}), which is shown in
the 3rd panels.

We note that the shapes of $\langle g_{tot}^{(I)}(t) \rangle_r$ (corresponding to the total input into the $I$th GR cluster) in the 3rd panels are nearly the same as those of $R_{\rm GR}^{(I)}(t)$ (corresponding to the output of the $I$th GR cluster) in the bottom panels. It is thus expected that well-matched (ill-matched) inputs into the GR clusters may lead to generation of well-matched (ill-matched) outputs (i.e., responses) in the GR clusters. To confirm this point clearly,
as in case of the firing patterns [$R_{\rm GR}^{(I)}(t)$] in the GR clusters, we introduce the matching index for the total synaptic input of the $I$th GR cluster  between $\langle g_{tot}^{(I)}(t) \rangle_r$ (conductance of total synaptic input into the $I$th GR cluster) and
the (airpuff) US signal $f_{\rm US}(t)$ for the desired timing. Similar to the matching index ${\cal M}^{(I)}$ for the firing patterns (i.e. outputs) in the $I$th GR cluster [see Eq.~(\ref{eq:CI})], the matching index ${\cal{M}}^{(I)}_{\rm in}$ for the total synaptic input is given by the cross-correlation at the zero-time lag (i.e., $Corr_{\rm in}^{(I)}(0)$) between $\langle g_{tot}^{(I)}(t) \rangle_r$ and $f_{\rm US}(t)$:
\begin{equation}
Corr_{\rm in}^{(I)} (\tau) = \frac{\overline{ \Delta f_{\rm US}(t+\tau) \Delta \langle g_{tot}^{(I)}(t) \rangle_r}}
{\sqrt{\overline{ {\Delta f_{\rm US}^{2}(t)}}} \sqrt{\overline{ {\Delta \langle g_{tot}^{(I)}(t) \rangle_r}^2}}},
\label{eq:INCI}
\end{equation}
where $\Delta f_{\rm US}(t) = f_{\rm US}(t) -\overline{f_{\rm US}(t)}$, $\Delta \langle g_{tot}^{(I)}(t) \rangle_r = \langle g_{tot}^{(I)}(t) \rangle_r-\overline{ \langle g_{tot}^{(I)}(t) \rangle_r}$, and the overline represents the time average.
Thus, we have two types of matching indices, ${\cal M}^{(I)}$ [output matching index: given by $Corr_{\rm GR}^{(I)}(0)$] and ${\cal M}^{(I)}_{\rm in}$
[input matching index: given by $Corr_{\rm in}^{(I)}(0)$] for the output and the input in the $I$th GR cluster, respectively.

Figure \ref{fig:Origin}(b) shows the plot of fraction of input matching indices $\{ {\cal M}^{(I)}_{\rm in} \}$ in the whole GR clusters.
We note that the distribution of input matching indices in Fig.~\ref{fig:Origin}(b) is nearly the same as that of
output matching indices in Fig.~\ref{fig:Char}(a). ${\cal{M}}^{(I)}_{\rm in}$ increases slowly from the negative value to the peak at 0.35, and then it decreases rapidly. In this distribution of $\{ {\cal{M}}^{(I)}_{\rm in} \}$, the range is (-0.49, 0.79), the mean is 0.3332, and the standard deviation is 0.6137.
Then, we get the variety degree ${\cal{V}}_{\rm in}$ for the total synaptic inputs $\{ \langle g_{tot}^{(I)}(t) \rangle_r \}$ of all the GR clusters:
\begin{eqnarray}
   {\cal{V}}_{\rm in} &=& {\rm Relative~ Standard ~ Deviation}\nonumber \\
& & ~{\rm for ~the~Distribution~of~} \{ {\cal{M}}^{(I)}_{\rm in} \}.
\label{eq:INDD}
\end{eqnarray}
Hence, ${\cal{V}}_{\rm in} \simeq 1.842$ for the synaptic inputs, which is nearly the same as ${\cal V}^*~(\simeq 1.842)$ for the firing patterns of GR cells.
Consequently, various synaptic inputs into the GR clusters results in generation of various outputs (i.e., firing patterns) of the GR cells.

\begin{figure}
\includegraphics[width=\columnwidth]{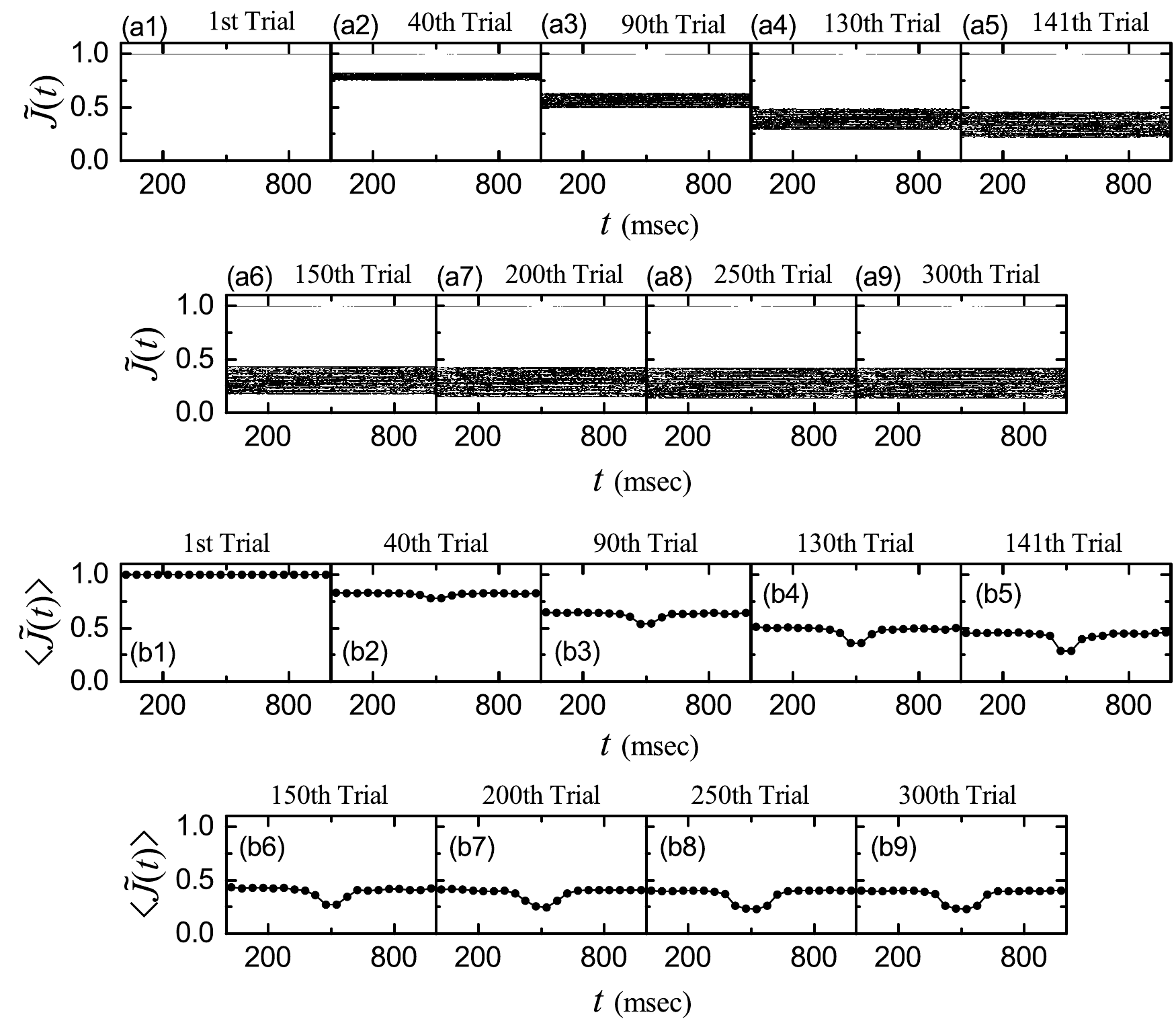}
\caption{Change in synaptic weights of active PF-PC synapses during learning trials in the optimal case of $p_c^* = 0.029$. (a1)-(a9) Trial-evolution of distribution of normalized synaptic weights $\tilde{J}(t)$ of active PF signals. (b1)-(b9) Trial-evolution of bin-averaged (normalized) synaptic weights $\langle {\tilde J(t)} \rangle$ of active PF signals. Bin size: $\Delta t =50$ msec.
}
\label{fig:SW}
\end{figure}

\subsection{Influence of Various Temporal Recoding on Synaptic Plasticity at PF-PC Synapses}
\label{subsec:ESP}
Based on dynamical classification of firing patterns of GR clusters, we study the influence of various temporal recoding in the GR clusters on
synaptic plasticity at PF-PC synapses. As shown in the preceding subsection, MF context input signals for the EBC are variously recoded in the granular layer (corresponding to the input layer of the cerebellar cortex). The variously-recoded well- and ill-matched PF (student) signals (coming from the GR cells) are fed into the PCs (i.e., principal cells of the cerebellar cortex) and the BCs in the Purkinje-molecular layer. The PCs also receive well-matched error-teaching CF (instructor) signals from the IO, together with the inhibitory inputs from the BCs. Then, the synaptic weights at the PF-PC synapses vary depending on the
matching degree between the PF and the CF signals.

We first consider the change in normalized synaptic weights $\tilde J$ of active PF-PC synapses during the learning trials in the optimal case of $p_c^* = 0.029$;
\begin{equation}
{\tilde {J}}_{ij}(t) = \frac {J_{ij}^{\rm (PC,PF)}(t)} {J_{0}^{\rm (PC,PF)}}.
\label{NSW}
\end{equation}
Here, the initial synaptic strength ($J_{0}^{\rm (PC,PF)}=0.006$) is the same for all PF-PC synapses.
Figures \ref{fig:SW}(a1)-\ref{fig:SW}(a9) show trial-evolution of distribution of $\tilde J$ of active PF-PC synapses.
As the learning trial is increased, normalized synaptic weights $\tilde J$ change due to synaptic plasticity at PF-PC synapses.
We note that the distribution of $\tilde J$ in each trial is composed of two markedly separated structures (i.e.,
a combination of separate top horizontal line with a central gap and lower band). Here, the top horizontal line with a central gap has
no essential change with the trials, while the lower bands come down with the trials and their vertical widths increase.
This kind of distribution of $\tilde J$ becomes saturated at about the 250th trial.

The top horizontal line with a central gap arises from the ill-matched firing group (with negative matching indices).
In the case of ill-matched PF signals, practically no LTD occurs because most of them have no associations with the error-teaching CF signals which are strongly localized in the middle of trial (i.e., near $t=500$ msec). As shown in Fig.~\ref{fig:WP}(c2), the activation degree $A^{(i)}(t)$ (denoted by the dotted line) of GR cells in the ill-matched firing group has a central ``zero-bottom'' where $A^{(i)}(t) \simeq 0$ (i.e., negligibly small number of spikes in the middle part of trial). In the initial and the final parts of the trial (outside the middle part), practically no LTD takes place due to no practical conjunctions with the
strongly-localized CF signals. Thus, the normalized synaptic weights $\tilde J$ of the active GR cells in the ill-matched firing group forms the top
horizontal line with a central gap which is nearly invariant with the trial.

On the other hand, lower bands arise from the well-matched firing group (with positive matching indices).
In the case of well-matched PF signals, they are strongly depressed by the error-teaching CF signals (i.e., strong LTD occurs) in each trial
due to good association between the well-matched PF and CF signals. As a result, a lower band is formed, it comes down with the trial, and eventually
becomes saturated.

To more clearly examine the above trial evolutions, we obtain the bin-averaged (normalized) synaptic weight in each $i$th bin (bin size: $\Delta t=$ 50 msec):
\begin{equation}
{\langle {\tilde J}(t) \rangle_i} = {\frac {1} {N_{s,i}}} \sum_{f=1}^{N_{s,i}} {\tilde J}_{i,f}(t),
\label{eq:BSW}
\end{equation}
where ${\tilde J}_{i,f}$ is the normalized synaptic weight of the $f$th active PF signal in the $i$th bin, and $N_{s,i}$ is the total number of
active PF signals in the $i$th bin.
Figures \ref{fig:SW}(b1)-\ref{fig:SW}(b9) show trial-evolution of bin-averaged (normalized) synaptic weights $\langle {\tilde J}(t) \rangle$
of active PF signals. In each trial,  $\langle {\tilde J}(t) \rangle$ forms a step-well-shaped curve.
As the trial is increased, the step-well curve comes down, its width and depth increase, and saturation seems to occur at about the 250th cycle.

\begin{figure}
\includegraphics[width=0.9\columnwidth]{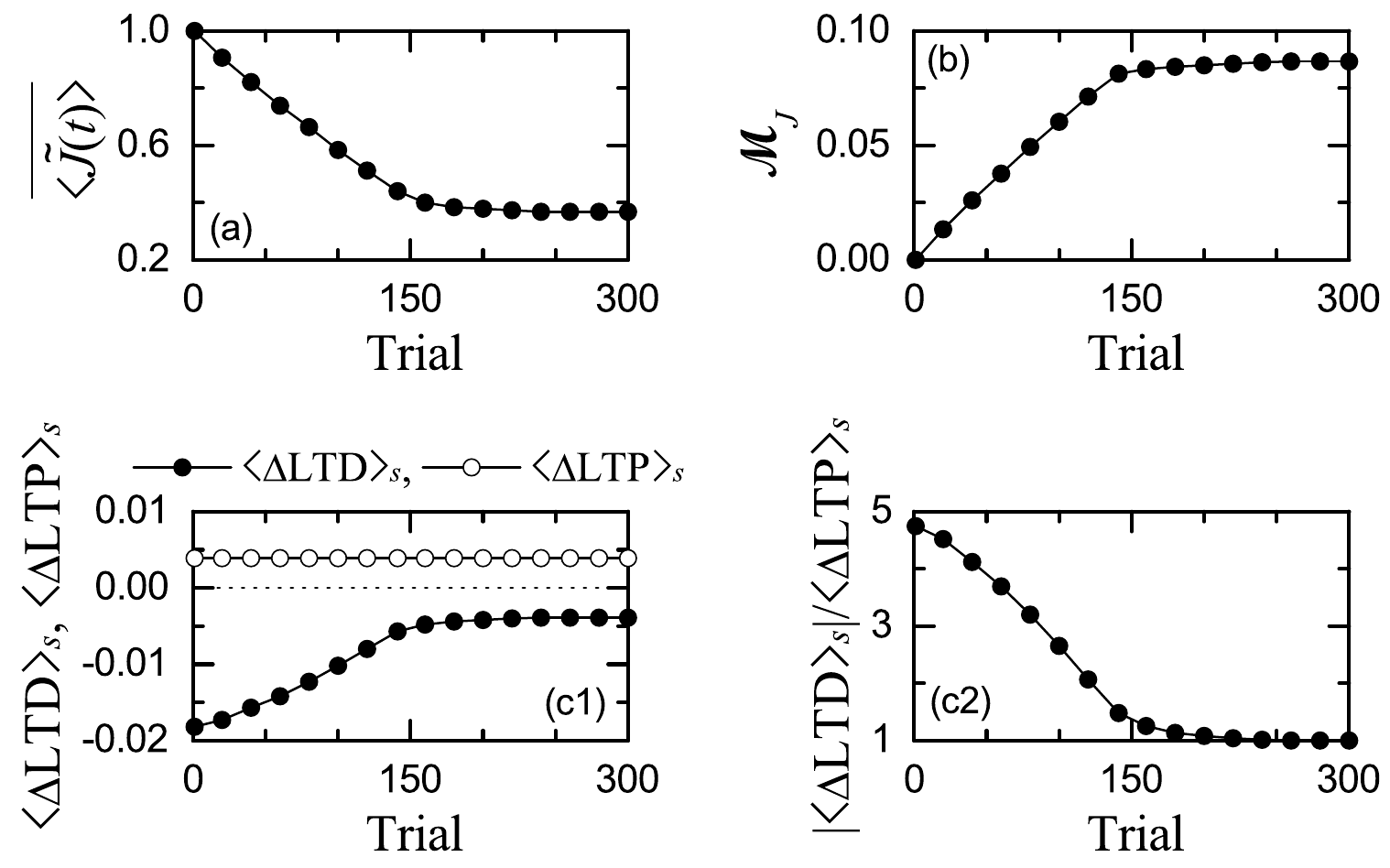}
\caption{
Plots of (a) trial-averaged mean $\overline{ \langle {\tilde J} (t) \rangle }$ and (b) modulation ${\cal{M}}_J$ for
the bin-averaged (normalized) synaptic weights $\langle {\tilde J} (t) \rangle$ in Figs.~\ref{fig:SW}(b1)-\ref{fig:SW}(b9) versus trial.
Dynamical balance between $\Delta {\rm LTD}$ and $\Delta {\rm LTP}$ during learning trials. Plots of (c1) trial-summed
$\Delta{\rm LTD}$ [$\langle \Delta{\rm LTD} \rangle_s$ (solid circles)] and $\Delta{\rm LTP}$ [$\langle \Delta{\rm LTP} \rangle_s$
(open circles)] of active PF signals and (c2) ratio of $|\langle \Delta{\rm LTD} \rangle_s|$ to $\langle \Delta{\rm LTP} \rangle_s$ versus trial.
}
\label{fig:Bal}
\end{figure}

We also obtain the trial-averaged mean $\overline{ \langle {\tilde J}(t) \rangle }$ via time average of $\langle {\tilde J}(t) \rangle$ over a trial:
\begin{equation}
\overline{ \langle {\tilde J} \rangle } = {\frac {1} {N_b}} \sum_{i=1}^{N_b} \langle {\tilde J}(t) \rangle_i.
\label{eq:MSW}
\end{equation}
Here, $N_b$ is the number of bins for trial averaging, and the overbar represents the time average.
Figures \ref{fig:Bal}(a) and \ref{fig:Bal}(b) show plots of the trial-averaged mean $\overline{ \langle {\tilde J}(t) \rangle }$
and the modulation [=(maximum - minimum)/2] ${\cal {M}}_J$ for $\langle {\tilde J}(t) \rangle$  versus trial.
The trial-averaged mean $\overline{ \langle {\tilde J}(t) \rangle }$ decreases monotonically from 1.0 due to LTD at PF-PC synapses, and it
becomes saturated at 0.367 at about the 250th trial.

However, strength of the LTD varies depending on the parts of the trial.
In the middle part without practical contribution of ill-matched firing group, strong LTD occurs, due to contribution of only well-matched active PF signals.
On the other hand, at the initial and the final parts, somewhat less LTD takes place, because both the ill-matched firing group (with practically no
LTD) and the well-matched firing group make contributions together. Consequently, with increasing trial, the middle-stage part comes down more rapidly than the initial and final parts. Hence, the modulation ${\cal {M}}_J$ increases monotonically from 0, and it gets saturated at 0.0867 at about the 250th trial.

We investigate the dynamical origin for the saturation of trial-averaged mean $\overline{ \langle {\tilde J}(t) \rangle }$
in Fig.~\ref{fig:Bal}(a). As explained in the Subsec.~\ref{subsec:SP}, a major LTD occurs when a CF firing makes association with earlier PF signals [see the $\Delta{\rm LTD}_{ij}^{(1)}$ in Eq.~(\ref{eq:LTD1})], while a minor LTD takes place when the CF firing makes conjunction with later CF signals [see the
$\Delta{\rm LTD}_{ij}^{(2)}$ in Eq.~(\ref{eq:LTD2})]. We consider the trial-summed $\Delta{\rm LTD}$ ($\langle \Delta {\rm LTD} \rangle_s$) in each trial, given by the sum of ${\Delta{\rm LTD}}_{ij}$ occurring at all active PF-PC synapses during the whole trial ($0<t<1000$ msec), where ${\Delta{\rm LTD}}_{ij} =
\Delta{\rm LTD}_{ij}^{(1)} + \Delta{\rm LTD}_{ij}^{(2)}$.
Similarly, we also consider the trial-summed $\Delta{\rm LTP}$ ($\langle \Delta {\rm LTP} \rangle_s$) in each trial, given by the sum of ${\Delta{\rm LTP}}_{ij}$ occurring at all active PF-PC synapses during the whole trial. Here, LTP occurs for the PF signals alone without association with CF signals; ${\Delta{\rm LTP}}_{ij}$ is given in Eq.~(\ref{eq:LTP}).

Figures \ref{fig:Bal}(c1) shows trial-evolution of trial-summed $\Delta{\rm LTD}$ [$\langle \Delta {\rm LTD} \rangle_s$ (solid circles)] and $\Delta{\rm LTP}$ [$\langle \Delta {\rm LTP} \rangle_s$ (open circles)] occurring at all active PF-PC synapses.
Here, $\langle \Delta {\rm LTP} \rangle_s~(\simeq 0.00386)$, independently of the trial.
In the beginning trials, $|\langle \Delta {\rm LTD} \rangle_s|$ (i.e., magnitude of $\langle \Delta {\rm LTD} \rangle_s$) is larger than $\langle \Delta {\rm LTP} \rangle_s$, and hence $\Delta {\rm LTD}$ is dominant. However, with increasing the trial, $|\langle \Delta {\rm LTD} \rangle_s|$ decreases monotonically, and becomes
saturated at 0.00386 at about the 250th trial. Thus, the saturated value of $|\langle \Delta {\rm LTD} \rangle_s|$ becomes the same as that of
$\langle \Delta {\rm LTP} \rangle_s$. This process may be well seen in Fig.~\ref{fig:Bal}(c2). With increasing the trial, the ratio of $|\langle \Delta {\rm LTD} \rangle_s|$ to $\langle \Delta {\rm LTP} \rangle_s$ decreases from 4.71, and it converges to 1 at about the 250th trial.
Thus, trial-level balance between $\Delta {\rm LTD}$ and $\Delta {\rm LTP}$ occurs
[i.e., $|\langle \Delta {\rm LTD} \rangle_s| = \langle \Delta {\rm LTP} \rangle_s~(\simeq 0.00386)]$ at about the 250th trial, which results in
the saturation of trial-averaged mean $\overline{ \langle {\tilde J}(t) \rangle }$.

\subsection{Influence of PF-PC Synaptic Plasticity on Subsequent Learning Process in The PC-CN-IO System}
\label{subsec:LP}
As a result of various recoding in the GR clusters, well- and ill-matched firing groups appear.
In the case of well-matched PF signals, they are strongly depressed by the CF signals due to good association between the
PF and CF signals. On the other hand, in the case of ill-matched PF signals, practically no LTD occurs because most of them have no conjunctions with the error-teaching CF signals. In this subsection, we investigate the influence of this kind of effective PF-PC synaptic plasticity on the subsequent learning process in the PC-CN-IO system.

\begin{figure}
\includegraphics[width=0.9\columnwidth]{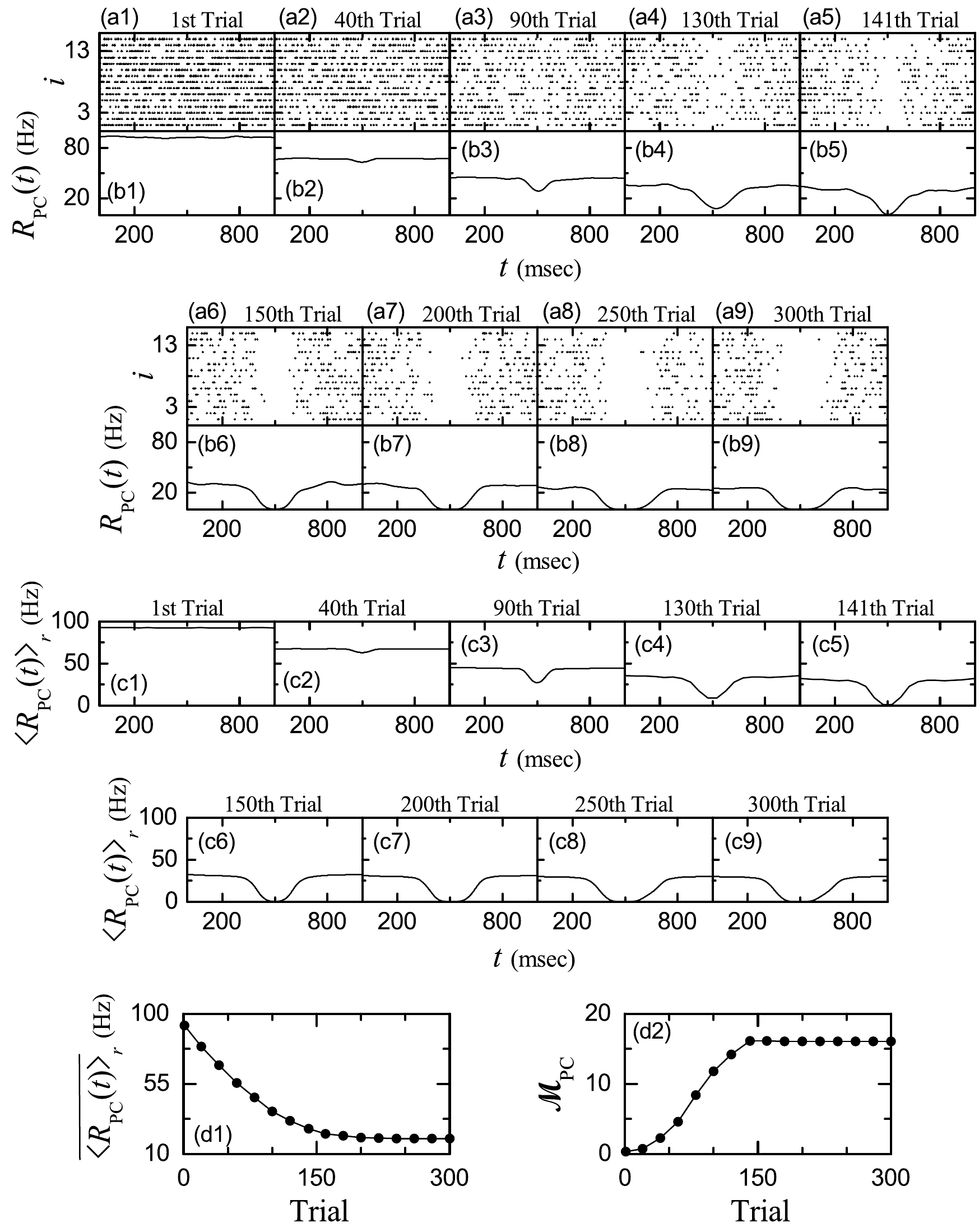}
\caption{Change in firing activity of PCs during learning trial in the optimal case of $p_c^* = 0.029$. (a1)-(a9) Raster plots of spikes of PCs and (b1)-(b9) instantaneous population spike rates $R_{\rm PC}(t)$. (c1)-(c9) Realization-averaged instantaneous population spike rates $\langle R_{\rm PC} (t) \rangle_r$; number of realizations is 100. Plots of (d1) trial-averaged mean $\overline{ \langle R_{\rm PC} (t) \rangle_r }$ and (d2) modulations ${\cal M}_{\rm PC}$ for $\langle R_{\rm PC} (t) \rangle_r$ versus trial.
}
\label{fig:PC}
\end{figure}

Figure \ref{fig:PC} shows change in firing activity of PCs during learning trial in the optimal case of $p_c^* = 0.029$.
Trial-evolutions of raster plots of spikes of 16 PCs and the corresponding instantaneous population spike rates $R_{\rm PC}(t)$ are shown
in Figs.~\ref{fig:PC}(a1)-\ref{fig:PC}(a9) and Figs.~\ref{fig:PC}(b1)-\ref{fig:PC}(b9), respectively.
Realization-averaged smooth instantaneous population spike rates $\langle R_{\rm PC}(t) \rangle_r$ are also shown in
Figs.~\ref{fig:PC}(c1)-\ref{fig:PC}(c9). Here, $\langle \cdots \rangle_r$ denotes realization average and the number of realizations is 100.
$\langle R_{\rm PC}(t) \rangle_r$ seems to be saturated at about the 250th cycle.

As shown in Figs.~\ref{fig:SW}(b1)-\ref{fig:SW}(b9), bin-averaged normalized synaptic weights $\langle {\tilde J}(t) \rangle$ form a step-well-shaped curve.
In the middle part of each trial, strong LTD occurs due to contribution of only well-matched firing group. On the other hand, at the initial and the final parts, somewhat less LTD takes place because both the ill-matched firing group (with practically no LTD) and the well-matched firing group make contributions together. As a result of this kind of effective depression at the (excitatory) PF-PC synapses, with increasing the learning trial, raster plots of spikes of all the 16 PCs become more and more sparse in the middle part of each trial (i.e, near $t=500$ msec), which may be clearly seen in the instantaneous population spike rate $\langle R_{\rm PC}(t) \rangle_r$. $\langle R_{\rm PC}(t) \rangle_r$ becomes lower in the middle part than at the initial and the final parts. Thus, $\langle R_{\rm PC}(t) \rangle_r$ also forms a step-well-shaped curve with a minimum in the middle part.

\begin{figure}
\includegraphics[width=\columnwidth]{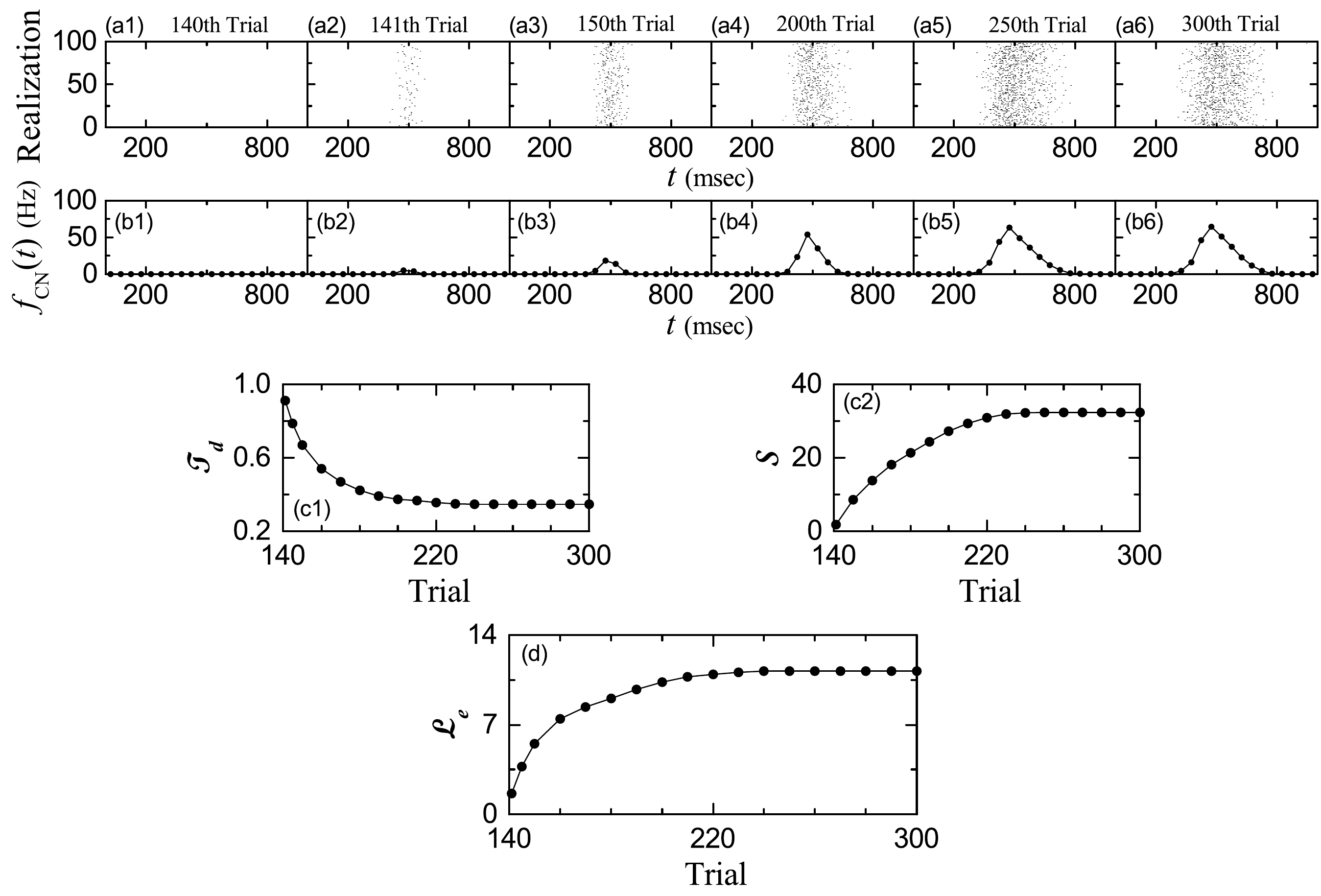}
\caption{Change in firing activity of the CN neuron during learning trial in the optimal case of $p_c^* = 0.029$.
(a1)-(a6) Raster plots of spikes of the CN neuron (i.e., collection of spike trains for all the realizations; number of realizations is 100) and (b1)-(b6) bin-averaged instantaneous individual firing rate $f_{\rm CN}(t)$; the bin size is $\Delta t=$ 50 msec. Plots of (c1) timing degree ${\cal T}_d$ and (c2)
strength $\cal S$ of CR versus trial. (d) Plot of (overall) learning efficiency degree ${\cal L}_e$ for the CR versus trial.
}
\label{fig:CN}
\end{figure}

As the trial is increased, such step-well-shaped curve for $\langle R_{\rm PC}(t) \rangle_r$ comes down and the (top) width and the depth of the well increase.
Eventually, at the 141th trial, a ``zero-bottom'' is formed in the step-well in the middle part of the trial (i.e., $\langle R_{\rm PC}(t) \rangle_r \simeq 0$ for $468 < t < 532$ msec). Appearance of the zero-bottom in the step-well is the prerequisite condition for acquisition of CR. At the zero-bottom of the step-well, PCs stop inhibition completely. This process may be seen well in Figs.~\ref{fig:PC}(c1)-\ref{fig:PC}(c5). Thus, from the 141th threshold trial, the CN neuron may fire spikes which evoke CR, which will be seen in Fig.~\ref{fig:CN}. With increasing the trial from the 141th trial, both the (top) width of the step-well and the zero-bottom width are increased, although the depth of the well remains unchanged [see Figs.~\ref{fig:PC}(c6)-\ref{fig:PC}(c9)].  As a result, the strength $\cal S$ of CR increases, while its timing degree ${\cal T}_d$ is decreased; the details will be given in Fig.~\ref{fig:CN}. The (overall) learning efficiency degree ${\cal L}_e$, taking into consideration both ${\cal T}_d$  and $\cal S$, increases with the trial, and becomes saturated at about the 250th trial.

Figures \ref{fig:PC}(d1) and \ref{fig:PC}(d2) show plots of trial-averaged mean $\overline{ \langle R_{\rm PC}(t) \rangle_r }$
(i.e., time average of $\langle R_{\rm PC}(t) \rangle_r$ over the trial) and modulation ${\cal M}_{\rm PC}$ of $\langle R_{\rm PC}(t) \rangle_r$ versus
trial, respectively. Due to effective LTD at the PF-PC synapses, the trial-averaged mean $\overline{ \langle R_{\rm PC}(t) \rangle_r }$ decreases monotonically from 92.47 Hz, and it gets saturated at 19.91 Hz at about the 250th cycle. On the other hand, the modulation ${\cal M}_{\rm PC}$ increases monotonically from 0.352 Hz, and it becomes saturated at 16.11 Hz at about the 141th cycle. After the 141th threshold trial, ${\cal M}_{\rm PC}$ remains unchanged due to
no change in the depth of the step-well, unlike the case of $\overline{ \langle R_{\rm PC}(t) \rangle_r }$. These PCs (principal cells of the cerebellar cortex) exert effective inhibitory coordination on the CN neuron which evokes the CR (i.e., learned eyeblink).

Figure \ref{fig:CN} shows change in firing activity of the CN neuron which produces the final output of the cerebellum during learning trial in the optimal case of $p_c^* = 0.029$. Trial-evolutions of raster plots of spikes of the CN neuron (i.e., collection of spike trains for all the realizations; number of realizations is 100) and  the bin-averaged instantaneous individual firing rates $f_{\rm CN}(t)$ (i.e., the number of spikes of the CN neuron in a bin with the bin width
$\Delta t = 50$ msec) are shown in Figs.~\ref{fig:CN}(a1)-\ref{fig:CN}(a6) and Figs.~\ref{fig:CN}(b1)-\ref{fig:CN}(b6), respectively.
At the 140th trial, the CN neuron does not fire due to strong inhibition from the PCs, and thus it is silent (i.e., it lies in the silent period) during the whole trial stage ($0<t<1000$ msec). However, as a result of appearance of the zero-bottom in the step-well for $\langle R_{\rm PC}(t) \rangle_r$ at the 141th threshold trial, the CN neuron begins to fire spikes in the middle part of the trial. In this case, as the time is increased from $t=0,$ the CN neuron first lies in the silent period, then a transition to the firing state occurs in the middle part, and finally another transition to the silent state also takes place. With increasing the trial, raster plots of spikes of the CN neuron become more and more dense in the middle part of each trial, in contrast to the case of PCs.

This process may be clearly seen in the instantaneous individual firing rates $f_{\rm CN}(t)$.
Due to the effective inhibitory coordinations of PCs on the CN neuron, $f_{\rm CN}(t)$ begins to increase from 0 in the middle part of the trial, it reaches a peak, and then it decreases to 0 relatively slowly. The peak of $f_{\rm CN}(t)$ appears a little earlier than the US presentation ($t=500$ msec) which may denote the anticipatory CR \cite{EB6,BN5}. Thus, $f_{\rm CN}(t)$ forms a bell-shaped curve. As the trial is increased, the bottom-base width and the peak height of the bell are increased, and $f_{\rm CN}(t)$ seems to be saturated at about the 250th trial.

Figures \ref{fig:CN}(c1) and \ref{fig:CN}(c2) show plots of the timing degree ${\cal T}_d$ and the strength $\cal S$ of the CR
versus trial, respectively. The timing degree ${\cal T}_d,$ representing the matching degree between the firing activity of the CN neuron [$f_{\rm CN}(t)$] and the US signal $f_{\rm US}(t)$, is given by the cross-correlation $Corr_{\rm T} (0)$ at the zero lag between $f_{\rm CN}(t)$ and $f_{\rm US}(t)$:
\begin{equation}
Corr_{\rm T} (\tau) = \frac {\overline {\Delta f_{\rm US}(t+\tau) \Delta f_{\rm CN}(t)}} { \sqrt{\overline{\Delta f_{\rm US}^{2}(t)}}
\sqrt{\overline{\Delta f_{\rm CN}^2(t)}}},
\label{eq:TD}
\end{equation}
where $\Delta f_{\rm US}(t) = f_{\rm US}(t)-\overline{f_{\rm US}(t)}$, $\Delta f_{\rm CN}(t) = f_{\rm CN}(t)-\overline{f_{\rm CN}(t)}$, and the overline
denotes the time average. Practically, ${\cal T}_d$ reflects the width of the bottom base of the bell curve.
With increasing the trial, the width of the bottom base increases, due to increase in the (top) width of the step-well curve for the PCs.
As a result, ${\cal T}_d$ decreases monotonically from 0.912 at the 141th trial, and it becomes saturated at 0.346 at about the 250th trial.
On the other hand, as the trial is increased, the peak height of the bell increases.
Thus, the strength $\cal S$ of CR (representing the amplitude of eyelid closure), given by the modulation [(maximum - minimum)/2] of $f_{\rm CN}(t),$ increases monotonically from 1.803 at the 141th trial, and it gets saturated at 32.38 at about the 250th trial.

\begin{figure}
\includegraphics[width=\columnwidth]{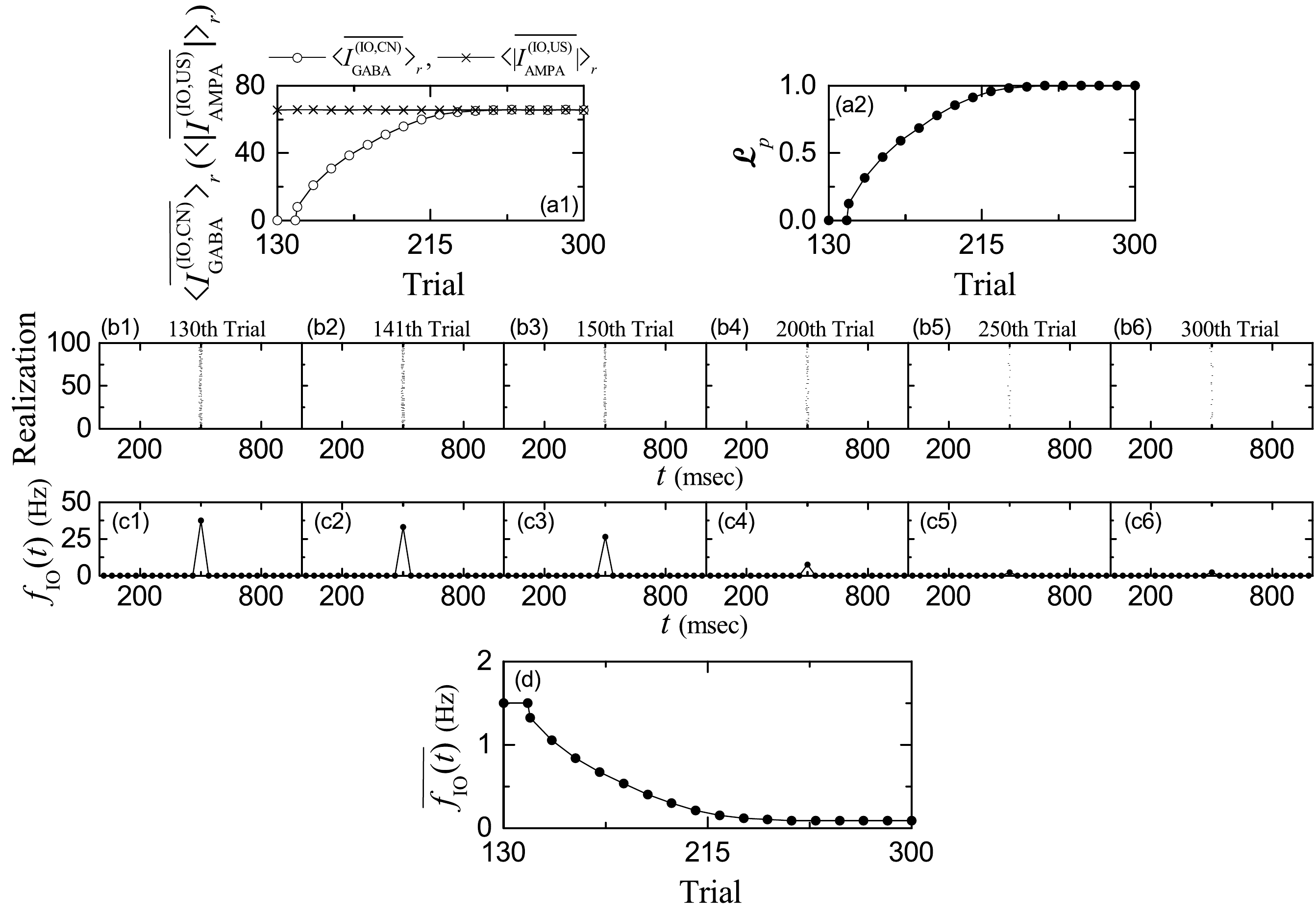}
\caption{Change in firing activity of the IO neuron during learning trial in the optimal case of $p_c^* = 0.029$.
Plots of (a1) realization-average for the time-averaged inhibitory synaptic current from the CN neuron ($\langle \overline {I_{\rm GABA}^{\rm (IO,CN)}} \rangle_r$) (open circles) and realization-average for the time-averaged excitatory synaptic current via the (airpuff) US signal ($\langle \overline {| I_{\rm AMPA}^{\rm (IO,US)}|} \rangle_r$) versus trial; number of realizations $\langle \cdots \rangle_r$ is 100.
(a2) Plot of learning progress degree ${\cal L}_p$ versus trial. (b1)-(b5) Raster plots of spikes of the IO neuron (i.e., collection of spike trains for all the realizations; number of realizations is 100) and (c1)-(c5) bin-averaged instantaneous individual firing rate $f_{\rm IO}(t)$; the bin size is $\Delta t=$ 40 msec.
(d) Plot of trial-averaged individual firing rate $\overline{ f_{\rm IO}(t) }$ versus trial.}
\label{fig:IO}
\end{figure}

Then, the (overall) learning efficiency degree ${\cal L}_e$ for the CR, taking into consideration both the timing degree ${\cal T}_d$ and the strength
$\cal S$ of CR, is given by their
product:
\begin{equation}
  {\cal L}_e = {\cal T}_d \cdot {\cal S}.
\label{eq:LED}
\end{equation}
Figure \ref{fig:CN}(d) shows a plot of ${\cal L}_e$ versus trial.
${\cal L}_e$ increases monotonically from 1.645 at the 141th trial, and it becomes saturated at about the 250th cycle.
Thus, we get the saturated learning efficiency degree ${\cal L}_e^*$ $(\simeq 11.19)$.
As will be seen in the next subsection, ${\cal L}_e^*$ is the largest one among the available ones.
Hence, in the optimal case of $p_c^*=0.029$ where firing patterns of GR clusters with the variety degree ${\cal{V}}^*~(\simeq 1.842)$
are the most various, motor learning for the EBC with the saturated learning efficiency degree
${\cal L}_e^*$ is the most effective.

Learning progress can be clearly seen in the IO system. During the learning trial, the IO neuron receives both the excitatory (airpuff) US
signal for the desired timing and the inhibitory signal from the CN neuron (representing a realized eye-movement).
Then, the learning progress degree ${\cal{L}}_p$ is given by the ratio of the time-averaged inhibitory input from the CN neuron to the magnitude of the
time-averaged excitatory input of the desired US timing signal:
\begin{equation}
{\cal L}_p = \frac {\overline{ I_{\rm GABA}^{\rm (IO,CN)} }} {|\overline{ I_{\rm AMPA}^{\rm (IO,US)} }|}.
\label{eq:LPD}
\end{equation}
Here, $\overline{ I_{\rm GABA}^{\rm (IO,CN)} }$ is the trial-averaged inhibitory GABA receptor-mediated current from the CN to the IO, and
$\overline{ I_{\rm AMPA}^{\rm (IO,US)} }$ is the trial-averaged excitatory AMPA receptor-mediated current into the IO via the desired US timing signal;
no (excitatory) NMDA receptors exist on the IO. [Note that the 4th term in Eq.~(\ref{eq:GE}) is given by $- I_{syn,i}^{(X)}(t)$, because
$I_{\rm GABA}^{\rm (IO,CN)} >0$ and $I_{\rm AMPA}^{\rm (IO,US)} <0.$]

Figure \ref{fig:IO}(a1) shows plots of $\langle \overline{ I_{\rm GABA}^{\rm (IO,CN)} } \rangle_r$ (open circles) and
$\langle |\overline{ I_{\rm AMPA}^{\rm (IO,US)} }| \rangle_r$ (crosses) versus trial in the optimal case of $p_c^* = 0.029$; $\langle \cdots \rangle_r$ represents an average over 100 realizations. At the 141th threshold trial, acquisition of CR starts (i.e., the CN neuron begins to fire spikes).
Hence, before the threshold the trial-averaged inhibitory input from the CN neuron is zero, it begins to increase from the threshold and converges to the constant trial-averaged excitatory input through the US signal for the desired timing. Thus, with increasing the trial, ${\cal{L}}_p$ is zero before the threshold, it begins to increase from the threshold, and becomes saturated at ${\cal{L}}_p =1$, as shown well in Fig.~\ref{fig:IO}(a2). In this saturated case, the trial-averaged excitatory and inhibitory inputs to the IO are balanced.

We also investigate the firing activity of IO neuron during learning process. Figures~\ref{fig:IO}(b1)-\ref{fig:IO}(b6) and Figures~\ref{fig:IO}(c1)-\ref{fig:IO}(c6)
show trial-evolutions of raster plots of spikes of the IO neuron (i.e., collection of spike trains for all the realizations; number of realizations is 100)
and  the bin-averaged instantaneous individual firing rates $f_{\rm IO}$ (i.e., the number of spikes of the IO neuron in a bin with the bin width
$\Delta t = 40$ msec), respectively.
Before the 141th threshold trial, relatively dense spikes appear in the middle part of the trial in the raster plot of spikes, due to the effect of excitatory US signal. However, with increasing the trial from the threshold, spikes in the middle part become sparse, because of increased inhibitory input from the CN neuron.
In this case, the bin-averaged instantaneous individual firing rate $f_{\rm IO}(t)$ of the IO neuron forms a bell-shaped curve due to the US
signal into the IO neuron. With increasing the trial from the 141th threshold, the amplitude of $f_{\rm IO}(t)$ begins to decrease due to the increased inhibitory input from the CN neuron, and it becomes saturated at about the 250th trial. Thus, the trial-averaged individual firing rate $\overline{ f_{\rm IO}(t) }$ is constant (=1.5 Hz) before the threshold without the inhibitory input from the CN neuron. However, with increasing the trial from the threshold, it is decreased from 1.326 Hz due to increase in the inhibitory input from the CN neuron, and gets saturated at 0.0902 Hz at about the 250th trial, as shown in Fig.~\ref{fig:IO}(d).

\begin{figure}
\includegraphics[width=0.8\columnwidth]{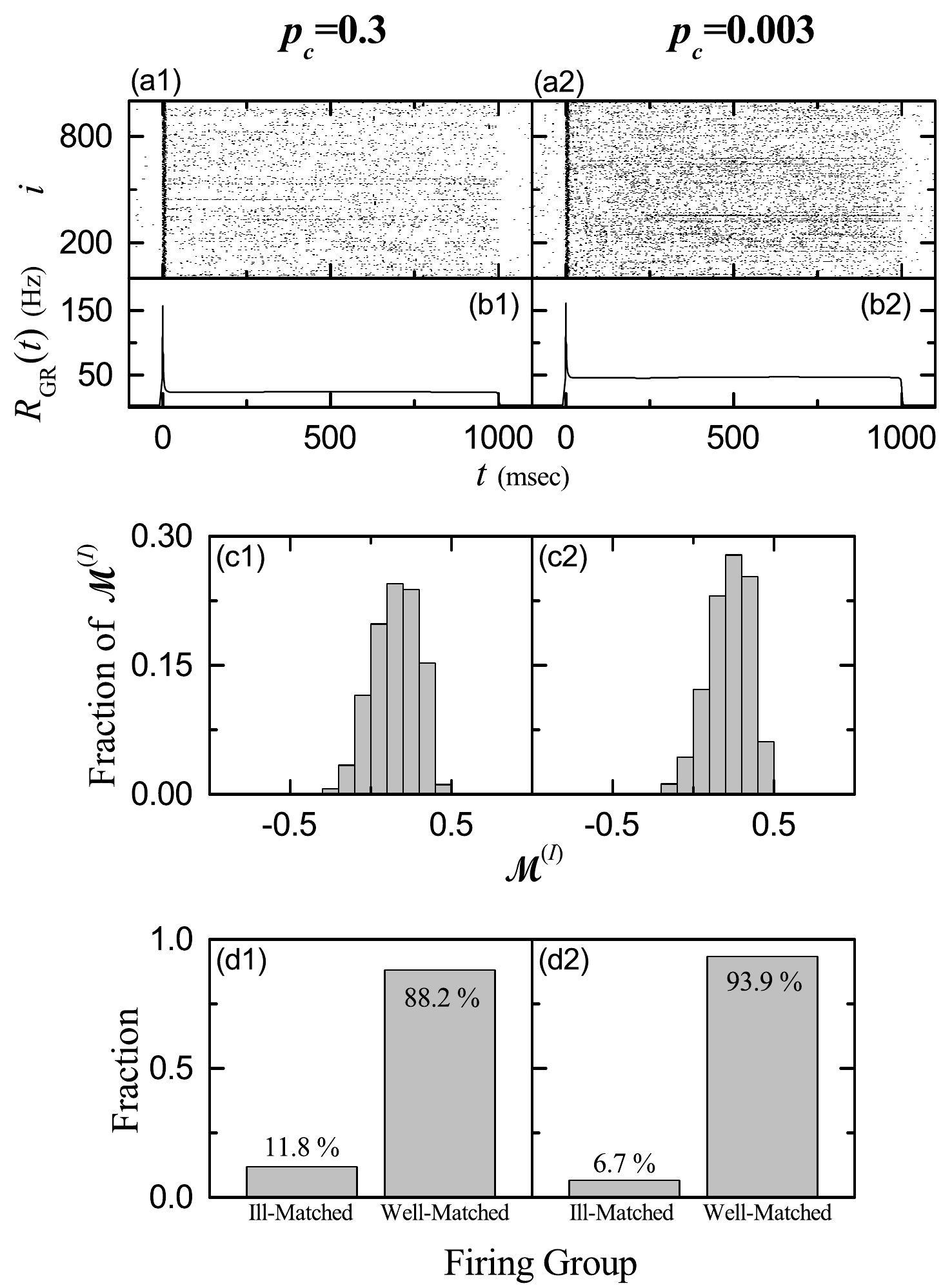}
\caption{Highly-connected ($p_c=0.3$) and lowly-connected ($p_c=0.003$) cases.
Raster plots of spikes of $10^3$ randomly chosen GR cells for (a1) $p_c=0.3$ and (a2) $p_c=0.003$.
Instantaneous whole-population spike rates $R_{\rm GR}(t)$ in the whole population of GR cells for (b1) $p_c=0.3$ and (b2) $p_c=0.003$.
Band width for $R_{\rm GR}(t)$: $h=10$ msec. Distributions of matching indices $\{ {\cal{M}}^{(I)} \}$ of the firing patterns in
the GR clusters in the whole population for (c1) $p_c=0.3$ and (c2) $p_c=0.003$. Bin size for the histograms in (c1) and (c2) is 0.1.
Fractions of well-matched and ill-matched firing groups for (d1) $p_c=0.3$ and (d2) $p_c=0.003$.
}
\label{fig:HL1}
\end{figure}

The firing output of the IO neuron is fed into the PCs via the CFs. Hence, with increasing the trial from the threshold,
the error-teaching CF instructor signals become weaker and saturated at about the 250th cycle.
While the saturated CF signals are fed into the PCs, trial-level balance between $\Delta {\rm LTD}$ and $\Delta {\rm LTP}$ occurs at the PF-PC synapses, as shown in
Figs.~\ref{fig:Bal}(c1) and \ref{fig:Bal}(c2). As a result, saturation for the trial-averaged bin-averaged synaptic weights $\overline{ \langle {\tilde J}(t) \rangle }$ appears [see Fig.~\ref{fig:SW}(c)]. Then, the subsequent learning process in the PC-CN system also gets saturated, and we obtain the saturated learning efficiency degree ${\cal L}_e^*$ $(\simeq 11.19)$, which is shown in Fig.~\ref{fig:CN}(d).

\subsection{Variation of The Connection Probability $p_c$ and Strong Correlation between Variety Degree $\cal{V}$ and Efficiency Degree ${\cal{L}}_e$ of CR}
\label{subsec:PC}
In the above subsections, we consider only the optimal case of $p_c^*=0.029$ (i.e., $2.9\%$) where the firing patterns of the GR clusters are the most various.
From now on, we change the connection probability $p_c$ from GO to GR cells, and study dependence of the variety degree $\cal V$ for the firing patterns in the GR clusters and the learning efficiency degree ${\cal L}_e$ of CR on $p_c$.

We first consider both the highly-connected case of $p_c=0.3$ (i.e., $30\%$) and the lowly-connected case of $p_c=0.003$ (i.e., $0.3\%$).
Figures \ref{fig:HL1}(a1) and \ref{fig:HL1}(a2) show the raster plots of spikes of $10^3$ randomly chosen GR cells for $p_c=0.3$ and 0.003, respectively.
The population-averaged firing activities in the whole population of GR cells may be well seen in the instantaneous whole-population spike rates $R_{\rm GR}(t)$ in Figs.~\ref{fig:HL1}(b1) and \ref{fig:HL1}(b2) for $p_c=0.3$ and 0.003, respectively.

As shown in Fig.~\ref{fig:RN}(b), each GR cluster is bounded by four glomeruli (corresponding to the terminals of the MFs) at both ends.
Each glomerulus receives inhibitory inputs from nearby 81 GO cells with the connection probability $p_c$. In the highly-connected case of $p_c=0.3$, on average, about 24 GO cell axons innervate each glomerulus. Then, each GR cell in a GR cluster receives about 97 inhibitory inputs via four dendrites which contact the four glomeruli at both ends. In this highly-connected case, inhibitory inputs from the pre-synaptic GO cells are increased, in comparison with the optimal case of $p_c^*=0.029.$ As a result, spike density in the raster plot of spikes is decreased (i.e., spikes become sparse) due to decreased individual firing rates, and
hence the flat top part of $R_{\rm GR}(t)$ becomes lowered, in comparison to the optimal case in Fig.~\ref{fig:WP}(b).

In the highly-connected case of $p_c=0.3,$ differences between total inhibitory synaptic inputs from pre-synaptic GO cells to each GR cells are decreased due to increase in the number of pre-synaptic GO cells. In addition, the excitatory inputs into each GR cells via MFs are Poisson spike trains with the same firing rates, and hence they are essentially the same. Hence, differences between the total synaptic inputs (including both the inhibitory and the excitatory inputs) into each GR cells become reduced. These less different inputs into GR cells produce less different outputs (i.e. firing activities) of GR cells, which become more similar to
the population-averaged firing activity $R_{\rm GR}(t)$ with a flat top in Fig.~\ref{fig:HL1}(b1). Thus, GR cells tend to exhibit relatively regular firings during the whole trial stage ($0<t<1000$ msec), in comparison with optimal case of $p_c^*=0.029$. Consequently, the raster plot of sparse spikes for $p_c=0.3$ becomes relatively uniform [compare Fig.~\ref{fig:HL1}(a1) with Fig.~\ref{fig:WP}(a)].

On the other hand, in the lowly-connected case of $p_c=0.003$, the inhibitory inputs from GO cells into GR cells are so much reduced, and
the excitatory MF signals into the GR cells become dominant inputs. Hence, spike density in the raster plot of spikes is increased (i.e., spikes become dense), because of increased individual firing rate, and the flat top part of $R_{\rm GR}(t)$ becomes raised, in comparison to the optimal case in Fig.~\ref{fig:WP}(b).
Furthermore, differences between the total synaptic inputs into each GR cells become reduced, because the dominant excitatory MF signals, generated by the Poisson spike trains with the same firing rates, are essentially the same, and thus firing activities of GR cells become more similar to $R_{\rm GR}(t)$ with a flat top in Fig.~\ref{fig:HL1}(b2). Hence, as in the highly-connected case, GR cells tend to show relatively regular firings during the whole trial stage.
As a result, the raster plot of dense spikes for $p_c=0.003$ also becomes relatively uniform, as shown in Fig.~\ref{fig:HL1}(a2), in comparison with the optimal case  in Fig.~\ref{fig:WP}(a).

Figures \ref{fig:HL1}(c1) and \ref{fig:HL1}(c2) show the distributions of matching indices $\{{\cal M}^{(I)} \}$ for $p_c=0.3$ and 0.003, respectively.
The ranges in the distributions of $\{ {\cal M}^{(I)} \}$ for $p_c=0.3$ and 0.003 are (-0.21, 0.44) and (-0.18, 0.48), respectively.
In both cases, their ranges are narrowed from both the positive and the negative sides, in comparison with the range (-0.49, 0.79) in the optimal case
of $p_c^*=0.029$. As explained above, in both the highly- and the lowly-connected cases of $p_c=0.3$ and 0.003, GR cells tend to exhibit relatively
regular firings in the whole trial stage, due to decrease in the differences in the total synaptic inputs from GO cells into each GR cells, which is in contrast to the optimal case of $p_c^*=0.029$ where random repetitions of transitions between bursting and silent states (both of which are
persistent long-lasting ones) occur. Then, in both the highly- and the lowly-connected cases, highly well-matched firing patterns with higher ${\cal M}^{(I)}$ and highly ill-matched firing patterns with higher magnitude $| {\cal M}^{(I)} |$ disappear, which leads to reduction in the ranges of the distributions of $\{ {\cal M}^{(I)} \}$ arise.

Due to the narrowed distribution of $\{ {\cal M}^{(I)} \}$, both the mean ($\simeq 0.239$) and the standard deviation ($\simeq 0.360$) in the highly-connected case of $p_c=0.3$ are decreased, in comparison to the optimal case where the mean and the standard deviation are 0.333 and 0.614, respectively.
Then, the variety degree ${\cal V}$ for the firing patterns [$R_{\rm GR}^{(I)}(t)$] in the GR clusters, denoting a quantitative measure for various recoding
in the granular layer, is given by the relative standard deviation for the distribution of $\{ {\cal M}^{(I)} \}$ [see Eq.~(\ref{eq:DD})].
For $p_c=0.3$, its variety degree is ${\cal V} \simeq 1.506$  which is smaller than ${\cal V}^*~(\simeq 1.842)$ in the optimal case of $p_c^*=0.029$.
Similar to the highly-connected case, for $p_c=0.003$ both the mean ($\simeq 0.272$) and the standard deviation ($\simeq 0.315$) for the distribution
of $\{ {\cal M}^{(I)} \}$ are also decreased. In this lowly-connected case, the variety degree ${\cal V}$ for the firing patterns [$R_{\rm GR}^{(I)}(t)$] in the GR clusters is ${\cal V} \simeq 1.157$  which is much smaller than ${\cal V}^*~(\simeq 1.842)$ in the optimal case. We also note that the variety degree
${\cal V}$ for $p_c=0.003$ is smaller than that for $p_c=0.3$; ${\cal V}^*=1.842~(p_c^*=0.029) >  {\cal V}=1.506~(p_c=0.3) >  {\cal V}=1.157~(p_c=0.003).$

Figures \ref{fig:HL1}(d1) and \ref{fig:HL1}(d2) show fractions of well-matched ($\{ {\cal M}^{(I)} >0 \}$) and ill-matched ($\{ {\cal M}^{(I)} <0 \}$) firing groups for $p_c=0.3$ and 0.003, respectively. In the highly-connected case of $p_c=0.3,$ the well-matched firing group is a major one with fraction 88.2$\%$, while the ill-matched firing group is a minor one with fraction 11.8$\%$. In comparison with the optimal case where the fraction of well-matched firing group is 82.1$\%$, the fraction of well-matched firing group for $p_c=0.3$ is increased. In this highly-connected case, the firing-group ratio, given by the ratio of the fraction of the well-matched firing group to that of the ill-matched firing group, is ${\cal{R}}_{\rm sp} \simeq 7.48$ which is larger than that (${\cal R}_{sp}^* \simeq 4.59$) in the optimal case. In the lowly-connected case of $p_c=0.003,$ the fraction of well-matched firing group is more increased to 93.9$\%$. Hence, the firing-group ratio is ${\cal{R}}_{\rm sp} \simeq 13.8$ which is much larger than that in the optimal case.

\begin{figure}
\includegraphics[width=\columnwidth]{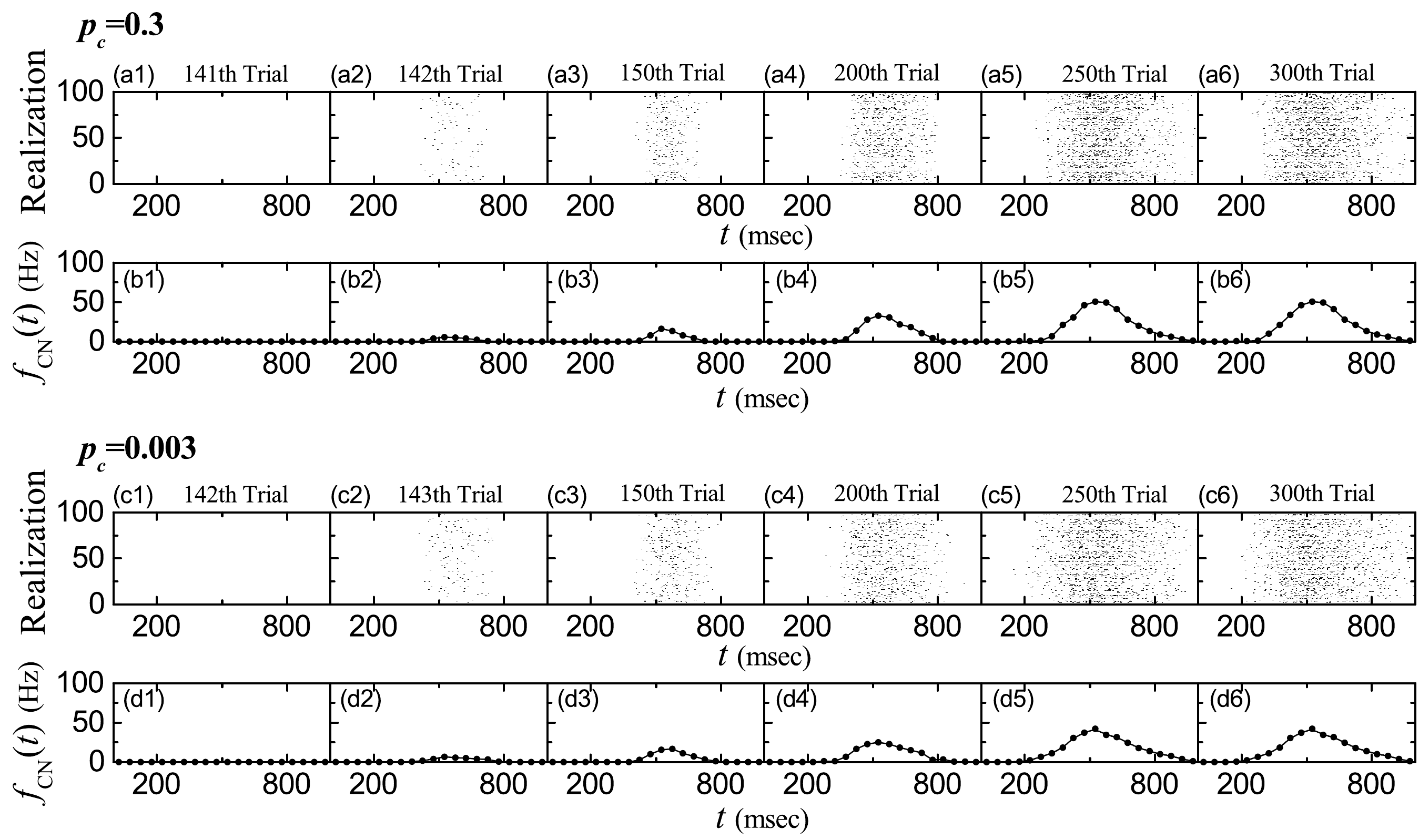}
\caption{Change in firing activity of the CN neuron in the highly-connected ($p_c=0.3$) and the lowly-connected ($p_c=0.003$) cases.
Case of $p_c=0.3$: (a1)-(a6) raster plots of spikes of the CN neuron (i.e., collection of spike trains for all the realizations; number of realizations is 100) and (b1)-(b6) bin-averaged instantaneous individual firing rate $f_{\rm CN}(t)$; the bin size is $\Delta t=50$ msec.
Case of $p_c=0.003$: (c1)-(c6) raster plots of spikes of the CN neuron and (d1)-(d6) bin-averaged instantaneous individual firing rate $f_{\rm CN}(t)$.
}
\label{fig:HL2}
\end{figure}

Due to decrease in differences between the total synaptic inputs into each GR cells, firing activities of GR cells for $p_c=0.3$ and 0.003 become more similar to the population-averaged firing activity $R_{\rm GR}(t)$, in comparison with the optimal case of $p_c^*=0.029$. We note that $R_{\rm GR}(t)$ is well-matched with the US signal $f_{\rm US}(t)$ [i.e, $R_{\rm GR}(t)$ has a positive conjunction index with respect to $f_{\rm US}(t)$], which results in increase in the fraction of  well-matched firing group for $p_c=0.3$ and 0.003. In contrast to the case of $p_c=0.3,$ in the lowly-connected case of $p_c=0.003,$ inhibitory inputs into each GR cells are so much reduced, and hence the dominant inputs are just the excitatory MF signals which are well-matched with the US signal $f_{\rm US}(t)$.
Thus, the fraction of well-matched firing group for $p_c=0.003$ becomes larger than that for $p_c=0.3$.

\begin{figure}
\includegraphics[width=0.85\columnwidth]{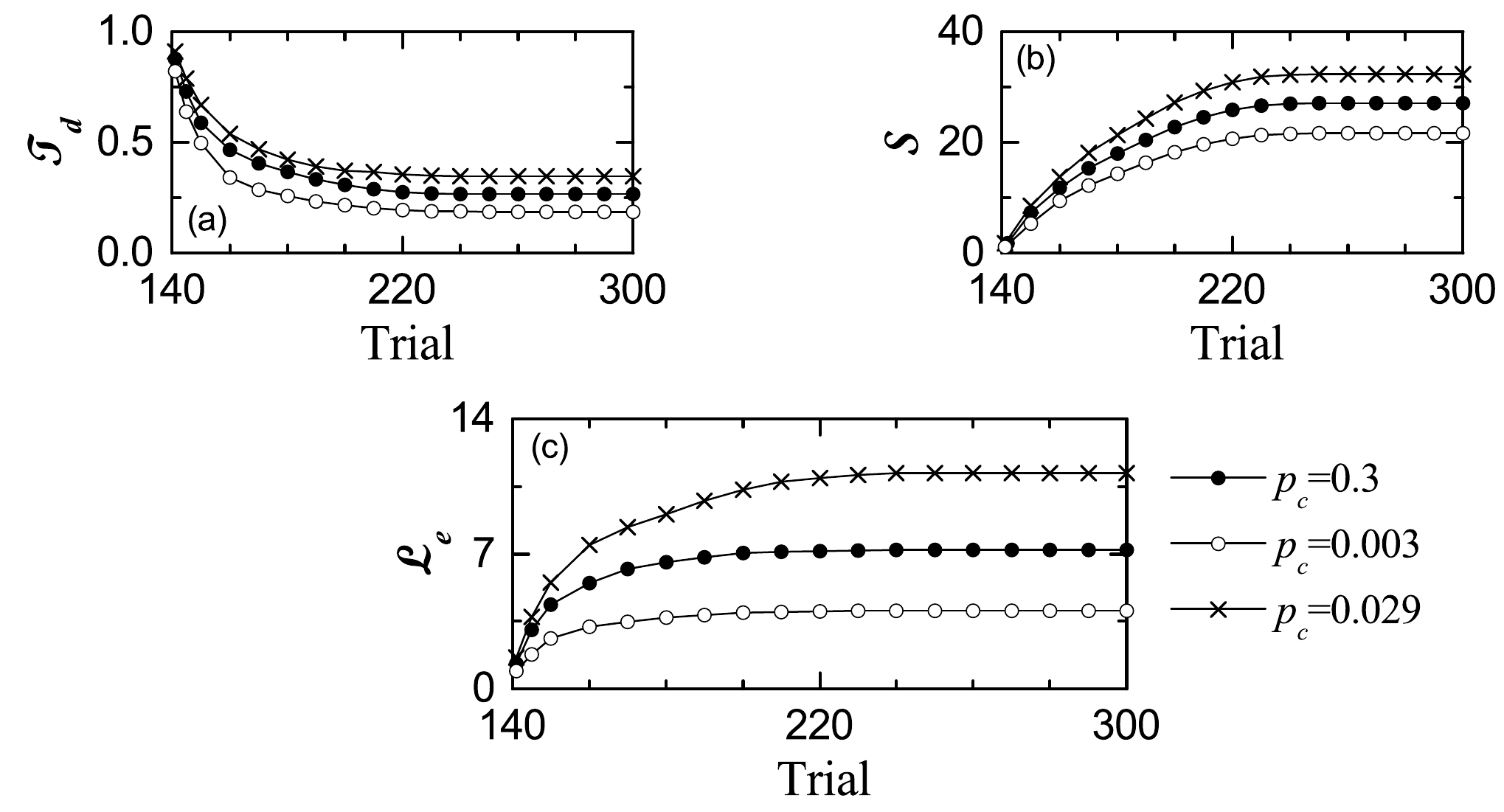}
\caption{Timing degree, strength, and learning efficiency degree of the CR in the highly-connected ($p_c=0.3$) and the lowly-connected ($p_c=0.003$) cases.
(a) Plots of timing degree ${\cal T}_d$ of the CR versus trial. (b) Plots of strengths $\cal S$ of the CR versus trial.
(c) Plots of learning efficiency degree ${\cal L}_e$ for the CR versus trial.
Solid circles, open circles, and crosses represent data in the cases of $p_c=$ 0.3, 0.003, and 0.029, respectively.
}
\label{fig:HL3}
\end{figure}

These changes in the variety degree $\cal V$ of the firing patterns in the GR clusters have direct effect on the synaptic plasticity at the PF-PC synapses and the subsequent learning process in the PC-CN system. As shown in the optimal case of $p_c^*=0.029$, the ill- and the well-matched firing groups play their own roles for the CR. The ill-matched firing group plays a role of protection barrier for the timing of CR, while the strength of CR is determined by strong LTD in the well-matched firing group. Due to break-up of highly well-matched and highly ill-matched firing patterns, the distributions of $\{ {\cal M}^{(I)} \}$ for the highly- and the lowly-connected cases of $p_c=0.3$ and 0.003 are narrowed. Hence, both the timing degree ${\cal T}_d$ and the strength $\cal S$ of CR are decreased for both $p_c=0.3$ and 0.003, which are well shown in Figs.~\ref{fig:HL2} and \ref{fig:HL3}.

Figure \ref{fig:HL2} shows change in the firing activity of the CN neuron which generates the final output of the cerebellum during learning trial in the highly- and the lowly-connected cases of $p_c=0.3$ and 0.003. For $p_c=0.3,$ trial-evolutions of the raster plots of spikes of the CN neuron (i.e., collection of spike trains for all the realizations; number of realizations is 100) and  the bin-averaged instantaneous individual firing rates $f_{\rm CN}(t)$ (i.e., the number of spikes of the CN neuron in a bin with the bin width $\Delta t = 50$ msec) are shown in Figs.~\ref{fig:HL2}(a1)-\ref{fig:HL2}(a6) and Figs.~\ref{fig:HL2}(b1)-\ref{fig:HL2}(b6), respectively. In this highly-connected case, at the 142th threshold trial, the CN neuron begins to fire in the middle part of the trial. Thus, acquisition of CR occurs a little later, in comparison with the optimal case of $p_c^*=0.029$ with the 141th threshold. In this case, $f_{\rm CN}(t)$  forms a bell-shaped curve.
With increasing the trial from the threshold, raster plots of spikes of the CN neuron become more and more dense in the middle part of each trial, and bottom-base width and peak height of the bell curve for $f_{\rm CN}(t)$ increase. At about the 250th trial, $f_{\rm CN}(t)$  seems to become saturated.

For $p_c=0.3,$ the bottom-base width of the bell curve is wider and its peak height is shorter, in comparison with the optimal case (see Fig.~\ref{fig:CN}). Due to break-up of highly ill-matched firing patterns (which play the role of protection barrier for timing of CR), bottom-base width (associated with the reciprocal of timing degree of CR) of the bell increases. Also, peak height of the bell (related to the strength of CR) decreases because of break-up of highly well-matched firing patterns (which induce strong LTD and determine the strength of CR). Consequently, as the variety degree ${\cal V}$ of the firing patterns in the GR clusters is deceased from 1.842 ($p_c^*=0.029$) to 1.507 ($p_c=0.3$), the bottom-base width of the bell curve is increased, and the peak height is decreased.

For $p_c=0.003,$ trial-evolutions of the raster plots of spikes of the CN neuron and the bin-averaged instantaneous individual firing rates $f_{\rm CN}(t)$
are shown in Figs.~\ref{fig:HL2}(c1)-\ref{fig:HL2}(c6) and Figs.~\ref{fig:HL2}(d1)-\ref{fig:HL2}(d6), respectively. In this lowly-connected case,
acquisition of CR occurs at the 143th trial which is a little later in comparison with the 142th threshold for $p_c=0.3$.
Similar to the highly-connected case, with increasing the trial from the threshold, raster plots of spikes of the CN neuron become more and more dense in the middle part of each trial, and the bottom-base width and the peak height of the bell curve for $f_{\rm CN}(t)$ increase. Eventually, saturation occurs at about the 250th trial. In comparison to the highly-connected case, the bottom-base width of the bell curve is wider and its peak height is shorter, because of more break-up of highly ill-matched firing patterns and highly well-matched firing patterns (which results in more decrease in the variety degree of firing patterns).
As a result, with decreasing the variety degree ${\cal V}$ from 1.507 ($p_c=0.3$) to 1.157 ($p_c=0.003$), the bottom-base width of the bell curve increases, and the peak height decreases (i.e., less variety in the firing patterns results in decrease in the timing degree and the strength of CR).

Figures \ref{fig:HL3}(a) and \ref{fig:HL3}(b) show plots of the timing degree ${\cal T}_d$ and the strength $\cal S$ of CR
versus trial for $p_c=0.3$ (solid circles), 0.003 (open circles), and 0.029 (crosses), respectively. The timing degree ${\cal T}_d,$ denoting the matching degree between the firing activity of the CN neuron [$f_{\rm CN}(t)$] and the US signal $f_{\rm US}(t)$ for a timing, is given by the cross-correlation $Corr_{\rm T} (0)$ at the zero lag between $f_{\rm CN}(t)$ and $f_{\rm US}(t)$ in Eq.~(\ref{eq:TD}). This timing degree ${\cal T}_d$ reflects the width of the bottom base of the bell curve. With increasing the trial, the width of the bottom base increases, as shown in Fig.~\ref{fig:HL2}, and hence ${\cal T}_d$ decreases monotonically, and it becomes saturated at about the 250th trial. We note that, as the variety degree ${\cal V}$ of the firing patterns is decreased ($\cal V=$ 1.507, 1.157, and 1.842
for $p_c=0.3,$ 0.003, and 0.029, respectively), ${\cal T}_d^*$ (saturated value of ${\cal T}_d$) decreases; ${\cal T}_d^* = 0.346~(p_c^*=0.029) > {\cal T}_d^*=0.266~(p_c=0.3) > {\cal T}_d^*=0.187~(p_c=0.003)$.

On the other hand, as the trial is increased, the peak height of the bell increases. Hence, the strength $\cal S$ of CR, given by the modulation of $f_{\rm CN}(t),$ increases monotonically, and it becomes saturated at about the 250th trial. In this case, ${\cal S}^*$ (saturated value of $\cal S$) also is decreased with decrease in the variety degree ${\cal V}$; ${\cal S}^*= 32.382~(p_c^*=0.029) > {\cal S}^*= 27.099 ~(p_c=0.3) > {\cal S}^*= 21.656~(p_c=0.003)$.

We then consider the learning efficiency degree ${\cal L}_e$ for the CR, given by product of the timing degree ${\cal T}_d$ and the strength $\cal S$ in
Eq.~(\ref{eq:LED}). Figure \ref{fig:HL3}(c) shows a plot of ${\cal L}_e$ versus trial.
${\cal L}_e$ increases monotonically from the threshold trial, and it becomes saturated at about the 250th cycle.
Thus, we obtain the saturated learning efficiency degree ${\cal L}_e^*$, the values of which are 7.216, 4.054, and 11.19 for $p_c=$ 0.3, 0.003, and 0.029, respectively. Among the three cases, ${\cal L}_e^*~(=11.19)$ in the optimal case is the largest one, and ${\cal L}_e^*$ is decreased with decrease
in the variety degree $\cal V$; ${\cal L}_e^* = 11.19~(p_c^*=0.029) > {\cal L}_e^*=7.216~(p_c=0.3) > {\cal L}_e^*=4.054 ~(p_c=0.003)$.

\begin{figure}
\includegraphics[width=0.8\columnwidth]{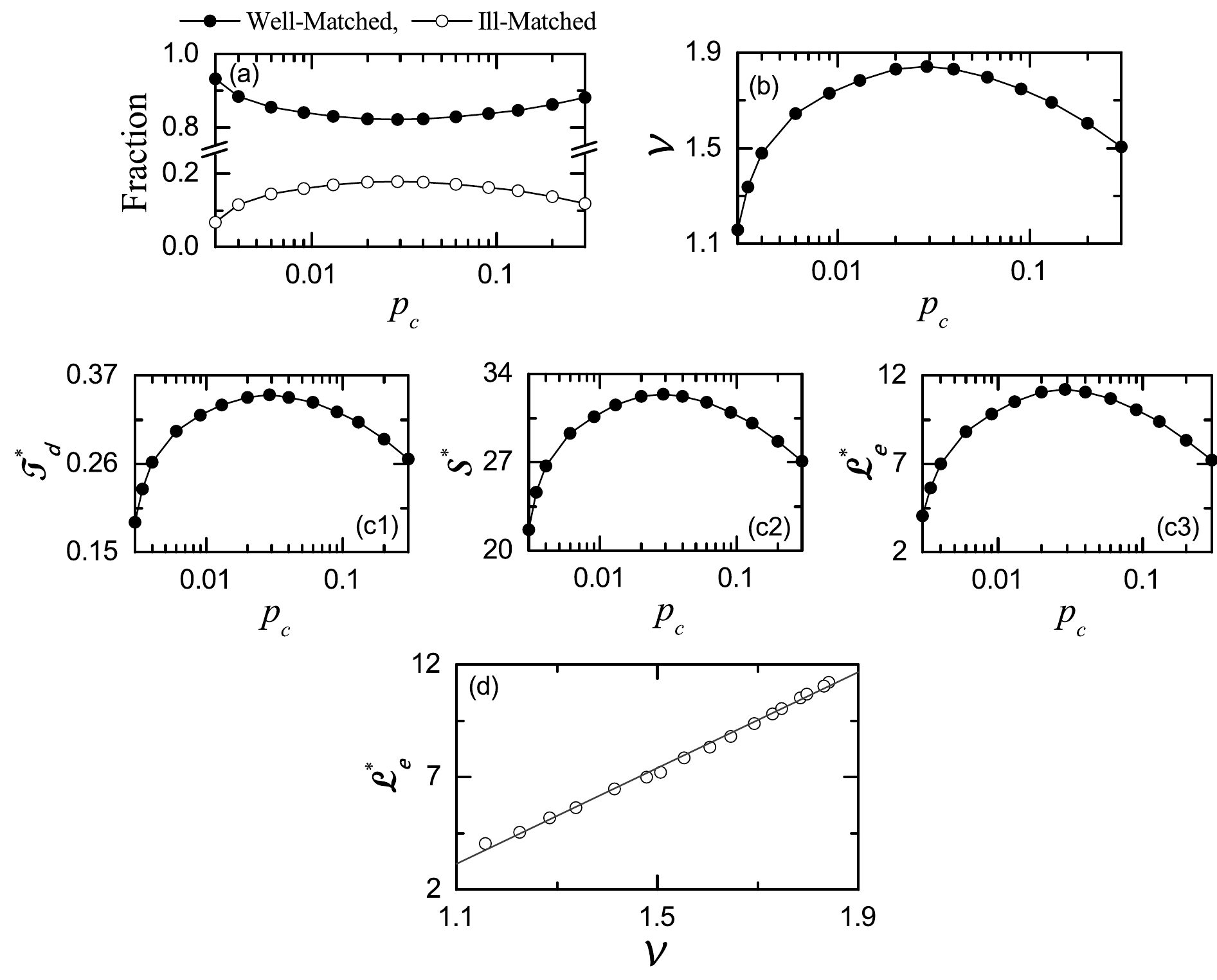}
\caption{Strong correlation between the variety degree $\cal {V}$ and the saturated learning efficiency degree ${\cal L}_e^*$.
(a) Fractions of well-matched (solid circles) and ill-matched (open circles) firing groups versus the connection probability $p_c$.
(b) Plot of variety degree $\cal {V}$ of the firing patterns in the GR clusters versus $p_c$.
Plots of saturated (c1) timing degree ${\cal T}_d^*$, (c2) strengths $\cal S^*$, and (c3) learning efficiency degree ${\cal L}_e^*$
of the CR versus $p_c$. (d) Plot of ${\cal L}_e^*$ versus $\cal {V}$.
}
\label{fig:Final}
\end{figure}

Finally, based on the above two examples for the highly- and the lowly-connected cases, we investigate dependence of the variety degree $\cal V$ for the firing patterns of the GR clusters and the saturated learning efficiency degree ${\cal L}_e^*$ on $p_c$ by varying it from the optimal value ($p_c^*=0.029$).
Figure \ref{fig:Final}(a) shows plots of fractions of the well- and the ill-matched firing groups versus $p_c$.
The fraction of the well-matched firing group (solid circles) forms a well-shaped curve with a minimum in the optimal case of $p_c^*~(=0.029)$, while
the fraction of the ill-matched firing group (open circles) forms a bell-shaped curve with a maximum in the optimal case.
In the optimal case of $p_c^*=0.029$, the firing-group ratio (i.e., ratio of fraction of the well-matched firing group to fraction of the ill-matched firing group) is ${\cal R}_{\rm sp}^* = 4.59$. As $p_c$ is changed (i.e., increased or decreased) from $p_c^*$, the fraction of the well-matched firing group increases, and then the firing-group ratio ${\cal R}_{\rm sp}$ increases from ${\cal R}_{sp}^* $ in the optimal case.

Figure \ref{fig:Final}(b) show a plot of the variety degree $\cal V$ for the firing patterns in the GR clusters.
The variety degrees $\cal V$ forms a bell-shaped curve with a maximum ${\cal V}^*~(\simeq 1.842)$ in the optimal case of $p_c^*=0.029$.
With changing $p_c$ from $p_c^*$, $\cal V$ decreases from ${\cal V}^*$. Hence, in the optimal case, temporal recoding of GR cells is the most various.
Figures \ref{fig:Final}(c1)-\ref{fig:Final}(c3) show plots of the saturated timing degree ${\cal T}_d^*$, the saturated strength ${\cal S}^*$, and the
saturated learning efficiency degree ${\cal L}_e^*$ of CR, respectively. All of them form bell-shaped curves with maxima ${\cal T}_d^*~(\simeq 0.346)$, ${\cal S}^*~(\simeq 32.38)$, and ${\cal L}_e^*~(\simeq 11.19)$ in the optimal case of $p_c^*=0.029$. As $p_c$ is changed from $p_c^*$, ${\cal T}_d^*$, ${\cal S}^*$, and ${\cal L}_e^*$ are decreased. Figure \ref{fig:Final}(d) shows a plot of ${\cal L}_e^*$ versus $\cal V$. As shown clearly in Fig.~\ref{fig:Final}(d), both
${\cal L}_e^*$ and $\cal V$ have a strong correlation with the Pearson's correlation coefficient $r\simeq 0.9982$.
Consequently, the more various in temporal recoding of the GR cells, the more effective in learning for the Pavlovian EBC.

\section{Summary and Discussion}
\label{sec:SUM}
We are concerned about the Pavlovian EBC. Various works on the EBC have been done experimentally in
many mammalian species such as humans, monkeys, dogs, ferrets, rabbits, rats, and mice \cite{Men,GA2,GA1,GA5,GA3,GA4,Mice,Rabbit1,Rabbit2}.
Also, computational works reproduced some features (e.g., representation of time passage) of the EBC in artificial models \cite{AN1,AN2,AN3,AN4,AN5,AN6}, a realistic biological model \cite{BN1,BN2,BN3}, a rate-coding model \cite{BN4}, and a spiking neural network model \cite{BN5}.
However, more clarification is necessary for influences of various temporal recoding in GR clusters on the Pavlovian EBC.

To the best of our knowledge, for the first time, we made complete quantitative classification of various firing patterns in the GR clusters in terms of the newly-introduced matching index $\{ {\cal M}^{(I)} \}$ and variety degree $\cal V$ in the case of Pavlovian EBC. Each firing pattern is characterized in terms of its matching index ${\cal M}^{(I)}$ between the firing pattern and the US signal for the desired timing. Then, the whole firing patterns are clearly decomposed into the well-matched (${\cal M}^{(I)}>0$) and the ill-matched (${\cal M}^{(I)}<0$) firing groups. Furthermore, the degree of various recoding of the GR cells may be quantified in terms of the variety degree $\cal V$, given by the relative standard deviation in the distribution of $\{ {\cal M}^{(I)} \}$.
Thus, $\cal V$ provides a quantitative measure for various temporal recoding of GR cells. It has also been shown that various total synaptic inputs (including both the excitatory inputs via MFs and the inhibitory inputs from the pre-synaptic GO cells) into the GR clusters result in generation of various firing patterns (i.e. outputs) in the GR clusters.

Based on the above dynamical classification of various firing patterns in the GR clusters, we made clear investigations on the influence of various recoding
of GR cells on the Pavlovian EBC (i.e., their effect on the synaptic plasticity at the PF-PC synapses and the subsequent learning process in the PC-CN-IO system).
To the best of our knowledge, this kind of approach, based on the well- and the ill-matched firing groups, is unique in studying the Pavlovian EBC.
The well-matched firing patterns are strongly depressed by the error-teaching CF signals, and they make dominant contributions in the middle part of
each trial. In contrast, for the ill-matched firing patterns with central gaps in the middle part, practically no LTD occurs due to no matching with the CF signals. Thus, in the middle part of each trial, strong LTD occurs via dominant contributions of well-matched firing group. On the other hand, at the initial and the final parts of each trial, less LTD takes place because both the ill-matched firing group with practically no LTD and the well-matched firing group
make contributions together. As a result, a big modulation in bin-averaged synaptic weight $\langle {\tilde J} \rangle$ arises via interplay of the well- and ill-matched firing groups.

Due to this type of effective synaptic plasticity at the PF-PC synapses, the (realization-averaged) population spike rate $\langle R_{\rm PC}(t)  \rangle$ of PCs
forms a step-well-shaped curve with a minimum in the middle part of each trial. When passing a threshold trial, a ``zero-bottom'' (where $\langle R_{\rm PC}(t)
\rangle \simeq 0$) appears in the central well. At this threshold trial, the CN neuron begins to fire  in the middle part of trial. Hence, appearance of the zero-bottom in the step-well for $\langle R_{\rm PC}(t)  \rangle$ is a prerequisite condition for acquisition of CR.
In the subsequent trials, the individual firing rate $f_{\rm CN}(t)$ of the CN neuron forms a bell-shaped curve with a maximum in the middle part (which is up-down reversed with respect to  $\langle R_{\rm PC}(t) \rangle$). Outside the bottom of the bell, the CN neuron cannot fire, due to inhibition of ill-matched firing group
(with practically no LTD). Hence, the ill-matched firing group plays a role of protection barrier for timing, and the timing degree of CR is
reciprocally associated with the bottom width of the bell. In this case, the peak of the bell in the middle part is formed due to strong LTD in the well-matched firing group, and its height is directly related to the strength of CR (corresponding to the amplitude of eyelid closure). In this way, both the well- and the ill-matched firing groups play their own roles for the timing and the strength of CR, respectively.

\begin{table}
\caption{Parameter values for LIF neuron models with AHP currents for the granule (GR) cell and the Golgi (GO) cell in the granular layer, the Purkinje cell (PC) and the basket cell (BC) in the Purkinje-molecular layer, and the cerebellar nucleus (CN) and the inferior olive (IO) neurons.
}
\label{tab:SingleParm}
\begin{tabular}{|c|c|c|c|c|c|c|c|}
\hline
\multicolumn{2}{|c|}{} & \multicolumn{2}{c|}{\multirow{2}{*}{Granular}} & \multicolumn{2}{c|}{Purkinje} &  &  \\
\multicolumn{2}{|c|}{\multirow{3}{*}{$X$-population}} & \multicolumn{2}{c|}{\multirow{2}{*}{Layer}} & \multicolumn{2}{c|}{-Molecular} & \multirow{2}{*}{CN} &  \multirow{2}{*}{IO} \\
\multicolumn{2}{|c|}{} & \multicolumn{2}{c|}{} & \multicolumn{2}{c|}{Layer} & \multirow{2}{*}{neuron} &  \multirow{2}{*}{neuron} \\ \cline{3-6}
\multicolumn{2}{|c|}{} & GR & GO & \multirow{2}{*}{PC} & \multirow{2}{*}{BC} & & \\
\multicolumn{2}{|c|}{} & cell & cell & & & & \\
\hline
\multicolumn{2}{|c|}{$C_X$} & 3.1 & 28.0 & 107.0 & 107.0 & 122.3 & 10.0 \\ \hline
\multirow{2}{*}{$I_L^{(X)}$} & $g_L^{(X)}$ & 0.43 & 2.3 & 2.32 & 2.32 & 1.63 & 0.67 \\ \cline{2-8}
& $V_L^{(X)}$ & -58.0 & -55.0 & -68.0 & -68.0 & -56.0 & -60.0 \\ \cline{2-8}
\hline
\multirow{4}{*}{$I_{AHP}^{(X)}$} & $\bar{g}_{AHP}^{(X)}$ & 1.0 & 20.0 & 100.0 & 100.0 & 50.0 & 1.0 \\ \cline{2-8}
& $\tau_{AHP}^{(X)}$ & 5.0 & 5.0 & 5.0 & 2.5 & 2.5 & 10.0 \\ \cline{2-8}
& $V_{AHP}^{(X)}$ & -82.0 & -72.7 & -70.0 & -70.0 & -70.0 & -75.0 \\ \cline{2-8}
& $v_{th}^{(X)}$ & -35.0 & -52.0 & -55.0 & -55.0 & -38.8 & -50.0 \\ \cline{2-8}
\hline
\multicolumn{2}{|c|}{$I_{ext}^{(X)}$} & 0.0 & 0.0 & 250.0 & 0.0 & 0.0 & 0.0 \\
\hline
\end{tabular}
\end{table}

By changing $p_c,$ we investigated dependence of the variety degree $\cal V$ of the firing patterns and the saturated learning efficiency degree ${\cal L}_e^*$ for the CR (given by the product of the timing degree and the strength of CR) on $p_c.$  Both $\cal V$ and ${\cal L}_e^*$ have been found to form bell-shaped curves with peaks (${\cal{V}}^* \simeq 1.842$ and ${{\cal{L}}_e}^*(\simeq 11.19)$) at the same optimal value of $p_c^*~=0.029$. In Refs.~\cite{Yama1,BN5} where the parameter values were taken, based on physiological data, the average number of nearby GO cell axons innervating each GR cell is about 8, which is very close to that in the optimal case. Thus, we hypothesize that the granular layer in the cerebellar cortex has evolved toward the goal of the most various recoding.
Moreover, Both $\cal V$ and ${\cal L}_e^*$ have also been found to have a strong correlation with the Pearson's correlation coefficient $r \simeq 0.9982$.
Hence, the more various in the firing patterns of GR cells, the more efficient in learning for the Pavlovian EBC, which is the main result in our work.

To examine our main result, we also suggest a real experiment for the EBC. To control $p_c$ in a given species of animals (e.g., a species of rabbit, rat, or mouse) in an experiment seems to be practically difficult, in contrast to the case of computational neuroscience where $p_c$ may be easily varied.
Instead, we may consider an experiment for several species of animals (e.g., 3 species of rabbit, rat, and mouse).
In each species, a large number of randomly chosen GR cells ($i=1, \cdots, M$) are considered.
Then, through many CS-US trials, one may get the peristimulus time histogram (PSTH) for each $i$th GR cell [i.e., (bin-averaged) instantaneous individual firing rate $f_{\rm GR}^{(i)}(t)$ of the $i$th GR cell]. GR cells are expected to exhibit various PSTHs.
Then, in the case of each $i$th GR cell, one can get its matching index ${\cal M}_i$ between its PSTH $f_{\rm GR}^{(i)}(t)$ and the CF signal for the desired timing [i.e., the PSTH of the IO neuron $f_{\rm IO}(t)$]. In this case, the conjunction index ${\cal M}_i$ is given by the cross-correlation at the zero-time lag between $f_{\rm GR}^{(i)}(t)$ and $f_{\rm IO}(t)$. Thus, one may get the variety degree ${\cal V}$ of PSTHs of GR cells, given by the relative standard deviation in the distribution of $\{ {\cal M}_i \}$, for the species.

\begin{table*}
\caption{Parameter values for synaptic currents $I_R^{(T,S)}(t)$ into the granule (GR) and the Golgi (GO) cells in the granular layer, the Purkinje cells (PCs) and the basket cells (BCs) in the Purkinje-molecular layer, and  the cerebellar nucleus (CN) and the inferior olive (IO) neurons in the other parts. In the granular layer, the GR cells receive excitatory inputs via mossy fibers (MFs) and inhibitory inputs from GO cells, and the GO cells receive excitatory inputs via parallel fibers (PFs) from GR cells. In the Purkinje-molecular layer, the PCs receive two types of excitatory inputs via PFs from GR cells and through climbing fibers (CFs) from the IO and one type of inhibitory inputs from the BCs. The BCs receive excitatory inputs via PFs from GR cells. In the other parts, the CN neuron receives excitatory inputs via MFs and inhibitory inputs from PCs, and the IO neuron receives excitatory input via the US signal and inhibitory input from the CN neuron.
}
\label{tab:SynParm}
\begin{tabular}{|c|c|c|c|c|c|c|c|c|c|c|c|c|c|c|}
\hline
& \multicolumn{5}{c|}{Granular Layer} & \multicolumn{4}{c|}{Purkinje-Molecular Layer} & \multicolumn{5}{c|}{Other Parts} \\
\hline
Target & \multicolumn{3}{c|}{\multirow{2}{*}{GR}} & \multicolumn{2}{c|}{\multirow{2}{*}{GO}} & \multicolumn{3}{c|}{\multirow{2}{*}{PC}} & \multirow{2}{*}{BC} & \multicolumn{3}{c|}{\multirow{2}{*}{CN}} & \multicolumn{2}{c|}{\multirow{2}{*}{IO}} \\
Cells ($T$) & \multicolumn{3}{c|}{} & \multicolumn{2}{c|}{} & \multicolumn{3}{c|}{} & \multirow{2}{*}{} & \multicolumn{3}{c|}{} & \multicolumn{2}{c|}{}  \\
\hline
Source & \multirow{2}{*}{MF} & \multirow{2}{*}{MF} & \multirow{2}{*}{GO} & \multirow{2}{*}{PF} & \multirow{2}{*}{PF} & \multirow{2}{*}{PF} & \multirow{2}{*}{CF} & \multirow{2}{*}{BC} & \multirow{2}{*}{PF} & \multirow{2}{*}{MF} & \multirow{2}{*}{MF} & \multirow{2}{*}{PC} & \multirow{2}{*}{US} & \multirow{2}{*}{CN} \\
Cells ($S$) &  &  &  &  &  & & & & & & & & & \\
\hline
Receptor & \multirow{2}{*}{AMPA} & \multirow{2}{*}{NMDA} & \multirow{2}{*}{GABA} & \multirow{2}{*}{AMPA} & \multirow{2}{*}{NMDA} & \multirow{2}{*}{AMPA} & \multirow{2}{*}{AMPA} & \multirow{2}{*}{GABA} & \multirow{2}{*}{AMPA} & \multirow{2}{*}{AMPA} & \multirow{2}{*}{NMDA} & \multirow{2}{*}{GABA} & \multirow{2}{*}{AMPA} & \multirow{2}{*}{GABA} \\
$(R)$ & & & & & & & & & & & & & &  \\
\hline
$\bar{g}_{R}^{(T)}$ & 0.18 & 0.025 & 0.028 & 45.5 & 30.0 & 0.7 & 0.7 & 1.0 & 0.7 & 50.0 & 25.8 & 30.0 & 1.0 & 0.18\\
\hline
$J_{ij}^{(T,S)}$ & 8.0 & 8.0 & 10.0 & 0.00004 & 0.00004 & 0.006 & 1.0 & 5.3 & 0.006 & 0.002 & 0.002 & 0.008 & 1.0 & 5.0 \\
\hline
$V_{R}^{(S)}$ & 0.0 & 0.0 & -82.0 & 0.0 & 0.0 & 0.0 & 0.0 & -75.0 & 0.0 & 0.0 & 0.0 & -88.0 & 0.0 & -75.0\\
\hline
$\tau_{R}^{(T)}$ & 1.2 & 52.0 & 7.0, 59.0 & 1.5 & 31.0, 170.0 & 8.3 & 8.3 & 10.0 & 8.3 & 9.9 & 30.6 & 42.3 & 10.0 & 10.0\\
\hline
$A_1$, $A_2$ & & & 0.43, 0.57 & & 0.33, 0.67 & & & & & & & & & \\
\hline
\end{tabular}
\end{table*}

In addition to the PSTHs of GR cells, under the many CS-US trials, one can also obtain a bell-shaped PSTH of a CN neuron [(bin-averaged) instantaneous individual firing rate $f_{\rm CN}(t)$ of the CN neuron]. The reciprocal of bottom-base width and the peak height of the bell curve correspond to timing degree ${\cal T}_d$ and strength $\cal S$ for the EBC, respectively. In this case, the (overall) learning efficiency degree ${\cal L}_e$ for the EBC is given by the product of ${\cal T}_d$ and $\cal S$. In this way, a set of $({\cal V}, {\cal L}_e)$ may be experimentally obtained for each species, and depending on the species, the set of
$({\cal V}, {\cal L}_e)$ may change. Then, for example in the case of 3 species of rabbit, rat, and mouse, with the three different data sets for $({\cal V}, {\cal L}_e)$, one can examine our main result (i.e., whether more variety in PSTHs of GR cells leads to more efficient learning for the EBC).

Finally, we make discussion on limitations of our present work and future works.
In the present work where the ISI between the onsets of CS and US was set at 500 msec, we investigated the effect of various temporal recoding of GR cells on the Pavlovian EBC. The acquisition rate and the timing degree and strength of CR have been known to depend on the ISI \cite{BN5,ISI}. Hence, in a future work, it would be interesting to study dependence of EBC on the ISI. Based on the results of our work, it would also be interesting to study extinction of CR, as a future work.
After acquisition of CR, we turn off the airpuff US. Then, the CR is expected to become gradually extinct via LTP at the PF-PC synapses \cite{EB3}.
In this work, we considered only the PF-PC synaptic plasticity. In the cerebellum, synaptic plasticity takes place at various synapses
\cite{Hansel,Gao} (e.g., MF-CN and PC-CN synapses \cite{Zheng}, PF-BC and BC-PC synapses \cite{Lennon}, and MF-GR cells synapses \cite{Dangelo2}).
Hence, it would be interesting to make a future study on the influence of various synaptic plasticity at various synapses on the cerebellar learning for the Pavlovian EBC; particularly, we are interested in the effect of the synaptic plasticity at the MF-CN synapse on the EBC \cite{EB3}.
In addition to change in $p_c$ (i.e., connection probability from GO to GR cells), one can vary synaptic inputs into the GR cells by changing NMDA receptor-mediated maximum conductances $\bar{g}_{\rm NMDA}^{\rm (GR)}$ and $\bar{g}_{\rm NMDA}^{\rm (GO)}$ \cite{BN5}.
Hence, as a future work, It would also be interesting to study the influence of NMDA receptor-mediated synaptic inputs on various recoding of GR cells and learning for the EBC by varying $\bar{g}_{\rm NMDA}^{\rm (GR)}$ and $\bar{g}_{\rm NMDA}^{\rm (GO)}$.

\section*{Acknowledgments}
This research was supported by the Basic Science Research Program through the National Research Foundation of Korea (NRF) funded by the Ministry of Education (Grant No. 20162007688).

\appendix*
\section{Parameter Values for The LIF Neuron Models and The Synaptic Currents}
In this appendix, we list two tables which show parameter values for the LIF neuron models in Subsec.~\ref{subsec:LIF} and the synaptic currents in   Subsec.~\ref{subsec:SC}. These values are adopted from physiological data \cite{Yama1,BN5}.

For the LIF neuron models,  the parameter values for the capacitance $C_X$, the leakage current $I_L^{(X)}$, the AHP current $I_{AHP}^{(X)}$, and the external constant current $I_{ext}^{(X)}$ are shown in Table \ref{tab:SingleParm}.

For the synaptic currents, the parameter values for the maximum conductance  $\bar{g}_{R}^{(T)}$, the synaptic weight $J_{ij}^{(T,S)}$, the synaptic reversal potential $V_{R}^{(S)}$, the synaptic decay time constant $\tau_{R}^{(T)}$, and the amplitudes $A_1$ and $A_2$ for the type-2 exponential-decay function in the granular layer, the Purkinje-molecular layer, and the other parts for the CN and IO are shown in Table \ref{tab:SynParm}.

\end{document}